\documentclass[12pt]{article}
\usepackage[margin = 1in, footskip = 0.25in]{geometry}
\usepackage{cite}  % BibTeX

\usepackage{amsmath}  % tfrac
\usepackage{amsfonts}  % mathbb
\usepackage[mathscr]{euscript}  % Euler script

\usepackage{graphicx}
\usepackage{subcaption}
\usepackage[font=small]{caption}

\usepackage{hyperref}

\newcommand{\half}{\tfrac{1}{2}}
\newcommand{\eighth}{\tfrac{1}{8}}

\newcommand{\del}{\delta}
\newcommand{\s}{\sigma}
\newcommand{\g}{\gamma}

\newcommand{\ket}{\rangle}
\newcommand{\bra}{\langle}

\newcommand{\la}{\mathscr{L}}

\title{Path-Integral Description of Stochastic Mechanics}
\author{Yoni BenTov}
\date{}

\numberwithin{equation}{section}

\begin{document}

% Title
\maketitle
\thispagestyle{myheadings}  % override the default header style in the title page

% Abstract
\begin{abstract}
I review the Feynman-Wiener path-integral formalism for diffusion with drift and jumps.
\end{abstract}

% Table of Contents
\tableofcontents
\newpage
\clearpage
\pagenumbering{arabic}
\section{Introduction}
The fundamental assumption of theoretical quantitative finance is that percentage changes in asset prices behave as if they were stochastic variables.\footnote{The fundamental assumption of \textit{empirical} quantitative finance is that one can get by using only tick data. Jim Simons verified that hypothesis to the tune of \$30 billion \cite{simons}.}
\\\\
Stochastic variables that evolve in time are called stochastic processes. A stochastic process can be described either by following an individual trajectory as that trajectory evolves randomly, or by following the distribution of possible values as that distribution evolves deterministically---the classic duality of fluid mechanics. Both pictures are unified in the incremental update rule for the distribution of \textit{paths}. 
\\\\
The path-integral formalism of Feynman \cite{feynman} and Wiener \cite{wiener} provides the foundation on which all of theoretical physics today is based.\footnote{Special relativity and quantum mechanics are unified: The action is manifestly Lorentz-invariant, and the path integral with fixed boundary conditions evolves by the Schrodinger equation. Renormalization is understood as the successive decoupling of high-energy or short-distance degrees of freedom. And sectors of different topology are to be weighted according to their topological invariants. I am not aware of any other formulation of fundamental physics that exhibits all of those features.} That formalism should also be the starting point for modeling stochastic processes. 
\\\\
What I want to demonstrate is how to think about stochastic processes in the way that high-energy theorists think about particle physics, collective behavior, and gravity.
\\\\
This paper will not be about financial assets per se, but it is important in the development of theory to have an application in mind. For applications of path integrals to options theory, see Dash \cite{dash}; to understand that options theory is but one application of the general theory of optimal control, see Merton \cite{merton}. 
\\\\
I will assume proficiency with partial differential equations \cite{haberman} and a passing familiarity with stochastic processes \cite{gardiner}. I will assume no previous knowledge of path integrals. I was raised on Srednicki\cite{srednicki} and Zee\cite{zeeQFT}, and to complement those books I recommend Feynman and Hibbs \cite{feynmanHibbs} and Bastianelli and van Nieuwenhuizen \cite{vanN}.\footnote{References on path integrals in quantum mechanics can be adapted to stochastic mechanics under the recognition that the Schrodinger equation and the Kolmogorov equation are related by a rotation of time in the complex plane. In quantum mechanics, the object that evolves according to the Schrodinger equation is called a transition amplitude, and its complex squared-magnitude is what is interpreted as a conditional probability. That the ``sum rule'' of probability theory applies to one object (the amplitude) and the ``closure rule'' applies to a different object (the amplitude squared) demonstrates that quantum mechanics requires an extension of probability theory. The clearest explanation was given by Feynman \cite{feynman}. Bell demonstrated that quantum correlations are too strong to be reproduced by a local classical model \cite{bell'sTheorem}. I do not like to use quantum mechanics as a crutch for stochastic mechanics, or vice versa; it is just that the mathematics is largely the same, and techniques can and should be carried across disciplines. It is neither reasonable nor necessary to ask a trader to learn quantum mechanics before applying the path-integral formalism.\label{ft:stochasticVsQuantum}}
\newpage
\section{Preliminary Considerations}
I would like to anchor the formalism of path integrals on the following aspects of partial differential equations and stochastic processes. 
\subsection{Conditional Probability and Evolution Laws}
Let $U(x,t|x_0,t_0)$ be the conditional probability that a stochastic process will have the value $x$ at a time $t$, given that it had the value $x_0$ at an earlier time $t_0 < t$.
\\\\
One possible starting point of stochastic mechanics is to postulate an evolution law for that conditional probability,
\begin{equation}
\partial_t U(x,t|x_0,t_0) = H(x,\partial_x,t)\, U(x,t|x_0,t_0),
\end{equation}
which is called a Kolmogorov forward equation or a Fokker-Planck equation.
\\\\
The quantity $U(x,t|x_0,t_0)$ could equally well be interpreted as the probability that, given the value $x$ at $t$, the process might have had the value $x_0$ at $t_0$. One could instead postulate an evolution law in the other set of variables,
\begin{equation}
\partial_{t_0} U(x,t|x_0,t_0) = K(x_0,\partial_{x_0},t)\, U(x,t|x_0,t_0)\;,
\end{equation}
which is called a Kolmogorov backward equation.\footnote{In physics, I had encountered the forward equation many times, but never the backward equation; finance was the first time I had heard of the backward equation. Kolmogorov introduced both equations as a pair \cite{kolmogorov}, and Bellman emphasized that the equations should be considered together\cite{bellman-AdaptiveControlProcesses}. Moreover, I had heard of the Feynman-Kac formula but never understood that it is the path-integral representation of the backward equation \cite{schulman}. One lesson I have learned throughout my career is that it is always, whenever possible, worth at least trying to revisit the original sources.}
\\\\
If the conditional probability is to satisfy the Chapman-Kolmogorov composition law,
\begin{equation}\label{eq:chapmanKolmogorov}
U(x_2,t_2|x_0,t_0) = \int_{-\infty}^\infty\!\! dx_1\;U(x_2,t_2|x_1,t_1)\,U(x_1,t_1|x_0,t_0)\;,
\end{equation}
then the forward and backward operators must be related:
\begin{equation}
K = -H^\dagger\;,
\end{equation}
where the adjoint is defined by integration by parts, dropping boundary terms. 
\\\\
By definition, the conditional probability also satisfies the equal-time condition
\begin{equation}
U(x,t|x_0,t) = \delta(x-x_0)\;,
\end{equation}
which just expresses that the notation for conditional probability means what it says. The function $U(x,t|x_0,t_0)$ is therefore also the Green function, evolution operator, or propagator,\footnote{The word ``propagator'' is also used to describe the two-point function, or autocorrelation, $E[x(t)\,x(t')]$.\label{ft:propagator}} for the Kolmogorov equations. 
\subsubsection{Initial Condition: Forward Evolution}
The typical situation in physics is to specify an initial condition and evolve forward in time. Take the distribution of values for this situation to be
\begin{equation}
\psi(x,t) = \int_{-\infty}^\infty\!\!dx_0\;U(x,t|x_0,t_0)\;\psi(x_0,t_0)\;,
\end{equation}
with initial condition
\begin{equation}
\psi(x,t_0) = v(x)\;.
\end{equation}
One prepares or measures a system at time $t_0$, knows that it has the distribution of values $v(x)$, and wants to know the distribution at later times. 
\subsubsection{Terminal Condition: Backward Evolution}
The typical situation in finance is to specify a terminal condition and evolve backward in time. Take the distribution of values for this situation to be
\begin{equation}
\phi(x,t) = \int_{-\infty}^{\infty}\!\!dx_f\;U(x_f,t_f|x,t)\;\phi(x_f,t_f)\;,
\end{equation}
with terminal condition
\begin{equation}
\phi(x,t_f) = u(x)\;.
\end{equation}
In this case, one knows the system at later times, because the function $u(x)$ represents an objective. Accordingly, $u(x)$ is variously called an objective function, criterion function, return function, or utility function. The problem is to choose decision variables so as to maximize or minimize the conditional expectation of $u(x)$ \cite{bellman-DynamicProgramming, merton}. Here I consider only the evolution law itself, without any decision variables.\footnote{That said, the reinterpretation of the principle of least action as a decision problem is the essence of the Hamilton-Jacobi formalism \cite{bellman-DynamicProgramming}.}
\subsection{Method of Characteristics}\label{sec:characteristics}
Consider a function $f(x,t)$, and a trajectory $x(t)$. In contrast to the typical situation in calculus, and motivated by Mandelbrot \cite{mandelbrot-FractalGeometryOfNature}, I do not want to assume that $x(t)$ is differentiable, or even continuous. I do, however, want to consider the difference between $x(t)$ and $x(t+dt)$ as $dt$ gets arbitrarily small.\footnote{The usual convention is that the symbol $dt$ innately implies the limit $dt \to 0$, in which case the stipulation that ``$dt$ gets arbitrarily small'' is redundant. Consider my usage to be compatible with that convention, perhaps overemphasizing it to the point of redundancy, as in ``ATM machine.''\label{ft:dt}} Following Ito, define the forward difference
\begin{equation}
dx(t) \equiv x(t+dt) - x(t)\;.
\end{equation}
Evaluate the function $f$ on neighboring points of the trajectory $x(t)$, and compute the change according to the above prescription:
\begin{equation}
df(x(t),t) \equiv f(x(t+dt),t+dt)-f(x(t),t) = f(x(t)+dx(t),t+dt)-f(x(t),t)\;.
\end{equation}
Taylor-expanding in $dt$ can proceed as usual, but expanding in $dx(t)$ must proceed in a few stages. 
\\\\
First I must sequester the discontinuous part, $\Delta x(t)$, from the continuous part, $\delta x(t) \equiv dx(t) - \Delta x(t)$. 
\\\\
Then I must choose a convention for the ordering of operators.\footnote{The series expansion of $f(x+y)$ about $y = 0$ is $f(x) + y\,\partial_x f(x) + \half y^2 \partial_x^2 f(x) + \ldots$ . What is the corresponding series expansion of $f(x+y(x))$? The answer is ambiguous, since $y(x)\,\partial_x f(x) \neq \partial_x[y(x)f(x)]$. Physical considerations aside, the mathematical ambiguity of ordering $x$s and $p$s in quantum mechanics could have been introduced in introductory calculus.} The standard convention is to put the expansion parameter before the derivative, leading to:
\begin{align}
&f(x(t) + \Delta x(t)+\delta x(t),t+dt) - f(x(t),t) = \nonumber\\
&\quad\left\{1+\delta x(t)\,\partial_x + \half\left[\delta x(t)\right]^2\partial_x^2 + \ldots\right\}f(x(t)+\Delta x(t),t+dt) - f(x(t),t)\;.
\end{align}
The stipulation that $\delta x(t)$ is the continuous part of $dx(t)$ means that
\begin{equation}
\delta x(t) \to 0\;\;\text{ as }\;\;dt \to 0\;.
\end{equation}
However slow that limiting procedure might be, it does mean that I can feel free to drop cross terms between $dt\,\partial_t f$ and $[\delta x(t)]^n\, \partial_x^n f$, for any $n > 0$. Therefore,
\begin{align}
df(x(t),t) &= \left\{1+\delta x(t)\, \partial_x + \half\left[\delta x(t)\right]^2\partial_x^2 + \ldots\right\}f(x(t)+\Delta x(t),t) - f(x(t),t) \nonumber\\
&+dt\,\partial_t f(x(t),t) + o(dt)\;,
\end{align}
where the ``small-oh'' symbol is a catch all for terms that go to zero more quickly than $dt$ as $dt \to 0$. 
\\\\
That expression for the total derivative is the essence of the method of characteristics.
\subsubsection{Trajectories that are Continuous and Differentiable}
First consider the elementary case, familiar to any student of calculus:
\begin{equation}
\delta x(t) = A(x(t),t)\,dt\;\;,\;\;\Delta x(t) = 0\;.
\end{equation}
The total derivative will terminate (to first order in $dt$) after the first partial derivative in $x$:
\begin{equation}
\frac{df(x(t),t)}{dt} = \left[A(x(t),t)\,\partial_x + \partial_t \right]f(x(t),t)\;.
\end{equation}
This is the usual realm of the method of characteristics: The first-order \textit{partial} differential equation\footnote{The right-hand side could be generalized with a source term, $Q(x,t)$, and a potential term, $V(x,t)f(x,t)$. One could generalize even further to ``quasilinear'' equations. For details, see Haberman \cite{haberman}.} (PDE)
\begin{equation}
[A(x,t)\,\partial_x + \partial_t]f(x,t) = 0
\end{equation}
can be reduced to an \textit{ordinary} differential equation (ODE)
\begin{equation}
\frac{df(x(t),t)}{dt} = 0
\end{equation}
on a trajectory $x(t)$ defined by 
\begin{equation}
\frac{dx(t)}{dt} = A(x(t),t)\;.
\end{equation}
The method has a second part, which is to apply the initial condition. I am concerned only with reformulating the PDE as an ODE, so I will stop here. 
\subsubsection{Trajectories that are Continuous but Nondifferentiable}
Now consider a continuous trajectory that goes to zero more slowly than $dt$ as $dt \to 0$. In particular, consider the special case
\begin{equation}\label{eq:squareRootTrajectory}
\delta x(t) = B(x(t),t)\,(dt)^{1/2},\;\;\Delta x(t) = 0\;.
\end{equation}
In that case, $[\del x(t)]^2$ is of order $dt$, and one must keep the second-order term in the total derivative. As a preliminary step, I have:
\begin{equation}
df(x(t),t) = dt\left[\partial_t + \half B(x(t),t)^2\partial_x^2\right]f(x(t),t) + \sqrt{dt}\;B(x(t),t)\,\partial_x f(x(t),t)\;.
\end{equation}
Without further structure, that is the furthest I can go.
\\\\
This is where stochasticity enters the picture. As a physical model, one might imagine that the trajectory is only a particular realization from a distribution of paths, with the particular property that
\begin{equation}
B(x(t),t) = \sigma(x(t),t)\;Z(t)\;,\;\;\text{ with } Z(t) \text{ drawn from } \rho(Z(t)) = \frac{1}{\sqrt{2\pi}}\;e^{-\half Z(t)^2}\;.
\end{equation}
Instead of the function $f(x,t)$ on a particular realization $x(t)$, one might consider the conditional expectation
\begin{equation}
F(x,t) \equiv E\left[f(x(\cdot),\cdot)\,|\,x(t)\,=\,x\right]\;,
\end{equation}
where that expectation is taken over the variable $Z(t)$. Since
\begin{equation}
\int_{-\infty}^\infty dZ\;\rho(Z)\;Z = 0\;\;\text{ and }\;\;\int_{-\infty}^\infty dZ\;\rho(Z)\;Z^2 = 1\;,
\end{equation}
I have
\begin{equation}
E\left[B(x(\cdot),\cdot)\,|\,x(t)\,=\,x\right] = 0\;\;\text{ and }\;\;E\left[B(x(\cdot),\cdot)^2\,|\,x(t)\,=\,x\right] = \sigma(x(t),t)^2\;.
\end{equation}
Therefore, the average convective derivative is
\begin{equation}
E\left[\left.\frac{df(x(\cdot),\cdot)}{dt}\,\right|x(t)\,=\,x\right] = \left[\partial_t + \half\sigma(x,t)^2\partial_x^2\right]F(x,t)\;.
\end{equation}
In that way, the method of characteristics can be applied to a second-order partial differential equation. The PDE
\begin{equation}
\left[\half \sigma(x,t)^2\,\partial_x^{\,2} + \partial_t\right]F(x,t) = 0
\end{equation}
can be reduced to a \textit{stochastic} differential equation (SDE)
\begin{equation}
df(x(t),t) = 0
\end{equation}
on an ensemble of trajectories defined by 
\begin{equation}
dx(t) = \sigma(x(t),t)Z(t)\sqrt{dt}\;,\;\;\text{ with } Z(t) \text{ drawn from } \rho(Z(t)) = \frac{1}{\sqrt{2\pi}}\;e^{-\half Z(t)^2}\;,
\end{equation}
under the identification
\begin{equation}
F(x,t) = E\left[f(x(\cdot),\cdot)\,|\,x(t)\,=\,x\right]\;.
\end{equation}
That is one of the oldest tricks in the book: An ensemble of Brownian paths diffuses. 
\subsubsection{Trajectories that are Discontinuous}
Finally, any pretense of ordinary calculus can be dispensed with by considering a purely discontinuous increment:
\begin{equation}
\del x(t) = 0\;,\;\;\Delta x(t) = C(x(t),t)\;.
\end{equation}
In that case, the total derivative will take the form
\begin{equation}
df(x(t),t) = f(x(t)+C(x(t),t),\,t) - f(x(t),t) + dt\;\partial_tf(x(t) + C(x(t),t),\,t)\;.
\end{equation}
Once again, without further structure, that is what it is. 
\\\\
And once again, stochasticity can take me further. Introduce a Poisson increment,\footnote{I trust the reader not to confuse the notation $P$ for Poisson increment with the various uses of $P$ for conditional probability and distribution of paths.}
\begin{equation}\label{eq:incrementalChangeInCount}
dP(t) \equiv \begin{cases} 
               1 & \text{with probability } \lambda(t)\;dt \\
               0 & \text{with probability } 1-\lambda(t)\;dt
             \end{cases}
\end{equation}
and consider the case
\begin{equation}
C(x(t),t) = \gamma(x(t),t)\;dP(t)\;.
\end{equation}
Using conditional expectations as before, I arrive at the average total derivative
\begin{align}
E\left[\left.\frac{df(x(\cdot),\cdot)}{dt}\,\right| x(t)\,=\,x\right] &= \lambda(t)\left[F(x+\gamma(x,t))-F(x,t)\right]+\partial_t F(x,t) \\
&= \left\{ \lambda(t)\left[e^{\,\gamma(x,t)\,\partial_x}-1\right] + \partial_t\right\}F(x,t)\;.\label{eq:evolutionLawForPoissonCharacteristics}
\end{align}
\subsection{Drift, Diffusion, and Jumps}
What I have shown in Sec.~\ref{sec:characteristics} is that a convective derivative of the form
\begin{align}
DF(x,t) &= \left\{\partial_t + A(x,t)\,\partial_x + \half \sigma(x,t)^2\,\partial_x^{\,2} + \lambda(t)\left[e^{\,\gamma(x,t)\,\partial_x} - 1\right] \right\}F(x,t)
\end{align}
can be understood as an average total derivative along an ensemble of trajectories defined by the increment
\begin{align}
&dx(t) = A(x(t),t)\,dt + B(x(t),t)\,\sqrt{dt} + C(x(t),t)\;,\;\;\text{with}\\
&B(x(t),t) = \sigma(x(t),t)\;Z(t)\;,\;\; Z(t)\;\text{drawn from}\;\frac{1}{\sqrt{2\pi}}\;e^{-\half Z(t)^2}\;;\;\;\text{and}\\
&C(x(t),t) = \gamma(x(t),t)\;dP(t)\;,\;\; dP(t) \equiv \begin{cases} 
               1 & \text{with probability } \lambda(t)\;dt \\
               0 & \text{with probability } 1-\lambda(t)\;dt
             \end{cases}\;.
\end{align}
Taking $F(x,t) = U(x_f,t_f|x,t)$, I may use the above to study the Kolmogorov backward equation 
\begin{equation}
DU(x_f,t_f|x,t) = 0\;,
\end{equation}
with the understanding that $D$ acts on the variables $x,t$. By use of the adjoint, I may also study the corresponding Kolmogorov forward equation. 
\subsection{Remarks}
This more advanced treatment of the total derivative is instructive for two reasons.
\\\\
First, the method of characteristics is not restricted to first-order quasilinear partial differential equations.\footnote{Bucy and Joseph \cite{bucyJoseph} point this out as well, in their Ch.~X. They also state, in their Ch.~IV: ``The representation theorem is the fundamental part of the theory of filtering from which all the theory can be derived. It is analogous to the `path integral' approach of Feynman for the solution of the Schr\"odinger equation in quantum theory and mathematically it is the explicit formula for the Radon-Nikodym derivative of one function space measure with respect to another.'' Given their explicit linking of things I wanted to learn with things I already knew, it is unfortunate that I have not gotten around to working through their book.} It is only on trajectories that are continuous and differentiable that one faces such a restriction.
\\\\
Second, the generalization of calculus to such trajectories has nothing per se to do with stochasticity. One could, for example, define the convective derivative along a Koch curve or Cantor set. Perhaps the point is pedantic, but I think that the terminology ``stochastic calculus'' is not appropriate, except when explicitly introducing stochasticity. Instead, I would follow Mandelbrot and call this generalized calculus ``fractal calculus,'' to be used in concert with fractal geometry \cite{mandelbrot-FractalGeometryOfNature}.\footnote{Not to be confused with fractional calculus \cite{fractionalCalculus}, which is a specific type of fractal calculus.}
\\\\
For the experts, I would like to remark that the Euler-Lagrange equations are the characteristics of the Hamilton-Jacobi equation \cite{bellman-AdaptiveControlProcesses}, and the renormalization-group beta functions are the characteristics of the Callan-Symanzik equation \cite{zinnJustin}.
\newpage
\section{Path-Integral Description of Poisson Noise}\label{sec:poissonPathIntegral}
It is time for path integrals. Reversing the order from my presentation of characteristics, I will begin with discontinuous trajectories.\footnote{Jumps are usually treated as an afterthought, despite their phenomenological importance \cite{gatheral}. Moreover, which of the following is more elementary: Counting nonnegative integers, or integrating over Gaussians?}
\subsection{Method of Sliding Planks}\label{sec:planks}
\begin{figure}[h]
  \centering
  \begin{subfigure}[t]{0.5\textwidth}
    \centering
    \includegraphics[height=2in]{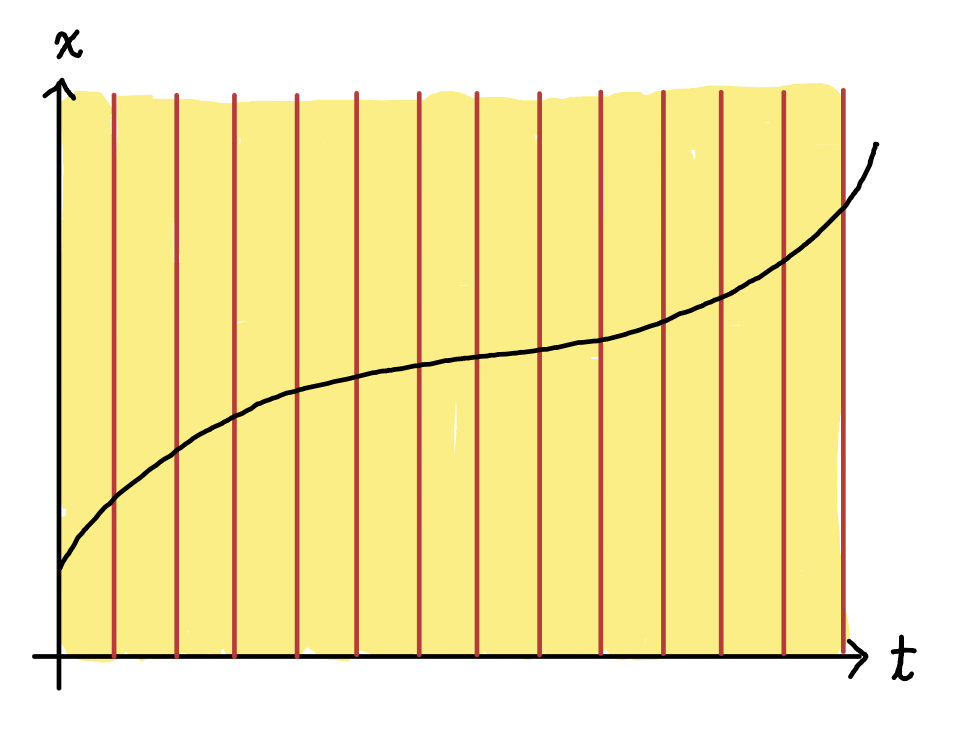}
    \caption{Continuous curve}
  \end{subfigure}~  % horizontal spacing
  \begin{subfigure}[t]{0.5\textwidth}
    \centering
    \includegraphics[height=2in]{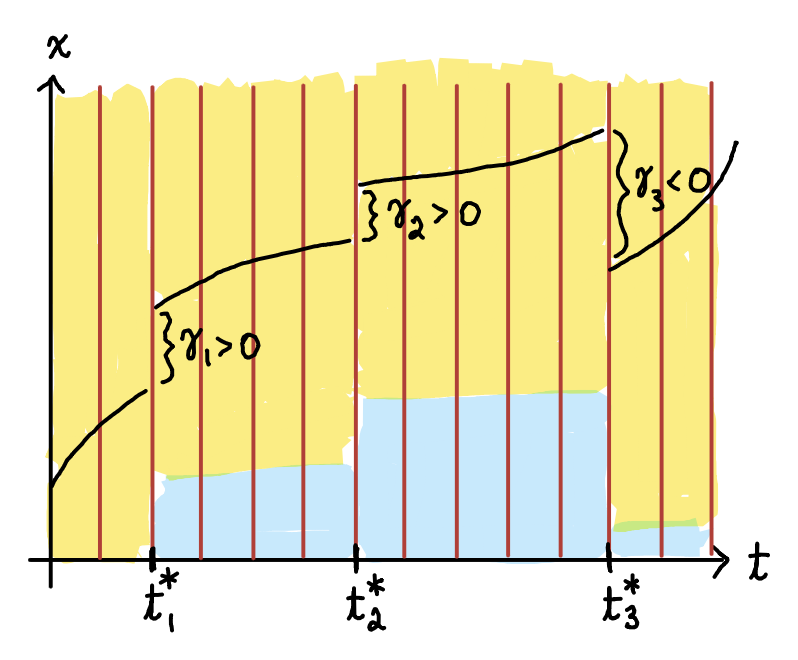}
    \caption{Curve with jumps}
  \end{subfigure}
\caption{Method of sliding planks. Slide all planks to the right of $t_1^*$ upward by an amount $|\gamma_1|$. Then slide all planks to the right of $t_2^*$ upward by an amount $|\gamma_2|$. Then slide all planks to the right of $t_3^*$ downward by an amount $|\gamma_3|$. And so on. The original continuous curve will be replaced by a new curve that experiences a jump condition across each $t_\alpha^*$.}\label{fig:planks}
\end{figure}
\phantom{}
\\
Picture a hardwood floor, and imagine painting a curve from left to right across the planks. Choose some plank in the interior, and imagine sliding every plank to the right of it by a fixed amount. Choose some other plank, and do the same. Look at what has become of the curve you painted: It is broken up by a sequence of discontinuous jumps. See Fig.~\ref{fig:planks}.
\\\\
 That is precisely Penrose's method of ``scissors-and-paste'' for gravitational shockwaves \cite{twistorQuantization, impulsiveWaves}. 
\\\\
So there are two complementary ways to think about jump processes: An ``active'' view, in which the process itself jumps as is ordinary conceived; and a ``passive view,'' in which the process goes on its merry way while the ground shifts beneath its feet. 
\\\\
What is so important about the passive view? As I explained in the introduction, what I set out to do for stochastic mechanics is analogous to what Feynman set out to do for quantum mechanics \cite{feynman}: Define the theory by the composition law for a distribution of paths. That means I want to write down actions in configuration space.\footnote{The terminology is borrowed from classical mechanics \cite{arnold}.} I know the action for Brownian motion (the Wiener measure, which I will get to in Sec.~\ref{sec:brownianPathIntegral})---what is the action for a Poisson process?
\\\\
The key insight of the passive view is that that question is actually two questions in one: (1) What is the action without jumps? (2) How do I couple the process to a broken floor?
\\\\
Formulating the question in that way brings me straight to invariance principles: If I insist on a formalism invariant under local translations, then I couple the process to a broken floor by situating the process on a background of translational defects. 
\\\\
A further invariance principle: Scale invariance. Insisting on a formalism invariant under local dilations and specializing to a constant background field will lead to mean-reversion. 
\\\\
Elaboration will have to wait until Sec.~\ref{sec:gaugeTheory}. Until you eat your spinach, you will not be ready for gauge invariance. 
%\\\\
\subsection{Distribution of Paths without Jumps}\label{sec:firstPathIntegral}
In the method of sliding planks, the horizontal axis is interpreted as time, labeled by $t$, in some fixed interval, $[t_0,t_f]$; and the vertical axis describes the value of the process under consideration, denoted by $x(t)$. 
\\\\
The simplest case to consider is that the process does nothing until it happens to encounter a cut-and-paste\footnote{Evidently the British prefer nouns, whereas Americans prefer verbs. The word ``paste'' is both.} boundary. I start by painting the curve 
\begin{equation}\label{eq:constantPath}
x(t) = \text{constant} \implies \dot x(t) = 0\;.
\end{equation}
The distribution of paths, $P(x(\cdot))$, in the absence of jump discontinuities, is 
\begin{equation}\label{eq:firstDistributionOfPaths}
P(x(\cdot)) = \delta\!\left(\dot x(\cdot)\right)\; \equiv \!\!\prod_{t\,=\,t_0\,+\,0^+}^{t_f\,-\,0^+}\!\! \delta\!\left(\dot x(t)\right)\;.
\end{equation}
There you go: Your first distribution of paths.
\subsubsection{Conditional Probability}\label{sec:conditionalProbabilityForConstantPath}
To instill some preliminary comfort with the mathematics, let me belabor the meaning of Eq.~(\ref{eq:firstDistributionOfPaths}) by turning the distribution of paths into a conditional probability. 
\\\\
Given the distribution of paths, $P(x(\cdot))$, one forms the conditional probability $U(x_f,t_f|x_0,t_0)$ by integrating over all paths subject to the specified boundary conditions:
\begin{equation}
U(x_f,t_f|x_0,t_0) = \int_{x(t_0)\,=\,x_0}^{x(t_f)\,=\,x_f}\!\!\!\!\mathscr Dx(\cdot)\;P(x(\cdot))\;.
\end{equation}
What does the integration over paths mean? In particular, what does the symbol $\mathscr Dx(\cdot)$ mean?
\\\\
The integration over paths is to be understood sequentially. You start at $x(t_0)\,=\,x_0$, consider a time step $\Delta t$, and end up with an intermediate value $x_1$ at the time $t_1 = t_0 + \Delta t$. That intermediate value could be anything in the domain of possible values of $x(t_1)$, weighted according to the specified distribution of paths $P(x(\cdot))$. 
\\\\
That construction is to be iterated $N$ times, until the specified terminal condition has been reached: $t_N \equiv t_f$, $x(t_N) = x(t_f) = x_f$. 
\\\\
The limiting procedure is that the time step, $\Delta t$, is taken arbitrarily small, and the number of steps, $N$, is taken arbitrarily large; with the product $N\Delta t$ held fixed to the total temporal interval of interest:\footnote{One could also consider unequal time steps, i.e., $\Delta t_i \neq \Delta t_j$ for $i \neq j$, provided that the largest of those is taken to zero in a modified version of the limit in Eq.~(\ref{eq:limit}). One reason to do that would be to preserve invariance under temporal reparametrizations (see Polyakov, Ch.~9 \cite{polyakov}). I will touch on temporal reparametrizations in Sec.~\ref{sec:temporalReparametrization}.}
\begin{equation}\label{eq:limit}
\Delta t \to 0\;,\;\;N \to \infty\;,\;\; N\Delta t = t_f-t_0\;.
\end{equation}
To obtain a finite answer, an overall normalization factor has to be fixed after the fact; in any case, the conditional probability is supposed to be normalized, so the cancellation of divergences such that the combined limit in Eq.~(\ref{eq:limit}) results in something finite provides a check on the formalism. 
\\\\
Time to put words into action:
\begin{align}
U(x_f,t_f|x_0,t_0) &= C(t_f,t_0)\int_{-\infty}^\infty \!\!dx_{N-1} \ldots \int_{-\infty}^\infty\!\! dx_1\;\delta(x_N-x_{N-1})\,\ldots\, \delta(x_2-x_1)\,\delta(x_1-x_0) \label{eq:firstPathIntegralCalculation}\\
&= C(t_f,t_0)\int_{-\infty}^\infty \!\!dx_{N-1} \ldots\int_{-\infty}^\infty\!\! dx_2\;\delta(x_N-x_{N-1})\,\ldots\,\delta(x_3-x_2)\,\delta(x_2-x_0) \\
&\ldots \nonumber\\
&= C(t_f,t_0)\int_{-\infty}^\infty \!\!dx_{N-1}\,\delta(x_N-x_{N-1})\,\delta(x_{N-1}-x_0) \\
&= C(t_f,t_0)\;\delta(x_f-x_0)\;.\qquad(x_f \equiv x_N)
\end{align}
The pertinent normalization condition is $\int_{-\infty}^{\infty}\!\! dx_f\,U(x_f,t_f|x_0,t_0) = 1$, so in this case the normalization factor is just $C(t_f,t_0) = 1$. 
\\\\
Note that in going from Eq.~(\ref{eq:firstDistributionOfPaths}) to Eq.~(\ref{eq:firstPathIntegralCalculation}), I factored out an overall $\Delta t$ from each delta function. If I had not done so, the normalization factor would have had to cancel them all out, leading to the same result.
\\\\
So you have computed your first path integral:
\begin{equation}\label{eq:firstPathIntegral}
U(x_f,t_f|x_0,t_0) = \int_{x(t_0)\,=\,x_0}^{x(t_f)\,=\,x_f}\!\!\!\!\mathscr Dx(\cdot)\;\delta\!\left(\dot x(\cdot)\right) = \delta(x_f-x_0)\;.
\end{equation}
As advertised, all of the probability is concentrated around a path that starts and ends at the same value. 
\\\\
On the one hand, Eq.~(\ref{eq:firstPathIntegral}) is trivial and obvious. On the other hand, deriving it from iterated integration helps train the mind to see functional integration as part of a workaday toolkit instead of vague abstraction.
\subsection{Distribution of Paths with a Fixed Collection of Jumps}\label{sec:planksAsMath}
%\\\\
The locations of a fixed number, $n$, of jump discontinuities are implemented by a collection of times, $\{t_\alpha^*\}_{\alpha\,=\,1}^n$, in a fixed ordering:
\begin{equation}\label{eq:fixedOrdering}
t_0 < t_1^* < t_2^* < \ldots < t_n^* < t_f\;.
\end{equation}
%\\\\
The painted curve of Eq.~(\ref{eq:constantPath}) will do nothing until it hits one of the $t_\alpha^*$, at which time the value of the curve will shift by some amount, $\gamma(x(t_\alpha^*),t_\alpha^*)$, called the jump size:\footnote{In gravity, the jump size has a fixed sign, related to the positivity of energy. In stochastic mechanics, to the best of my knowledge, there is no such condition. The function $\g(x,t)$ is an arbitrary real number for each value of $x$ and $t$.}
\begin{equation}\label{eq:pathWithJumps}
\dot x(t) = \sum_{\alpha\,=\,1}^n\gamma(x(t_\alpha^*),t_\alpha^*)\;\delta (t-t_\alpha^*)\;.
\end{equation}
Going forward, I will drop any possible dependence of the jump size on the value of the field, in which case I can just write 
\begin{equation}
\gamma(x(t_\alpha^*),t_\alpha^*) = \gamma_\alpha\;.
\end{equation}
The situation is as in Fig.~\ref{fig:purePoisson}.
\begin{figure}[h]
  \centering
  \begin{subfigure}[t]{0.5\textwidth}
    \centering
    \includegraphics[height=0.75in]{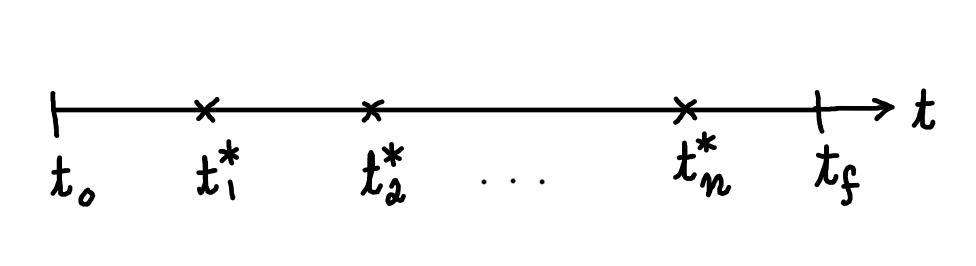}
    \caption{A configuration of the generating process is a collection of jump times in a fixed order.}
  \end{subfigure}~  % seriously
  \begin{subfigure}[t]{0.5\textwidth}
    \centering
    \includegraphics[height=2in]{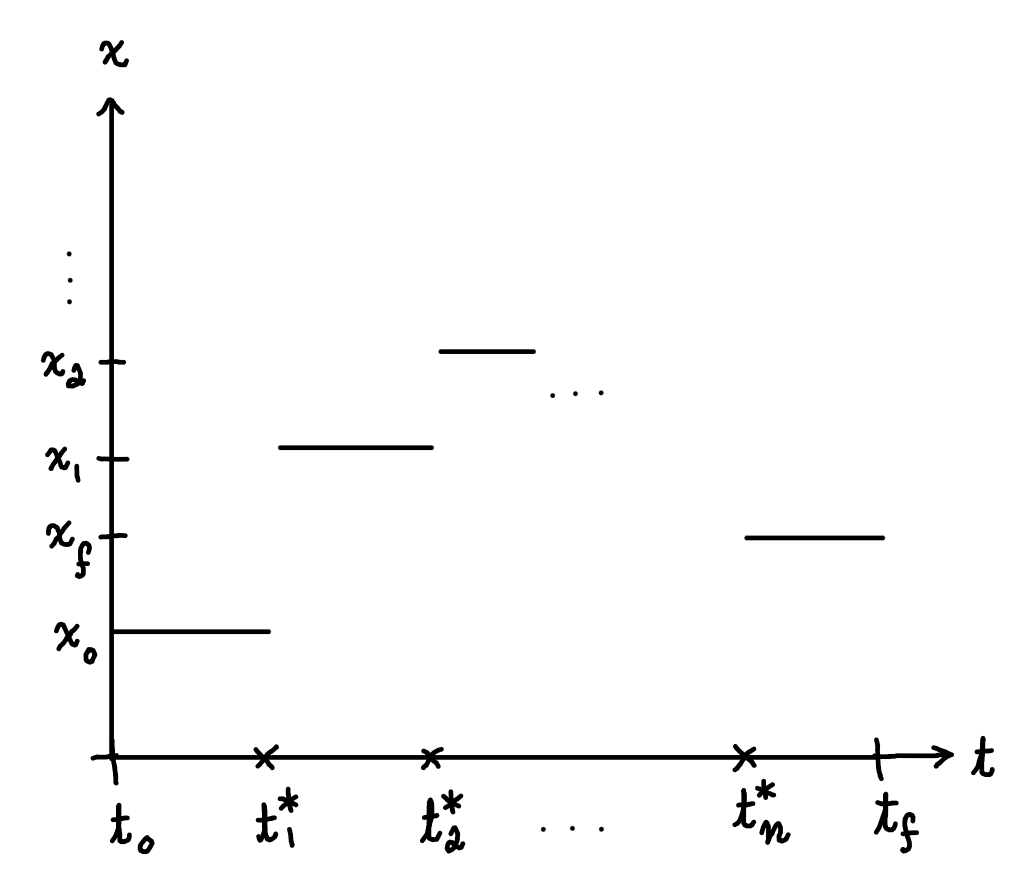}
    \caption{A path that does nothing until it hits a cut-and-paste boundary. Because the jump sizes are arbitrary and time-dependent, the final value may be larger or smaller than any preceding values.}
  \end{subfigure}
\caption{A stochastic process on a fixed realization of Poisson noise.}
\end{figure}\label{fig:purePoisson}
\\\\
The new distribution of paths will be:\footnote{It has only recently come to my attention that Feynman contemplated elementary path integrals of this nature. See Problem 2-6 (the ``zigzag'' paths) and Eq.~(12.19) (the characteristic functional for Poisson noise) in Feynman and Hibbs \cite{feynmanHibbs}.}
\begin{equation}\label{eq:distributionOfPathsForPureJumpProcess}
P(x(\cdot)\,|\,\{t_{\alpha}^*\}_{\alpha\,=\,1}^n) = \delta\!\left(\dot x(\cdot) - \sum_{\alpha\,=\,1}^n\gamma_\alpha\, \delta(\cdot-t_\alpha^*)\right)\;.
\end{equation}
Instead of trying to proceed with iterated integration as in Sec.~\ref{sec:conditionalProbabilityForConstantPath}, I will introduce the Fourier representation of the delta function. 
\begin{align}
P(x(\cdot)\,|\,\{t_{\alpha}^*\}_{\alpha\,=\,1}^n) &= \int\mathscr D k(\cdot)\;e^{\,i\int_{t_0}^{t_f}\!dt\;k(t)\left[\dot x(t) - \sum_{\alpha\,=\,1}^n\gamma_\alpha\,\delta(t-t_\alpha^*)\right]} \\
&= \int\mathscr D k(\cdot)\;e^{\,i\int_{t_0}^{t_f}\!dt\;k(t)\,\dot x(t)}\;e^{-i\sum_{\alpha\,=\,1}^n \gamma_\alpha\,k(t_\alpha^*)}\;.\label{eq:distributionOfPathsForPoissonAfterIntroducingFourierRepresentation}
\end{align}
The goal is now to make sense of that last factor, $e^{-i\sum_{\alpha\,=\,1}^n \gamma_\alpha\,k(t_\alpha^*)}$, upon averaging over the number of jumps and when those jumps occur. 
\subsection{Distribution of Paths with Random Jumps}\label{sec:randomJumps}
The fixed configuration in Sec.~\ref{sec:planksAsMath} has a fixed number of jumps at specified times in a fixed ordering. First I will average over the times, and then average over the number of jumps. 
%\\\\
\subsubsection{Average over Jump Times for a Fixed Number of Jumps}\label{sec:averageOverJumpTimes}
Suppose that the times $\{t_\alpha^*\}_{\alpha\,=\,1}^n$ are drawn from the joint distribution
\begin{equation}
p(t_1^*,\ldots,t_n^*)\;. 
\end{equation}
Consider the average value of an arbitrary function, $f(t_1^*,\ldots, t_n^*)$, given the fixed ordering in Eq.~(\ref{eq:fixedOrdering}):
\begin{align}
E_{\{t^*_\alpha \}_{\alpha\,=\,1}^n}\left[f(t_1^*,\ldots,t_n^*)\right] &\equiv \int_{t_0}^{t_f}\!\! dt_1^*\ldots \int_{t_0}^{t_f}\!\! dt_n^*\;\;p(t_1^*,\ldots,t_n^*)\; \times \nonumber\\
&\qquad\qquad f(t_1^*,\ldots,t_n^*)\;\Theta(t_n^*-t_{n-1}^*)\;\ldots\; \Theta(t_2^*-t_1^*)\;,
\end{align}
where $\Theta(t)$ is the step function. By relabeling integration variables, write the $n$-fold integral in a form manifestly symmetric under permutations:
\begin{align}
&E_{\{t^*_\alpha \}_{\alpha\,=\,1}^n}\left[f(t_1^*,\ldots,t_n^*)\right] = \frac{1}{n!}\int_{t_0}^{t_f}\!\! dt_1^*\ldots \int_{t_0}^{t_f}\!\! dt_n^* \;\times \nonumber\\
&\qquad\left[\phantom{\frac{}{}} p(t_1^*,\ldots,t_n^*)\;f(t_1^*,\ldots,t_n^*)\;\Theta(t_n^*-t_{n-1}^*)\;\ldots\; \Theta(t_2^*-t_1^*) \right. \nonumber\\
&\qquad\qquad\left.\phantom{\frac{}{}}+\; (\text{all permutations of $\{t_1^*,\ldots,t_n^*\}$)}\;\right]\;.
\end{align}
For the special case in which \textit{both} the distribution $p$ \textit{and} the function $f$ are invariant under permutations of their arguments, the average becomes
\begin{align}
&E_{\{t^*_\alpha \}_{\alpha\,=\,1}^n}\left[f(t_1^*,\ldots,t_n^*)\right] = \frac{1}{n!}\int_{t_0}^{t_f}\!\! dt_1^*\ldots \int_{t_0}^{t_f}\!\! dt_n^* \;p(t_1^*,\ldots,t_n^*)\;f(t_1^*,\ldots,t_n^*) \;\times \nonumber\\
&\qquad\left[\phantom{\frac{}{}} \Theta(t_n^*-t_{n-1}^*)\;\ldots\; \Theta(t_2^*-t_1^*)\;+\; (\text{all permutations of $\{t_1^*,\ldots,t_n^*\}$)}\; \right]\;.
\end{align}
But the product of step functions summed over all permutations of their arguments is just 1. Therefore, for the special case of permutation-symmetric $p$ and $f$, the average simplifies:
\begin{equation}
E_{\{t^*_\alpha \}_{\alpha\,=\,1}^n}\left[f(t_1^*,\ldots,t_n^*)\right] = \frac{1}{n!}\int_{t_0}^{t_f}\!\! dt_1^*\ldots \int_{t_0}^{t_f}\!\! dt_n^* \;p(t_1^*,\ldots,t_n^*)\;f(t_1^*,\ldots,t_n^*)\;.
\end{equation}
More than any particular analytic formula, the defining property of Poisson noise is that its events are independent and identically distributed.\footnote{A good summary of the Poisson process can be found in Cox \& Lewis \cite{coxLewis}.} If the jump times are independent, then their joint distribution factorizes into a product of individual distributions:
\begin{equation}\label{eq:productOfIndependentFactors}
p(t_1^*,\ldots,t_n^*) = p_1(t_1^*)\ldots p_n(t_n^*)\;.
\end{equation}
If the jump times are distributed identically, then all of those individual functions are the same:
\begin{equation}\label{eq:productOfIndependentAndIdenticalFactors}
p(t_1^*,\ldots,t_n^*) = p_1(t_1^*)\ldots p_1(t_n^*)\;.
\end{equation}
That is invariant under permutations of its arguments. 
\\\\
As for the function $f$, the particular case of interest in Sec.~\ref{sec:planksAsMath} was
\begin{equation}\label{eq:functionOfInterest}
f(t_1^*,\ldots,t_n^*) = e^{\,\kappa(t_1)}\;\ldots\;e^{\,\kappa(t_n)}\;,
\end{equation}
with $\kappa(t) = -i\gamma(t)k(t)$. That is also a permutation-invariant product of independent and identical factors, as in Eq.~(\ref{eq:productOfIndependentAndIdenticalFactors}).
\\\\
In this further specialized case, the average over jump times takes the form of a product:
\begin{align}
E_{\{t^*_\alpha \}_{\alpha\,=\,1}^n}\left[f(t_1^*,\ldots,t_n^*)\right] &= \frac{1}{n!}\int_{t_0}^{t_f}\!\! dt_1^*\ldots \int_{t_0}^{t_f}\!\! dt_n^* \;p(t_1^*,\ldots,t_n^*)\;f(t_1^*,\ldots,t_n^*) \nonumber \\
&= \frac{1}{n!}\int_{t_0}^{t_f}\!\! dt_1^*\;p_1(t_1^*)\;e^{\;\kappa(t_1^*)}\ldots \int_{t_0}^{t_f}\!\! dt_n^*\;p_1(t_n^*)\;e^{\;\kappa(t_n^*)} \\
&= \frac{1}{n!}\left[\int_{t_0}^{t_f}\!\! dt\;p_1(t)\;e^{\,\kappa(t)}\right]^n\;. \label{eq:averageOverSectorOfFixedNumberOfJumps}
\end{align}
\subsubsection{Average over Number of Jumps}\label{sec:averageOverNumberOfJumps}
In Sec.~\ref{sec:averageOverJumpTimes}, I averaged over all configurations of a fixed number of jumps. How should the sectors of fixed number be combined into a total average?
\\\\
Before answering that question, I need to be more precise about what a total average might mean. A particular function $f(t_1^*,\ldots,t_n^*)$ has a fixed number of arguments. So to average over the number of jumps, I need to consider a \textit{sequence} of functions of a \textit{variable} number of arguments:
\begin{equation}
f_0\;,\;\; f_1(t_1^*)\;,\;\; f_2(t_1^*,t_2^*)\;,\;\;\ldots\;,\;\; f_n(t_1^*,\ldots,t_n^*)\;,\;\;\ldots\;.
\end{equation}
Fortunately, the particular functions of interest are of the form in Eq.~(\ref{eq:functionOfInterest}). So I already have a sequence of functions to consider:
\begin{equation}
f_0 = 1\,,\,\, f_1(t_1^*) = e^{\,\kappa(t_1^*)}\,,\,\,f_2(t_2^*) = e^{\,\kappa(t_1^*)}e^{\,\kappa(t_2^*)}\,,\,\,\ldots\,,\,\,f_n(t_1^*,\ldots,t_n^*) = e^{\,\kappa(t_1^*)}\ldots e^{\,\kappa(t_n^*)}\,,\,\,\ldots\;.
\end{equation}
Given that, I can define a total average over sectors of different numbers of jumps as
\begin{equation}
E_{t^*}\!\left[f(\cdot)\right] \equiv \sum_{n\,=\,0}^\infty w_n\;E_{\{t_\alpha^*\}_{\alpha\,=\,1}^n}\left[f_n(t_1^*,\ldots,t_n^*)\right]\;,
\end{equation}
with some unspecified weights $\{w_n\}_{n\,=\,0}^\infty$, and with the average within a sector of fixed $n$ given by Eq.~(\ref{eq:averageOverSectorOfFixedNumberOfJumps})\;.
\\\\
What should those weights be?
\\\\
Once again, the defining property of Poisson noise is that its events are independent and identically distributed. If the sectors of fixed $n$ are to be combined without fear or favor, then the weights should be equal to a common constant:\footnote{I expect, given the gauge theory interpretation, that under some circumstances it would be more appropriate to try $w_n = (-1)^n$.}
\begin{equation}
w_0 = w_1 = w_2 = \ldots \equiv C\;.
\end{equation}
Therefore, the total average will exponentiate:
\begin{align}
E_{t^*}\!\left[f(\cdot)\right] &= C\sum_{n\,=\,0}^{\infty} \frac{1}{n!}\left[\int_{t_0}^{t_f}\!\!dt\;p_1(t)\;e^{\,\kappa(t)}\right]^n = C\;e^{\,\int_{t_0}^{t_f}\!dt\;p_1(t)\;e^{\,\kappa(t)}}\;.
\end{align}
The normalization factor, $C$, is to be fixed by demanding that $E_{t^*}(1) = 1$. Setting $\kappa(t) = 0$ in the above gives:
\begin{equation}
E_{t^*}(1) = C\;e^{\,\int_{t_0}^{t_f}\!p_1(t)\,dt} \equiv 1 \implies C = e^{-\int_{t_0}^{t_f}\!p_1(t)\,dt}\;.
\end{equation}
Therefore, 
\begin{equation}\label{eq:precursorToCampbellHardy}
E_{t^*}\!\left[f(\cdot)\right] = e^{\,\int_{t_0}^{t_f}\!dt\;p_1(t)\,\left( e^{\,\kappa(t)} - 1\right)}\;.
\end{equation}
\subsubsection{Campbell-Hardy Formula}\label{sec:campbellHardy}
With the $f$ of Eq.~(\ref{eq:functionOfInterest}) and $\kappa(t) = -i\gamma(t)k(t)$, Eq.~(\ref{eq:precursorToCampbellHardy}) is the average of the last factor in Eq.~(\ref{eq:distributionOfPathsForPoissonAfterIntroducingFourierRepresentation}). The only remaining wrinkle to iron out is the relation of the distribution $p_1(t)$ to the Poisson event rate, $\lambda(t)$, introduced back in Eq.~(\ref{eq:incrementalChangeInCount}). 
\\\\
The terminology ``event rate'' gives away the game: $p_1(t)$, as introduced in Sec.~\ref{sec:averageOverJumpTimes}, is the probability that a jump will be found between $t$ and $t+dt$. But the probability that a jump will be found between $t$ and $t+dt$ was defined by Eq.~(\ref{eq:incrementalChangeInCount}) to be $\lambda(t)\,dt$. So
\begin{equation}
p_1(t) = \lambda(t)\;,
\end{equation}
and Eq.~(\ref{eq:precursorToCampbellHardy}) becomes
\begin{equation}\label{eq:campbellHardy}
E_{t^*}\!\left(e^{-i\sum_{t^*}\gamma(t^*)\,k(t^*)}\right) = e^{\,\int_{t_0}^{t_f}\!\!dt\;\lambda(t)\;\left( e^{-i\gamma(t)k(t)} - 1\right)}\;.
\end{equation}
That expression is called the Campbell-Hardy formula, or Campbell's theorem \cite{kingman}.
\\\\
The phase-space distribution of paths for a process driven exclusively by Poisson noise is therefore
\begin{equation}\label{eq:phaseSpaceDistributionOfPathsForPoisson}
P(x(\cdot),\,k(\cdot)) \equiv e^{\,i\int_{t_0}^{t_f}dt\;k(t)\,\dot x(t)} E_{t^*}\!\left(e^{-i\sum_{t^*} \gamma(t^*)\,k(t^*)}\right) \equiv e^{\int_{t_0}^{t_f}\! dt\;\widetilde\la(x(t),\,k(t))}\;,
\end{equation}
with phase-space action density\footnote{I am unaware of a standard name for this quantity. The portion $\mathscr H(x(t),\,k(t)) \equiv -ik(t)\,\dot x(t)+\widetilde\la(x(t),\,k(t)) = \lambda(t)\,e^{-i\gamma(t)\,k(t)}$ is the Hamiltonian, or, better, the Kolmogorovian. The overall configuration-space action, defined by a Legendre transformation, has an additional, conventional, overall sign. See App.~\ref{sec:legendre} and footnote~\ref{ft:stochasticVsQuantum}.}
\begin{equation}\label{eq:phaseSpaceActionDensityForPoisson}
\widetilde\la(x(t),k(t)) = ik(t)\,\dot x(t) + \lambda(t)\;e^{-i\gamma(t)\,k(t)}\;,
\end{equation}
up to an overall field-independent additive constant.
%\\\\
\subsection{Phase-Space Path Integral from Poisson Increments}\label{sec:phaseSpacePoisson}
I derived Eq.~(\ref{eq:campbellHardy}) by specifying a fixed configuration of jumps over the entire interval $[t_0,t_f]$, then averaging over all possible configurations. I should be able to arrive at the same right-hand side by instead considering one arbitrarily small temporal increment at a time. 
\\\\
Over each arbitrarily small interval between $t$ and $t+\Delta t$, the counting process will either change by 1 or remain unchanged:
\begin{equation}\label{eq:poissonIncrement}
\Delta P(t) \equiv P(t+\Delta t)-P(t) = \begin{cases} 
              1 & \text{with probability } \lambda(t)\;\Delta t \\
              0 & \text{with probability } 1-\lambda(t)\;\Delta t
            \end{cases}\;+\; o(\Delta t)\;.
\end{equation}
On a fixed realization of the Poisson process, for any function $\kappa(t)$, 
\begin{equation}
\sum_{t^*} \kappa(t^*) = \lim_{\substack{\Delta t \,\to\, 0\,,\; N \,\to\, \infty \\ N\Delta t \,=\, t_f-t_0}}\sum_{i\,=\,0}^{N-1}\Delta P_i\,\kappa(t_i) \equiv \int_{t_0}^{t_f}\!\! dP(t)\;\kappa(t)\;.
\end{equation}
In subsequent manipulations the limit should be understood.
\\\\
The assertion of an alternative derivation of Eq.~(\ref{eq:campbellHardy}) is that averaging over configurations is equivalent to averaging over increments:
\begin{equation}\label{eq:poissonAssertion}
E_{t^*}\!\left(e^{\,\sum_{t^*}\kappa(t^*)}\right) = E_{P(\cdot)}\left(e^{\,\int_{t_0}^{t_f}\! dP(t)\,\kappa(t)}\right)\;,
\end{equation}
with the averaging over increments done as prescribed by Eq.~(\ref{eq:poissonIncrement}). Let me take nothing on faith and dutifully carry out the procedure:
\begin{align}
E_{P(\cdot)}\left[e^{\,\int_{t_0}^{t_f}\! dP(t)\;\kappa(t)}\right] &= \prod_{i\,=\,0}^{N-1}\sum_{\Delta P_i\,=\,0}^\infty\left[\lambda(t_i)\,\Delta t\,\delta_{\Delta P_i,1}+\left(1-\lambda(t_i)\Delta t\right)\delta_{\Delta P_i,0} + o(\Delta t)\right]\;e^{\,\Delta P_i\,\kappa(t_i)} \label{eq:incrementalEvolutionOperatorForPoissonNoise}\\
&= \prod_{i\,=\,0}^{N-1}\left[\lambda(t_i)\,\Delta t\,e^{\,\kappa(t_i)} + (1-\lambda(t_i)\,\Delta t)\,e^0 \right] \\
&= \prod_{i\,=\,0}^{N-1}\left[1+\lambda(t_i)\,\Delta t\left(e^{\,\kappa(t_i)}-1\right)\right] \\
&= e^{\sum_{i\,=\,0}^{N-1}\lambda(t_i)\,\Delta t\,\left(e^{\,\kappa (t_i)}-1\right) + o(\Delta t)} \\
&= e^{\,\int_{t_0}^{t_f} \!\lambda(t)\,dt\,\left(e^{\,\kappa(t)}-1\right)}\;. \label{eq:RHSofPoissonAssertion}
\end{align}
Inserting $\kappa(t) = -i\gamma(t)\,k(t)$ into Eq.~(\ref{eq:RHSofPoissonAssertion}), I indeed obtain the right-hand side of Eq.~(\ref{eq:campbellHardy}). 
\subsection{Phase-Space Path Integral from Kolmogorov Equation}\label{sec:phaseSpaceFromKolmogorov}
A third way to derive the phase-space path integral is to start from the pertinent Kolmogorov equation. In Eq.~(\ref{eq:evolutionLawForPoissonCharacteristics}) I arrived at the backward equation; the corresponding forward equation is
\begin{equation}
\partial_t\psi(x,t) = \lambda(t)\left(e^{-\gamma\,\partial_x}-1\right)\psi(x,t)\;,
\end{equation}
where for simplicity I specialized to a constant jump size. Compare the distribution at a time $t_N$ to its value at the immediately preceding time, $t_{N-1} = t_N-\Delta t$:
\begin{align}
\psi(x_N,t_N) &= \psi(x_N,t_{N-1}+\Delta t) = \left[1+\Delta t\,\partial_{t_{N-1}} + o(\Delta t)\right]\psi(x_N,t_{N-1}) \\
&= \left[1+\Delta t\,\lambda(t_{N-1})\left(e^{-\gamma\,\partial_{x_N}}-1\right)\right]\psi(x_N,t_{N-1})\;. \label{eq:preliminaryStepInPoissonComposition}
\end{align}
Introduce the Fourier transform, 
\begin{equation}
\psi(x_N,t_{N-1}) = \int_{-\infty}^\infty\frac{dk_{N-1}}{2\pi}\;e^{\,ik_{N-1}x_{N}}\;\widetilde\psi(k_{N-1},t_{N-1})\;,
\end{equation}
and have $e^{-\gamma\,\partial_{x_N}}$ act inside the $k_{N-1}$ integral:
\begin{equation}
e^{-\gamma\partial_{x_N}}\psi(x_{N},t_{N-1}) = \int_{-\infty}^{\infty}\frac{dk_{N-1}}{2\pi}\;e^{-i\gamma k_{N-1}}\;e^{\,ik_{N-1}x_{N}}\;\widetilde\psi(k_{N-1},t_{N-1})\;.
\end{equation}
Now undo the Fourier tranform using clever notation:
\begin{equation}
\widetilde\psi(k_{N-1},t_{N-1}) = \int_{-\infty}^\infty\!\! dx_{N-1}\;e^{-ik_{N-1}x_{N-1}}\psi(x_{N-1},t_{N-1})\;.
\end{equation}
Interchanging the order of integrations over $k_{N-1}$ and $x_{N-1}$, I find
\begin{align}
\psi(x_N,t_{N-1}) = \int_{-\infty}^\infty dx_{N-1}\left[\int_{-\infty}^{\infty}\frac{dk_{N-1}}{2\pi}\;e^{-i\gamma k_{N-1}}\;e^{\,ik_{N-1}(x_{N}-x_{N-1})} \right]\psi(x_{N-1},t_{N-1})\;,
\end{align}
which is starting to look familiar. Inserting that into Eq.~(\ref{eq:preliminaryStepInPoissonComposition}), I arrive at the incremental evolution equation
\begin{equation}
\psi(x_N,t_N) = \int_{-\infty}^\infty\!\!dx_{N-1}\;U(x_N,t_N|x_{N-1},t_{N-1})\;\psi(x_{N-1},t_{N-1})\;,
\end{equation}
with the incremental propagator 
\begin{align}
U(x_N,t_N|x_{N-1},t_{N-1}) &= \int_{-\infty}^{\infty}\frac{dk_{N-1}}{2\pi}\left[1+\Delta t\,\lambda(t_{N-1})\left(e^{-i\gamma k_{N-1}}-1\right)\right]e^{\,ik_{N-1}\left(x_N-x_{N-1}\right)} \\
&= \int_{-\infty}^{\infty}\frac{dk_{N-1}}{2\pi}\left[ e^{\,\Delta t\,\lambda(t_{N-1})\left(e^{-i\gamma k_{N-1}}-1\right) + o(\Delta t)} \right]e^{\,ik_{N-1}\left(x_N-x_{N-1}\right)} \\
&= \int_{-\infty}^{\infty}\frac{dk_{N-1}}{2\pi}\;e^{\,\Delta t\left[\lambda(t_{N-1})\left(e^{-i\gamma k_{N-1}}-1\right) + ik_{N-1}\left( \frac{x_N - x_{N-1}}{\Delta t}\right)  \right]}\;.
\end{align}
Repeating that over all increments from $t_0$ to $t_N = t_0 +  N\,\Delta t$ would give the propagator over a finite interval:
\begin{align}
U(x_N,t_N|x_0,t_0) &= \int_{-\infty}^\infty\frac{dk_{N-1}}{2\pi}\ldots\int_{-\infty}^\infty\frac{dk_0}{2\pi}\;e^{\,\sum_{j\,=\,0}^{N-1}\Delta t\left[\lambda(t_j)\left(e^{-i\gamma k_j}-1\right) + ik_j\left( \frac{x_{j+1} - x_j}{\Delta t} \right) \right]} \\
&= \int\mathscr Dk(\cdot)\;e^{\,\int_{t_0}^{t_N}\!dt\,\left[ \lambda(t)\,\left(e^{-i\gamma k(t)} - 1 \right) + i\, k(t)\,\dot x(t)\,\right]}\;. \label{eq:phaseSpacePathIntegralForPoissonVerification}
\end{align}
That agrees with Eq.~(\ref{eq:phaseSpaceActionDensityForPoisson}).
\subsection{More Elementary but Less Satisfactory Derivation}
There is an alternative way to arrive at Eq.~(\ref{eq:phaseSpaceActionDensityForPoisson}) using the moment-generating function for the Poisson distribution. Let
\begin{equation}\label{eq:poissonDistribution}
f(n,\Delta t) = \frac{1}{n!}(\lambda \Delta t)^n\, e^{-\lambda\Delta t}
\end{equation}
be the probability that $n$ jumps occur during the finite interval $\Delta t$. The function $f(n,\Delta t)$ is a distribution in $n$, in that it is normalized:
\begin{equation}
\sum_{n\,=\,0}^{\infty} f(n,\Delta t) = 1\;.
\end{equation}
The generating function is:
\begin{align}
\Omega(\kappa,\Delta t) &\equiv \sum_{n\,=\,0}^\infty e^{\,\kappa n} f(n,\Delta t) \\
&= \sum_{n\,=\,0}^{\infty}e^{\,\kappa n}\frac{1}{n!}(\lambda\Delta t)^n\,e^{-\lambda\Delta t} \\
&= e^{-\lambda\Delta t}\sum_{n\,=\,0}^{\infty}\frac{1}{n!}(\lambda\Delta t\,e^{\,\kappa})^n \\
&= e^{-\lambda\Delta t}\;e^{\,\lambda\Delta t\,e^{\kappa}} \\
&= e^{\,\lambda \Delta t\,\left(e^{\kappa}-1\right)}\;. \label{eq:characteristicFunction}
\end{align}
The first observation to make is that, with $\kappa = -i\gamma k$, one has
\begin{equation}\label{eq:characteristicFunctionWithImaginaryArgument}
\Omega(-i\gamma k,\Delta t) = e^{\,\lambda\Delta t\,\left(e^{-i\gamma k}-1\right)}\;.
\end{equation}
The second observation is that, if I were to stick a temporal label on $k$ and multiply an arbitrarily large number of $\Omega$s in the limit $\Delta t \to 0$, I would get:
\begin{equation}
\Omega(k(\cdot),T) \equiv \lim_{\substack{\Delta t\,\to\,0\,,\;N\,\to\, \infty \\ N\Delta t \,= \,T}}\prod_{i\,=\,0}^{N-1}\Omega(-i\gamma k_i,\Delta t) = e^{\,\int_0^T\!\lambda\, dt\,\left(e^{-i\gamma k(t)}-1\right)}\;,
\end{equation}
the jump contribution to the phase-space action density in Eq.~(\ref{eq:phaseSpaceActionDensityForPoisson}).
\\\\
But what really happened there? The moment that $k$, and therefore $\kappa$, becomes a general function of time, the exponentiation that led to Eq.~(\ref{eq:characteristicFunction}) no longer goes through. Instead, one has to consider infinitesimal $\Delta t$ and consider a time-ordered exponential. But as $\Delta t \to 0$, $f(n,\Delta t)$ approaches the increment $dP(t)$ in Eq.~(\ref{eq:poissonIncrement}), and one ends up repeating the construction of the path integral from Sec.~\ref{sec:campbellHardy}.
\subsection{Remarks}
Something that keeps me up at night is that jumps should be equivalent to fermions. 
\\\\
In Eq.~(\ref{eq:phaseSpaceActionDensityForPoisson}), the interaction term is the chiral half of the interaction potential of the sine-Gordon model \cite{bosonization-coleman, bosonization-mandelstam}. De Angelis et al. \cite{JLFermionsAsJumps} have essentially said that fermionization is possible, but I cannot follow their work. 
\\\\
Moreover, fermions coupled to gravity generate torsion \cite{hehlObukhov}, and the effective field theory of torsion is equivalent to the effective field theory of lattice dislocations \cite{tod, puntigam}.\footnote{Tod \cite{tod} also emphasizes the cut-and-paste method.}
\\\\
It would be useful to figure out how to reformulate discontinuous jumps as continuous interactions with latent fermions. Instead of Brownian motion with jumps, a better starting point for quantitative finance could end up being the supersymmetric sigma model \cite{superNLSM}.
\newpage
\section{Path-Integral Description of Brownian Noise}\label{sec:brownianPathIntegral}
This is the example treated in every discussion of path integrals, but I promised no previous familiarity, and I intend to deliver. 
\subsection{Action from Stochastic Differential Equation}\label{sec:actionFromSDE}
I will start with the increment:
\begin{equation}\label{eq:brownianIncrement}
dx(t) = \sigma\,Z(t)\,\sqrt{dt}\;,\;\;Z(t)\;\;\text{drawn from}\;\;\rho(Z(t))\,=\,\frac{1}{\sqrt{2\pi}}\;e^{-\half Z(t)^2}\;.
\end{equation}
On first pass, I will take $\sigma$ constant; I will defer consideration of time-dependent volatility until discussing local transformations in Sec.~\ref{sec:gaugeTheory}. I will treat field-dependent volatility in Sec.~\ref{sec:nonlinearSigmaModel}. The notation $Z(t)$ means that, for \textit{each} interval $[t,t+dt]$, one is instructed to draw a \textit{new} $Z$ from the distribution $\rho$. 
\\\\
The prescription to derive the distribution of paths is verbal: Integrate over all paths whose update rule is given by Eq.~(\ref{eq:brownianIncrement}). 
\\\\
As in the case of the Poisson increment, I can turn those words into math by starting at a fixed time, $t_0$, and applying the rule one $\Delta t$ at a time, a large number $N$ times, then taking the limits $\Delta t \to 0$ and $N \to \infty$ with $N \Delta t \equiv t_f-t_0$ held fixed.\footnote{Recall footnote~\ref{ft:dt} about $dt$ vs.~$\Delta t$. Because, on the one hand, the limiting procedure is always understood, and on the other hand, one has to invoke it repeatedly to do these calculations, I find myself switching notation constantly. My repeated attempts to be consistent end up overcomplicating the treatment instead of clarifying it, so forgive the occasional flip-flop. To be pedantic, every time I choose to write $\Delta t$ I should accompany it by $+\,o(\Delta t)$.} For ease of notation, write $Z_i = Z(t_i)$ and $x_i = x(t_i)$, $i = 0,\ldots,N-1$. 
\\\\
For a fixed draw $Z(t_0) = Z_0$, Eq.~(\ref{eq:brownianIncrement}) means that the first updated value of the path will be
\begin{equation}
x_1 = x_0 + \s\,Z_0\,\sqrt{\Delta t}\;,
\end{equation}
and the distribution of paths over one increment would simply be 
\begin{equation}
\delta(x_1 - (x_0+\s\,Z_0\,\sqrt{\Delta t}))\;.
\end{equation}
Averaging over all conceivable draws according to the distribution $\rho$ gives the incremental distribution of paths for the stochastic case:
\begin{align}
e^{-S(x(\cdot);\, t_1|t_0)} &\equiv \int_{-\infty}^\infty\!\! dZ_0\;\rho(Z_0)\;\delta\!\left(x_1-x_0 - \sigma Z_0\sqrt{\Delta t}\right) \\
&= \frac{1}{\sigma\sqrt{\Delta t}}\;\rho\!\left(\frac{x_1-x_0}{\sigma\sqrt{\Delta t}}\right) \\
&= \frac{1}{\sqrt{2\pi\sigma^2 \Delta t}}\;e^{-\frac{1}{2\sigma^2}\Delta t\,\left(\frac{x_1-x_0}{\Delta t}\right)^2}\;.
\end{align}
That last form makes the distribution of paths over a finite interval clear:
\begin{align}
e^{-S(x(\cdot);\,t_f|t_0)} &\equiv \int_{-\infty}^{\infty}\!\! dZ_{N-1}\;\rho(Z_{N-1})\,\ldots \int_{-\infty}^{\infty}\!\! dZ_0\;\rho(Z_0)\; \prod_{i\,=\,0}^{N-1}\delta\!\left(x_{i+1}-x_i-\s Z_i\sqrt{\Delta t} \right)\\
&= \frac{1}{(2\pi\sigma^2 \Delta t)^{N/2}}\;e^{-\sum_{i\,=\,0}^{N-1} \frac{1}{2\sigma^2}\Delta t\,\left(\frac{x_{i+1}-x_i}{\Delta t}\right)^2} \\
&= C\;e^{-\int_{t_0}^{t_f}\!dt\,\frac{1}{2\sigma^2}\,\dot x(t)^2}\;. \label{eq:wienerMeasure}
\end{align}
That is the shortest and most straightforward derivation of the Brownian path integral that I know of. Eq.~(\ref{eq:wienerMeasure}) is called the Wiener measure \cite{wiener}.
\\\\
The overall infinite factor can be considered part of the measure,\footnote{There is a slight mismatch between the vernacular of physics and math. In physics, $\mathscr Dx(\cdot)$ is called the measure, while $\int_{t_0}^{t_f}\! dt\,\frac{1}{2\sigma^2}\,\dot x(t)^2$ is said to be part of the action. But that part of the action sets the scale of fluctuations and equally well deserves to be called a measure.}
\begin{equation}
\mathscr Dx(\cdot) = C\prod_{t\,=\,t_0\,+\,0^+}^{t_f\,-\,0^+}dx(t)\;,
\end{equation}
or an additive constant in the action. Note that, although $C$ does not depend on the field, it does depend on time, through the variable $N = (t_f-t_0)/dt$. Typically, $C$ is fixed by hand at the end of a calculation.
\subsection{Distribution of Values from Distribution of Paths}
In the case of drift-free Brownian motion with volatility $\sigma$ (i.e., with variance $\sigma^2 T$, where $T = t_f-t_0$ the total temporal interval), I already know that the conditional probability is
\begin{equation}\label{eq:conditionalProbabilityOfBrownianMotion}
U(x_f,t_f|x_0,t_0) = \frac{1}{\sqrt{2\pi\sigma^2(t_f-t_0)}}\;e^{-\frac{1}{2\sigma^2(t_f-t_0)}(x_f-x_0)^2}\;.
\end{equation}
The goal of this section is to verify that the distribution of paths
\begin{equation}\label{eq:distributionOfBrownianPaths}
P(x(\cdot)) = e^{-\int_{t_0}^{t_f}\!\!dt\;\frac{1}{2\sigma^2}\;\dot x(t)^2}
\end{equation}
will correctly reproduce Eq.~(\ref{eq:conditionalProbabilityOfBrownianMotion}). (The overall normalization will be included when calculating the conditional probability.)
\subsubsection{Lattice Calculation: Iterated Integrals}\label{sec:iteratedIntegrals}
Once again, split the total time interval into $N$ steps of size $\Delta t$, understood according to the limiting procedure of Eq.~(\ref{eq:limit}). The conditional probability is 
\begin{align}
&U(x_f,t_f|x_0,t_0) = C_{N,0}\int_{-\infty}^{\infty}\!\! dx_{N-1}\ldots \int_{-\infty}^{\infty}\!\! dx_1\;e^{-\sum_{i\,=\,0}^{N-1}\Delta t\;\frac{1}{2\sigma^2}\left( \frac{x_{i+1}-x_i}{\Delta t} \right)^2} \\
&= C_{N,0}\int_{-\infty}^{\infty}\!\!dx_{N-1}\;e^{-\frac{1}{2\sigma^2\Delta t}\left(x_{N}-x_{N-1}\right)^2}\ldots\; \times \nonumber\\
&\qquad \int_{-\infty}^{\infty}\!\! dx_2\;e^{-\frac{1}{2\sigma^2\Delta t}\left(x_3-x_2\right)^2}\int_{-\infty}^{\infty}\!\! dx_1\;e^{-\frac{1}{2\sigma^2\Delta t}\left(x_2-x_1\right)^2}e^{-\frac{1}{2\sigma^2\Delta t}\left(x_1-x_0\right)^2}\;. \label{eq:brownianConditionalProbabilitySetup}
\end{align}
In contrast to the situation in Eq.~(\ref{eq:firstPathIntegralCalculation}), here one has to work. First, combine the two factors in the innermost exponents:
\begin{align}
(x_2-x_1)^2+(x_1-x_0)^2 &= x_2^2-2x_2x_1+x_1^2+x_1^2-2x_1x_0+x_0^2 \\
&= x_2^2+x_2^2+2\left[x_1^2 - (x_2+x_0)x_1\right]\\
&= x_2^2+x_2^2+2\left[\left(x_1-\frac{x_2+x_0}{2}\right)^2-\left(\frac{x_2+x_0}{2}\right)^2\right]\\
&= \frac{1}{2}\left[2(x_2^2+x_0^2) - (x_2+x_0)^2\right]+2\left(x_1-\frac{x_2+x_0}{2}\right)^2 \\
&= \half\underbrace{\left[2x_2^2+2x_0^2-(x_2^2+2x_2x_0+x_0^2)\right]}_{\,=\; x_2^2\,-\,2x_2x_2\,+\,x_0^2} + 2\left(x_1-\frac{x_2+x_0}{2}\right)^2  \\
&= \half\left(x_2-x_0\right)^2 + 2\left(x_1 + \text{constant}\right)^2\;.
\end{align}
By ``constant,'' I mean with respect to the integral over $x_1$---the integration bounds are $\pm\infty$, so an additive shift will drop out. Therefore, the innermost integral is
\begin{align}
I_1(x_2,x_0) &\equiv \int_{-\infty}^{\infty}\!\! dx_1\;e^{-\frac{1}{2\sigma^2\Delta t}\left(x_2-x_1\right)^2}e^{-\frac{1}{2\sigma^2\Delta t}\left(x_1-x_0\right)^2} \\
&= \int_{-\infty}^{\infty}\!\! dx_1\;e^{-\frac{1}{2\sigma^2\Delta t}\left[\half(x_2-x_0)^2+2(x_1+\text{constant})^2\right]} \\
&= e^{-\frac{1}{2\Delta t\cdot 2\sigma^2}(x_2-x_0)^2}\int_{-\infty}^{\infty}\!\! dx_1\;e^{-\frac{2}{2\sigma^2\Delta t}(x_1+\text{constant})^2} \\
&= e^{-\frac{1}{2\Delta t\cdot 2\sigma^2}(x_2-x_0)^2}\sqrt{\frac{1}{2}\cdot 2\pi\sigma^2\Delta t}\;.
\end{align}
I expressed the various numerical factors that way because I have done this calculation before and know where it is going. Note that the dependence on $x_2$ and $x_0$ is of the same form as the dependence of the original innermost exponent on $x_1$ and $x_0$, as required by composition.\footnote{See Sec.~\ref{sec:OUsourceFree} for the analogous property in the mean-reverting model.}
\\\\
Calculating the next integral, 
\begin{equation}
I_2(x_3,x_0) \equiv \int_{-\infty}^{\infty}\!\! dx_2\;e^{-\frac{1}{2\sigma^2\Delta t}\left(x_3-x_2\right)^2} I_1(x_2,x_0)\;,
\end{equation}
will make the pattern clear. Combine the two factors in the exponents, taking care to note the relative factor of 2 between them:
\begin{align}
2(x_3-x_2)^2 + (x_2-x_0)^2 &= 2(x_3^2-2x_3x_2+x_2^2)+x_2^2-2x_2x_0+x_0^2 \\
&= 2x_3^2+x_0^2 + 3x_2^2 - 2(2x_3+x_0)x_2 \\
&= 2x_3^2+x_0^2 + 3\left[x_2^2 - \frac{2}{3}(2x_3+x_0)x_2\right] \\
&= 2x_3^2+x_0^2 + 3\left[\left(x_2-\frac{2x_3+x_0}{3}\right)^2 - \left(\frac{2x_3+x_0}{3}\right)^2\right] \\
&= \frac{1}{3}\left[3(2x_3^2+x_0^2)-(2x_3+x_0)^2\right] + 3\left(x_2-\frac{2x_3+x_0}{3}\right)^2 \\
&= \frac{1}{3}\underbrace{\left[6x_3^2+3x_0^2-(4x_3^2+4x_3x_0+x_0^2)\right]}_{\,=\; 2x_3^2\,-\,4x_3x_0\,+\,2x_0^2} + 3\left(x_2 + \text{constant}\right)^2 \\
&= \frac{2}{3}\left(x_3-x_0\right)^2 + 3\left(x_2+\text{constant}\right)^2\;.
\end{align}
Therefore:
\begin{align}
I_2(x_3,x_0) &= \sqrt{\frac{1}{2}\cdot 2\pi\sigma^2\Delta t}\;\int_{-\infty}^{\infty}\!\! dx_2\;e^{-\frac{1}{2\Delta t\cdot 2\sigma^2} \left[2(x_3-x_2)^2+(x_2-x_0)^2\right]} \\
&= \sqrt{\frac{1}{2}\cdot 2\pi\sigma^2\Delta t}\;e^{-\frac{1}{2\Delta t\cdot 2\sigma^2}\frac{2}{3}(x_3-x_0)^2}\int_{-\infty}^{\infty}\!\! dx_2\;e^{-\frac{3}{2\Delta t\cdot 2\sigma^2}(x_2+\text{constant})^2} \\
&= \sqrt{\frac{1}{2}\cdot 2\pi\sigma^2\Delta t}\;\sqrt{\frac{2}{3}\cdot 2\pi\sigma^2\Delta t}\;e^{-\frac{1}{3\Delta t\cdot 2\sigma^2}(x_3-x_0)^2}\;.
\end{align}
Now it is apparent what will happen for arbitrary $N$. 
\\\\
First, inside the exponential function, composition will happen as required, and the factor multiplying $2\sigma^2$ in the denominator will be $N\Delta t = t_f-t_0$.
\\\\
Second, in the overall factor outside the exponential function, the numerator of each new fraction will cancel the denominator of the preceding fraction, leaving only an overall $N$.
\\\\
Finally, again regarding that overall factor, there will be $N-1$ powers of $\sqrt{2\pi\sigma^2\Delta t}$. To pair the $\Delta t$ with the aforementioned $N$, one should multiply and divide by one more power of $\sqrt{2\pi\sigma^2\Delta t}$, leaving the dangling powers for the overall normalization of the conditional probability. 
\\\\
Therefore:
\begin{align}
I_{N-1}(x_N,x_0) &= \sqrt{\frac{1}{2}\cdot 2\pi\sigma^2\Delta t}\;\sqrt{\frac{2}{3}\cdot 2\pi\sigma^2\Delta t}\;\ldots\;\sqrt{\frac{N-1}{N}\cdot 2\pi\sigma^2\Delta t}\; e^{-\frac{1}{N\Delta t\cdot 2\sigma^2}(x_N-x_0)^2} \\
&= \left(\sqrt{2\pi\sigma^2\Delta t}\right)^N\sqrt{\frac{1}{2\pi\sigma^2N\Delta t}}\;e^{-\frac{1}{N\Delta t\cdot 2\sigma^2}(x_N-x_0)^2}\;.
\end{align}
Returning to Eq.~(\ref{eq:brownianConditionalProbabilitySetup}) and setting the overall normalization as described, namely
\begin{equation}
C_{N,0} = \frac{1}{\left(\sqrt{2\pi\sigma^2\Delta t}\right)^N}\;,
\end{equation}
I arrive at the desired result:
\begin{align}
U(x_f,t_f|x_0,t_0) &= C_{N,0}\, I_{N-1}(x_N,x_0) = \sqrt{\frac{1}{2\pi\sigma^2N\Delta t}}\;e^{-\frac{1}{N\Delta t\cdot 2\sigma^2}(x_N-x_0)^2} \\
&= \frac{1}{\sqrt{2\pi\sigma^2\cdot(t_f-t_0)}}\;e^{\frac{1}{2\sigma^2\cdot(t_f-t_0)}(x_f-x_0)^2}\;.\qquad(x_f \equiv x_N)
\end{align}
\subsubsection{Continuum Calculation: Most-Probable Path}\label{sec:mostProbablePath}
This is one of the few path-integral calculations that can be done directly in the continuum, without an explicit choice of regularization.\footnote{It is unclear to me the extent to which it is important to find more such calculations. One takeaway of the history of effective field theory is to understand that models are typically tied to a regime of validity and have to be regularized in one form or another. In the present context, the reason for regularization is that the paths that enter the Brownian path integral are not differentiable; and the paths that enter the Poisson path integral are not even continuous. There is nothing wrong with that---it is just a more general calculus. On the other hand, $\mathscr N = 4$ supersymmetric Yang-Mills theory does not require regularization, and Feynman's footnote 10 does carry some truth: ``[O]ne feels as Cavalieri must have felt calculating the volume of a pyramid before the invention of calculus'' \cite{feynman}.} The goal is to evaluate the path integral
\begin{equation}\label{eq:brownianPathIntegralWithDrift}
U(x_f,t_f|x_0,t_0) = \int_{x(t_0)\,=\,x_0}^{x(t_f)\,=\,x_f}\!\!\!\!\mathscr Dx(\cdot)\;e^{-\int_{t_0}^{t_f}\!dt\;\frac{1}{2\sigma^2}\left[\dot x(t)-\mu\right]^2}\;
\end{equation}
by organizing the integral over paths around the saddle-point configuration, or most-probable path. The most-probable path is defined as follows. 
\\\\
Take the action functional 
\begin{equation}\label{eq:brownianActionWithDrift}
S(x(\cdot);t_f|t_0) = \int_{t_0}^{t_f}\!\!dt\;\frac{1}{2\sigma^2}\left[\dot x(t)-\mu\right]^2\;,
\end{equation}
and decompose the general path $x(t)$, subject to the boundary conditions $x(t_0) = x_0$ and $x(t_f) = x_f$, in the form
\begin{equation}\label{eq:decompositionOfGeneralPath}
x(t) = x_*(t) + y(t)\;,
\end{equation}
with
\begin{equation}\label{eq:boundaryConditions}
x_*(t_0) = x_0\;,\;\;x_*(t_f) = x_f\;,\;\; y(t_0) = y(t_f) = 0\;.
\end{equation}
The most-probable path, $x_*(t)$, is defined to be the solution of\footnote{As with any extremization procedure, there could in principle be multiple solutions; and one has to check whether any particular extremum is a local maximum or minimum.}
\begin{equation}\label{eq:definitionOfMostProbablePath}
S(x_*(\cdot)+y(\cdot);\,t_f|t_0) - S(x_*(\cdot);\,t_f|t_0) = 0\,\text{ to first order in $y$}\;.
\end{equation}
In Eq.~(\ref{eq:definitionOfMostProbablePath}), I eschewed asymptotic notation: The variation $y(t)$ does not have to be small. The variation is arbitrary, subject to the condition of fixed endpoints in Eq.~(\ref{eq:boundaryConditions}). 
\\\\
For the action in Eq.~(\ref{eq:brownianActionWithDrift}), the condition Eq.~(\ref{eq:definitionOfMostProbablePath}) will amount to
\begin{equation}\label{eq:newton'sLaw}
\ddot x_*(t) = 0\;.
\end{equation}
The solution subject to the boundary conditions in Eq.~(\ref{eq:boundaryConditions}) is
\begin{equation}
x_*(t) = x_0 + v\cdot (t-t_0)\;,\;\;v = \frac{x_f-x_0}{t_f-t_0}\;.
\end{equation}
Since $v$ is constant, the velocity on the most probable path is constant, and the action on that path evaluates to
\begin{equation}\label{eq:actionOnMostProbablePath}
S(x_*(\cdot);t_f|t_0) = \frac{t_f-t_0}{2\sigma^2}\left(v-\mu\right)^2 = \frac{1}{2\sigma^2\cdot(t_f-t_0)}\left[x_f-x_0 + \mu\cdot(t_f-t_0) \right]^2\;.
\end{equation}
In the decomposition of the general path as described by Eq.~(\ref{eq:decompositionOfGeneralPath}), the portion $x_*(t)$ is a fixed, deterministic solution of a prescribed differential equation, while the portion $y(t)$ describes an arbitrary fluctuation, with fixed endpoints. The path-integral measure under this decomposition becomes
\begin{equation}
\int_{x(t_0)\,=\,x_0}^{x(t_f)\,=\,x_f}\!\!\!\!\mathscr Dx(\cdot)\left(\ldots\right) = \int_{y(t_0)\,=\,0}^{y(t_f)\,=\,0}\!\!\!\!\mathscr Dy(\cdot)\left(\ldots\right)\;.
\end{equation}
The boundary conditions are eaten up by the deterministic path, leaving \textit{all paths that start and end at zero}. All paths. All of them. I will return to this point shortly.
\\\\
With all of the above manipulations, the path integral in Eq.~(\ref{eq:brownianPathIntegralWithDrift}) takes the form 
\begin{equation}\label{eq:brownianPathIntegralWithDrift--almostDone}
U(x_f,t_f|x_0,t_0) = C(t_f-t_0)\;e^{-S(x_*(\cdot);\,t_f|t_0)}\;,
\end{equation}
with $S(x_*(\cdot);\,t_f|t_0)$ from Eq.~(\ref{eq:actionOnMostProbablePath}), and 
\begin{equation}\label{eq:brownianFluctuationIntegral}
C(t_f-t_0) = \int_{y(t_0)\,=\,0}^{y(t_f)\,=\,0}\!\!\!\!\mathscr Dy(\cdot)\;e^{-\int_{t_0}^{t_f}\! dt\;\frac{1}{2\sigma^2}\,\dot y(t)^2}\;.
\end{equation}
The normalization factor in Eq.~(\ref{eq:brownianFluctuationIntegral}) can be evaluated in one of at least three ways. First, since the right-hand side is just a simplified Brownian path integral, one could apply the method of Sec.~\ref{sec:iteratedIntegrals}. Second, one could impose the Chapman-Kolmogorov composition law on Eq.~(\ref{eq:brownianPathIntegralWithDrift--almostDone}). Third, one could simply normalize Eq.~(\ref{eq:brownianPathIntegralWithDrift--almostDone}) directly. 
\\\\
Any of those methods will straightforwardly lead to
\begin{equation}
C(t_f-t_0) = \frac{1}{\sqrt{2\pi\sigma^2\cdot(t_f-t_0)}}\;,
\end{equation}
completing the calculation. 
\subsubsection{Remarks on the Most-Probable Path}
The steps in Sec.~\ref{sec:mostProbablePath} are relatively simple, but they merit some reflection.
\\\\
Perhaps the point of least analytical significance is the one of utmost computational significance: I do not understand how to tell a computer to evaluate the fluctuation integral in Eq.~(\ref{eq:brownianFluctuationIntegral}).
\\\\
I do not have a problem with the idea of finding alternative ways to express the fluctuation integral \cite{metropolis, numericalPathIntegral}. But I should be able to tell a computer to \textit{integrate over paths}. 
\\\\
What I suspect should be possible\footnote{Motivated, in part, by Jaynes, Sec.~17.7: The folly of randomization \cite{jaynes}.} is to select some basket of paths, evaluate the action on them; select another basket of paths, evaluate the action on them; and so on, until a sufficient number of paths has been selected to adequately sample the ensemble. There should be a way, for example, to systematically approach 
\begin{equation}
\pi^{-1} = 2\sigma^2\cdot(t_f-t_0)\;C(t_f-t_0)^2
\end{equation}
by evaluating the action on a larger and larger collection of paths that start and end at zero. 
\\\\
Now let me turn to the analytics. 
\\\\
The Brownian action in Eq.~(\ref{eq:brownianActionWithDrift}) is quadratic in the field, and the measure $\mathscr Dx(\cdot)$ involves only ordinary integrals over intermediate values, without any functionals of the field. Whereas in the general case an expansion around the most-probable path would require perturbation theory of some kind, in this case the series terminates at second order, and the expansion is exact. 
\\\\
It is important to recognize that the exact nature of the expansion is not a result of \textit{cancellation} but of \textit{factorization}.\footnote{In quantum-mechanical treatments of this procedure it is sometimes said that the nonclassical paths destructively interfere, leaving behind only the classical path. For Brownian motion, the action is real, the distribution of paths is positive, and nothing can interfere with anything---but the mathematics is the same. The interpretation as interference cannot be right.} Because the action is quadratic, the most-probable path and the fluctuations do not talk to each other, the former eating up the boundary conditions and the latter being sequestered into the overall normalization. Again, the nonextremal paths are still there---all of them, every last one---hiding in Eq.~(\ref{eq:brownianFluctuationIntegral}). 
\\\\
It is also important to recognize that, while the most-probable path between the boundary conditions is a straight line, no single realization of the process will ever be a straight line. In financial mechanics, one never bets on the extremal path. 
\subsection{Evolution Law from Distribution of Paths}\label{sec:evolutionLawFromPathIntegralForBrownianMotion}
And now I go in the other direction. The reason is, again, that I model processes by writing down actions.
\subsubsection{Forward Evolution}\label{sec:forwardEvolutionForBrownianMotion}
Write 
\begin{equation}\label{eq:evolutionFromInitialCondition}
\psi(x_N,t_N) = \int_{-\infty}^\infty\!\! dx_0\; U(x_N,t_N|x_0,t_0)\,\psi(x_0,t_0)\;,
\end{equation}
and consider the incremental forward evolution
\begin{equation}\label{eq:incrementalForwardComposition}
\psi(x_N,t_N) = C_{N,N-1}\int_{-\infty}^{\infty}\!\!dx_{N-1}\;e^{-dt\frac{1}{2\sigma^2}\left(\frac{x_{N}-x_{N-1}}{dt}\right)^2}\psi(x_{N-1},t_{N-1})\;.
\end{equation}
This is the case considered by Feynman. Define a dimensionless fluctuation:
\begin{equation}
\xi \equiv \frac{x_N-x_{N-1}}{\s\sqrt{dt}}\;.
\end{equation}
In terms of that, 
\begin{equation}
\psi(x_N,t_N) = C_{N,N-1}\;\sigma\sqrt{dt}\int_{-\infty}^{\infty}\!\!d\xi\;e^{-\half \xi^2}\psi(x_N-\s\sqrt{dt}\,\xi,t_{N-1})\;.
\end{equation}
Now expand the integrand:
\begin{equation}
\psi(x_N-\s\sqrt{dt}\,\xi,t_{N-1}) = \psi(x_N,t_{N-1})-\s\sqrt{dt}\,\frac{\partial}{\partial x_N}\psi(x_{N},t_{N-1}) + \half \s^2 dt\,\xi^2\frac{\partial^2}{\partial x_N^2}\psi(x_{N},t_{N-1}) + o(dt)\;.
\end{equation}
Using 
\begin{equation}
\int_{-\infty}^{\infty}\!\! d\xi\;e^{-\half\xi^2} = \sqrt{2\pi}\;,\;\;\int_{-\infty}^{\infty}\!\! d\xi\;e^{-\half\xi^2}\;\xi = 0\;,\;\;\int_{-\infty}^{\infty}\!\! d\xi\;e^{-\half\xi^2}\;\xi^2 = \sqrt{2\pi}\;,
\end{equation}
I find
\begin{equation}
\psi(x_N,t_N) = C_{N,N-1}\;\sigma\sqrt{2\pi\,dt}\left[\psi(x_N,t_{N-1}) + \half\sigma^2 dt\,\frac{\partial^2}{\partial x_{N}^2}\psi(x_{N},t_{N-1}) + o(dt)\right]\;.
\end{equation}
On the left-hand side, write $t_N = t_{N-1} + dt$, expand to first order in $dt$, and match the terms of order $(dt)^0 = 1$ and the terms of order $dt$. The first equality fixes the normalization of the incremental evolution equation:
\begin{equation}
C_{N,N-1}\s\sqrt{2\pi\,dt} = 1 \implies C_{N,N-1} = \frac{1}{\sqrt{2\pi\sigma^2 dt}}\;,
\end{equation}
recovering the result I had found previously. The second equality gives the diffusion equation with diffusion constant $D = \half\sigma^2$:
\begin{equation}
\frac{\partial}{\partial t_{N-1}}\psi(x_{N},t_{N-1}) = \half\sigma^2 \frac{\partial^2}{\partial x_{N}^2}\psi(x_N,t_{N-1})\;.
\end{equation}
\subsubsection{Backward Evolution}\label{sec:backwardEvolutionForBrownianMotion}
In analogy with Eqs.~(\ref{eq:evolutionFromInitialCondition}) and~(\ref{eq:incrementalForwardComposition}), write
\begin{equation}\label{eq:evolutionFromTerminalCondition}
\phi(x_0,t_0) = \int_{-\infty}^\infty\!\! dx_N\,U(x_N,t_N|x_0,t_0)\,\phi(x_N,t_N)\;,
\end{equation}
and consider the incremental backward evolution
\begin{equation}\label{eq:incrementalBackwardComposition}
\phi(x_0,t_0) = C_{1,0}\int_{-\infty}^{\infty}\!\! dx_1\,e^{-dt\frac{1}{2\sigma^2}\left(\frac{x_1-x_0}{dt}\right)^2}\phi(x_1,t_1)\;.
\end{equation}
Once again define a dimensionless fluctuation 
\begin{equation}
\xi \equiv \frac{x_1-x_0}{\sigma\sqrt{dt}}\;,
\end{equation}
in terms of which
\begin{equation}
\phi(x_0,t_0) = C_{1,0}\;\sigma\sqrt{dt}\int_{-\infty}^{\infty}\!\!d\xi\;e^{-\half\xi^2}\phi(x_0+\sigma\sqrt{dt}\,\xi,t_1)
\end{equation}
Expanding the integrand proceeds exactly as for forward evolution, leading to
\begin{equation}
\phi(x_0,t_0) = C_{1,0}\;\sigma\sqrt{2\pi\,dt}\left[\phi(x_0,t_1)+\half\sigma^2dt\;\frac{\partial^2}{\partial x_0^2}\phi(x_0,t_1) + o(dt)\right]\;.
\end{equation}
On the left-hand side, write $t_0 = t_1-dt$, expand to first order in $dt$, and match the terms of order $(dt)^0 = 1$ and the terms of order $dt$. The first equality gives the same normalization factor as before,
\begin{equation}
C_{1,0} = \frac{1}{\sqrt{2\pi\sigma^2dt}}\;.
\end{equation}
The second equality gives the backward diffusion equation with diffusion constant $D = \half\sigma^2$:
\begin{equation}
-\frac{\partial}{\partial t_1}\phi(x_0,t_1) = \half\sigma^2\frac{\partial^2}{\partial x_0^2}\phi(x_0,t_1)\;.
\end{equation}
\newpage
\section{Brownian Motion with Field-Dependent Parameters}\label{sec:nonlinearSigmaModel}
In this section I will recapitulate some of the previous derivations for a generalization of Brownian motion with field-dependent parameters. The distribution of paths is
\begin{equation}\label{eq:P=sqrtDetG_e^-S}
P(x(\cdot)) = \sqrt{g(x(\cdot))}\;e^{-S(x(\cdot))}\;,
\end{equation}
with action
\begin{equation}\label{eq:S=NLSM}
S(x(\cdot)) = \int \!dt\;\frac{1}{2\sigma^2}\,g(x(t))\left[\dot x(t)-\mu(x(t))\right]^2\;.
\end{equation}
When $\partial_x g(x) \neq 0$, this is called a nonlinear sigma model. The overall factor of
\begin{equation}
\sqrt{g(x(\cdot))} = \prod_t\sqrt{g(x(t))}
\end{equation}
is required for the Chapman-Kolmogorov composition law to hold for the conditional probability, 
\begin{equation}
U(x_1,t_1|x_0,t_0) = \int_{x(t_0)\,=\,x_0}^{x(t_1)\,=\,x_1}\mathscr Dx(\cdot)\, P(x(\cdot)).
\end{equation}
The parameter $\sigma^2$ counts the number of loops and is analogous to $\hbar$ in quantum mechanics. 
\\\\
The derivation of the forward-evolution equation is a nightmare. On first pass, you might want to set $g(x(t)) = 1$ and focus on $\mu(x(t)) \neq $ constant, then return for $g(x(t)) \neq $ constant, while setting $\mu(x(t)) = 0$. The derivation of the backward-evolution equation is simpler; you may want to skip to Sec.~\ref{sec:NLSMBackwardEquation} and try that first. 
\\\\
In both cases, I will continue to use Ito regularization. See Sec.~{\ref{sec:itoVsStratonovich}} for a justification. 
\subsection{Forward Equation}\label{sec:NLSMForwardEquation}
The goal is to show that the function
\begin{equation}\label{eq:NLSM_psi}
\psi(x,t) = \int_{-\infty}^\infty \!\!dx_0\,U(x,t|x_0,t_0)\,\psi(x_0,t_0)
\end{equation}
satisfies the Kolmogorov forward equation
\begin{equation}\label{eq:NLSM_psiEquation}
\partial_t\psi(x,t) = \frac{\sigma^2}{2}\partial_x^{\,2}\!\left[g(x)^{-1}\,\psi(x,t)\right] - \partial_x\!\left[\mu(x)\,\psi(x,t)\right]\;.
\end{equation}
Once again, I follow Feynman and consider incremental evolution: 
\begin{equation}\label{eq:incrementalCompositionForNLSM}
\psi(x_N,t_N) = C_{N,N-1}\int_{-\infty}^\infty \!\!dx_{N-1}\sqrt{g(x_{N-1})}\;e^{-S_{N,N-1}}\,\psi(x_{N-1},t_{N-1})\;,
\end{equation}
with the incremental action 
\begin{align}
S_{N,N-1} &= \Delta t\,\frac{1}{2\sigma^2}\,g(x_{N-1})\left[\frac{x_N-x_{N-1}}{\Delta t}-\mu(x_{N-1})\right]^2 \\
&= \frac{1}{2}g(x_{N-1})\left[\frac{x_N-x_{N-1}}{\sigma\sqrt{\Delta t}} - \frac{\sqrt{\Delta t}}{\sigma}\mu(x_{N-1})\right]^2\;.
\end{align}
As before, the fluctuations are of order $\sqrt{\Delta t}$:
\begin{equation}\label{eq:NLSM_fluctuation}
\xi \equiv \frac{x_{N}-x_{N-1}}{\sigma\sqrt{\Delta t}} \implies x_{N-1} = x_N-\sigma\sqrt{\Delta t}\,\xi\;.
\end{equation}
Both in the action and in the integration measure, all functions have to be expanded through $O(\Delta t)$. The novelty here is that the expansion will entail keeping fluctuations up to sixth order.
%\\\\
\subsubsection{Action to $O(\Delta t)$}\label{sec:actionToO(dt)}
First, the drift:
\begin{align}
\mu(x_{N-1}) &= \mu(x_N - \sigma\sqrt{\Delta t}\,\xi) \nonumber\\
&= \mu(x_N) - \sigma\sqrt{\Delta t}\,\xi \frac{\partial}{\partial x_N}\mu(x_N) + O(\Delta t)\;.
\end{align}
Since $\mu(x_{N-1})$ is already multiplied by $\sqrt{\Delta t}$, I can stop there. 
\\\\
The most important aspects of this calculation are to not get lost, and to not lose steam. Accordingly, I will use the somewhat compact notation
\begin{equation}
\mu(x_{N-1}) = \mu - \sigma \sqrt{\Delta t}\,\xi\,\partial\mu + O(\Delta t)\;. 
\end{equation}
That is, once the expansion has been carried out to the desired order, in which case all of the coefficients on the right-hand side are constant with respect to the integration, I will suppress the subsript $N$; and I will use the notation $\partial$ for derivatives with respect to $x_N$. I will also make liberal use of the ``big-oh'' and ``small-oh'' notations, so pay attention to the typecase. \\\\
In that notation, the coefficient in brackets reads
\begin{equation}
A \equiv \frac{x_N-x_{N-1}}{\sigma\sqrt{\Delta t}} - \frac{\sqrt{\Delta t}}{\sigma}\,\mu(x_{N-1}) = \xi - \frac{1}{\sigma}\sqrt{\Delta t}\,\mu + \Delta t\,\xi\,\partial\mu + o(\Delta t)\;,
\end{equation}
and its square is:
\begin{align}
A^2 &= \left(\xi - \frac{1}{\sigma}\sqrt{\Delta t}\,\mu + \Delta t\,\xi\,\partial\mu\right)^2 \\
&= \xi^2 - \frac{2}{\sigma}\sqrt{\Delta t}\,\mu\,\xi + 2\Delta t\,\partial\mu\,\xi^2+\left(\frac{1}{\sigma}\sqrt{\Delta t}\,\mu + O(\Delta t)\right)^2 \\
&= \xi^2 - \frac{2}{\sigma}\sqrt{\Delta t}\,\mu\,\xi + \Delta t\left(\frac{\mu^2}{\sigma^2} + 2\partial\mu\,\xi^2\right) + o(\Delta t)\;.
\end{align}
The expansion of the metric will read:
\begin{align}
g(x_{N-1}) &= g(x_N-\sigma\sqrt{\Delta t}\,\xi) \nonumber\\
&= g - \sigma\sqrt{\Delta t}\,\xi\,\partial g + \half\sigma^2\Delta t\,\xi^2\,\partial^2 g + O(\Delta t^{3/2})\;.
\end{align}
I have found it convenient to organize the algebra in terms of the function\footnote{The function in Eq.~(\ref{eq:Gamma1d}) transforms as an abelian gauge field under the nonlinear transformation $dx \to dx' = \Lambda(x)\,dx$. I have not managed to use that observation to simplify the calculation of the Kolmogorov equation from the path integral. See, however, Sec.~\ref{sec:logFieldLaplacian} and App.~\ref{sec:laplacianForDensities}.}
\begin{equation}\label{eq:Gamma1d}
\Gamma \equiv \partial \ln\!\sqrt{g} = \half g^{-1}\partial g
\end{equation}
and its derivative:
\begin{align}
\partial \Gamma &= \half\left[g^{-1}\partial^2 g - \left( g^{-1}\partial g\right)^2\right] = \half\left(g^{-1}\partial^2 g - 4\Gamma^2\right) \\
&\implies \half g^{-1}\partial^2 g = \partial\Gamma + 2\Gamma^2\;.
\end{align}
In terms of those, the expansion of the metric is:
\begin{equation}\label{eq:expansionOfMetric}
g(x_{N-1}) = g\left[1-2\sigma\sqrt{\Delta t}\,\xi\,\Gamma + \sigma^2\Delta t\,\xi^2\left(\partial\Gamma + 2\Gamma^2\right)\right]\;.
\end{equation}
The action is:
\begin{align}
S_{N,N-1} &= \half g(x_{N-1})A^2 \\
&= \half g\left[1-2\sigma\sqrt{\Delta t}\,\xi\,\Gamma + \sigma^2\Delta t\,\xi^2\left(\partial\Gamma + 2\Gamma^2\right)\right] \times \nonumber\\
&\qquad\quad \left[\xi^2 - \frac{2}{\sigma}\sqrt{\Delta t}\,\mu\,\xi + \Delta t\left(\frac{\mu^2}{\sigma^2} + 2\partial\mu\,\xi^2\right) \right] \\
&= \half g\left\{\xi^2-2\sqrt{\Delta t}\left(\frac{\mu}{\sigma}\,\xi + \sigma\Gamma\, \xi^3\right) \right. \nonumber\\
&\left.\quad + \Delta t\left[\frac{\mu^2}{\sigma^2} + 2\partial\mu\,\xi^2 + 4\mu\Gamma\,\xi^2 + \sigma^2\left(\partial\Gamma + 2\Gamma^2\right)\xi^4 \right] + o(\Delta t)\right\}\;.
\end{align}
So the $O(\Delta t)$ expansion of the action is fourth-order in the fluctuations. 
\subsubsection{Measure to $O(\Delta t)$}\label{sec:measureToO(dt)}
Write Eq.~(\ref{eq:expansionOfMetric}) in the form
\begin{equation}
g(x_{N-1}) = g\cdot(1+h)\;,\;\; h = -2\sigma\sqrt{\Delta t}\,\Gamma\,\xi + \sigma^2\Delta t\left(\partial\Gamma + 2\Gamma^2\right)\xi^2\;.
\end{equation}
Using the expansion $(1+h)^{1/2} = 1+\half h - \eighth h^2 + O(h^3)$, I can expand the integration measure:
\begin{align}
\sqrt{g(x_{N-1})} &= \sqrt{g}\left(1+h\right)^{1/2} = \sqrt{g}\left[1+\half h - \eighth h^2 + O(h^3)\right] \\
&= \sqrt{g}\left[1-\sigma\sqrt{\Delta t}\,\Gamma\,\xi + \half\sigma^2\Delta t\left(\partial\Gamma + 2\Gamma^2\right)\xi^2 - \eighth\left(-2\sigma\sqrt{\Delta t}\,\Gamma\,\xi\right)^2 + o(\Delta t)\right] \\
&= \sqrt{g}\left[1-\sigma\sqrt{\Delta t}\,\Gamma\,\xi+\half\sigma^2\Delta t\left(\partial\Gamma + \Gamma^2\right)\xi^2\right]\;.
\end{align}
\subsubsection{Distribution of Paths to $O(\Delta t)$}
Combine the results of Secs.~\ref{sec:actionToO(dt)} and~\ref{sec:measureToO(dt)}:
\begin{equation}\label{eq:sqrt(detg)e^(-S)}
\sqrt{g(x_{N-1})}\,e^{-S_{N,N-1}} = \sqrt{g}\,e^{-\half g\,\xi^2}e^{-\left(\sqrt{\Delta t}\, a+ \Delta t\,b\right)}\left[1-\sigma\sqrt{\Delta t}\,\Gamma\,\xi+\half\sigma^2\Delta t\left(\partial\Gamma + \Gamma^2\right)\xi^2\right]\;,
\end{equation}
with
\begin{align}
&a = -g\left(\frac{\mu}{\sigma}\,\xi + \sigma\Gamma\,\xi^3\right)\;,\\
&b = g\left[\frac{\mu^2}{2\sigma^2}+\left(\partial\mu + 2\mu\Gamma\right)\xi^2 + \frac{\sigma^2}{2}\left(\partial\Gamma + 2\Gamma^2\right)\xi^4\right]\;.
\end{align}
Expand the exponential in Eq.~(\ref{eq:sqrt(detg)e^(-S)}):
\begin{align}
e^{-\left(\sqrt{\Delta t}\, a+ \Delta t\,b\right)} &= 1-\left(\sqrt{\Delta t}\,a+\Delta t\,b\right) + \half\left(\sqrt{\Delta t}\,a + O(\Delta t)\right)^2 \nonumber\\
&= 1-\sqrt{\Delta t}\,a-\Delta t\left(b-\half a^2\right) + o(\Delta t)\;.
\end{align}
Note that
\begin{equation}
a^2 = g^2\left(\frac{\mu^2}{\sigma^2}\,\xi^2 + 2\mu\Gamma\,\xi^4 + \sigma^2\Gamma^2\xi^6\right)
\end{equation}
contains a term sixth-order in the fluctuation, as advertised.
\\\\
Combining terms gives
\begin{align}
\sqrt{g(x_{N-1})}\,e^{-S_{N,N-1}} &= \sqrt{g}\,e^{-\half g\,\xi^2} \;\times \nonumber\\
&\left\{1-\sqrt{\Delta t}\,a-\Delta t\left(b-\half a^2\right)\right. \nonumber\\
&-\sigma\sqrt{\Delta t}\,\Gamma\xi\left[1-\sqrt{\Delta t}\,a\right] \nonumber\\
&\left.+\half\sigma^2\Delta t\left(\partial\Gamma + \Gamma^2\right)\xi^2 + o(\Delta t) \right\} \\
& \nonumber\\
&= \sqrt{g}\,e^{-\half g\,\xi^2} \;\times \nonumber\\
&\left\{ 1 - \sqrt{\Delta t}\left(a+\sigma\Gamma\,\xi\right) \right. \nonumber\\
&\left. +\Delta t\left[-\left(b-\half a^2\right) + \sigma\Gamma\,a\,\xi + \half\sigma^2\left(\partial\Gamma + \Gamma^2\right)\xi^2 \right]\right\}\;.
\end{align}
Simplify the $O(\sqrt{\Delta t})$ terms:
\begin{align}
p \equiv a + \sigma\Gamma\,\xi &= -g\left[\left(\frac{\mu}{\sigma} - \sigma g^{-1}\Gamma\right)\xi + \sigma\Gamma\,\xi^3\right]\;.
\end{align}
Simplify the $O(\Delta t)$ terms, in stages:
\begin{align}
-b+\half a^2 + \sigma\Gamma a\,\xi &= g\left\{-\frac{\mu^2}{2\sigma^2}-\left(\partial\mu+2\mu\Gamma\right)\xi^2 - \frac{\sigma^2}{2}\left(\partial\Gamma + 2\Gamma^2\right)\xi^4 \right. \nonumber\\
&+\frac{\mu^2}{2\sigma^2}\,g\,\xi^2 + \mu\Gamma\,g\,\xi^4 + \half \sigma^2\Gamma^2\,g\,\xi^6 \nonumber\\
&\left. -\sigma\Gamma\,\xi\left( \frac{\mu}{\sigma}\,\xi + \sigma\Gamma\,\xi^3 \right)\phantom{\frac{\mu^2}{\sigma^2}}\!\!\!\!\!\!\!\!\right\} \\
&\nonumber\\
&= g\left\{ \frac{\mu^2}{2\sigma^2}\left(-1+g\,\xi^2\right) -\partial\mu\,\xi^2 + \mu\Gamma\left(-3\xi^2 + g\,\xi^4\right)\right. \nonumber\\
&\left. -\frac{\sigma^2}{2}\,\partial\Gamma\,\xi^4 + \half\sigma^2\Gamma^2\left(-4\xi^4 + g\,\xi^6\right)   \phantom{\frac{\mu^2}{\sigma^2}}\!\!\!\!\!\!\!\!\right\}
\end{align}
Finally, the $O(\Delta t)$ terms are:
\begin{align}
q &\equiv -b+\half a^2 + \sigma\Gamma a\,\xi + \half\sigma^2\left(\partial\Gamma + \Gamma^2\right)\xi^2 \\
&= g\left\{ \frac{\mu^2}{2\sigma^2}\left(-1+g\,\xi^2\right) -\partial\mu\,\xi^2 + \mu\Gamma\left(-3\xi^2 + g\,\xi^4\right)\right\} \nonumber\\
&\quad+\half\sigma^2\partial\Gamma\left(\xi^2 - g\,\xi^4\right) + \half\sigma^2\Gamma^2\left(\xi^2-4g\,\xi^4 + g^2\,\xi^6\right)\;.
\end{align}
\subsubsection{Initial Condition to $O(\Delta t)$}
Last but not least, I have to expand the distribution of the previous value, which serves as the initial condition for the current value:
\begin{align}
\psi(x_{N-1},t_{N-1}) &= \psi(x_N-\sigma\sqrt{\Delta t}\,\xi,\, t_{N-1}) \nonumber\\
&= \psi - \sigma\sqrt{\Delta t}\,\xi\,\partial\psi + \half\sigma^2\Delta t\,\xi^2\, \partial^2\psi + o(\Delta t)\;.
\end{align}
\subsubsection{Simplification}
Now is where everything comes together. Recalling and repeating Eq.~(\ref{eq:incrementalCompositionForNLSM}), I find:
\begin{align}
&\psi(x_N,t_N) = C_{N,N-1}\int_{-\infty}^\infty \!\!dx_{N-1}\sqrt{g(x_{N-1})}\;e^{-S_{N,N-1}}\,\psi(x_{N-1},t_{N-1}) \nonumber\\
&= C_{N,N-1}\,\sigma\sqrt{\Delta t}\int_{-\infty}^\infty\!\!d\xi\;\sqrt{g}\,e^{-\half g\,\xi^2}\left(1-\sqrt{\Delta t}\,p + \Delta t\,q\right)\left(\psi - \sigma\sqrt{\Delta t}\,\xi\,\partial\psi + \half\sigma^2\Delta t\,\xi^2\,\partial^2\psi\right) \\
&= C_{N,N-1}\,\sigma\sqrt{\Delta t}\int_{-\infty}^\infty\!\!d\xi\;\sqrt{g}\,e^{-\half g\,\xi^2}\times \nonumber\\
&\qquad\left[\psi - \sqrt{\Delta t}\left(p\,\psi + \sigma\xi\,\partial\psi\right) + \Delta t\left(\half\sigma^2\xi^2\,\partial^2\psi + \sigma\,p\,\xi\,\partial\psi + q\,\psi \right) + o(\Delta t) \right]\;.
\end{align}
A few remarks are in order before sailing into the abyss. 
\begin{enumerate}
\item The measure of fluctuations, $e^{-\half g\xi^2}$, is even in $\xi$, and the domain of integration is symmetric. All of the $O(\sqrt{\Delta t})$ terms are odd in $\xi$ and therefore integrate to zero.
\item Recall the basic Gaussian integral:
\begin{equation}\label{eq:basicGaussianIntegral}
G \equiv \int_{-\infty}^{\infty}\!\!d\xi\;e^{-\half g\,\xi^2} = \sqrt{\frac{2\pi}{g}}\;.
\end{equation}
\item Apply the magic numbers: 1, 3, 15. What are they? The values of the ratios of Gaussian integrals up to the order needed: 
\begin{equation}\label{eq:otherGaussianIntegrals}
I_n \equiv \frac{1}{G}\int_{-\infty}^{\infty}\!\!d\xi\;e^{-\half g\,\xi^2} \xi^{2n} = \frac{1}{g},\frac{3}{g^2},\frac{15}{g^3} \;\text{ for }\;n = 1, 2, 3\;.
\end{equation}
\end{enumerate}
Time to apply those remarks. To avoid displaying Gaussian integrals over and over again, I will adopt an arrow notation. For example, the integration multiplying $\partial^2\psi$ gives
\begin{equation}
\xi^2 \to \frac{1}{g}\;.
\end{equation}
The integration multiplying $\partial\psi$ is
\begin{align}
p\,\xi &\to -g\left[\left(\frac{\mu}{\sigma}-\sigma g^{-1}\Gamma\right)\frac{1}{g} + \sigma\Gamma\,\frac{3}{g^2}\right] \\
&= -\left(\frac{\mu}{\sigma} +2\sigma g^{-1}\Gamma\right)\;.
\end{align}
The integration multiplying $\psi$ is a jolly old time:
\begin{align}
q &\to g\left\{ \frac{\mu^2}{2\sigma^2}(-1+1) - \partial\mu\, g^{-1} + \mu\Gamma\left[-3g^{-1}+g(3g^{-2}) \right] \right\} \nonumber\\
& + \half\sigma^2\partial\Gamma\left[g^{-1} - g(3g^{-2})\right] + \half\sigma^2\Gamma^2\left[g^{-1}-4g(3g^{-2}) + g^2(15g^{-3})\right] \\
&\nonumber\\
&= -\partial\mu -\sigma^2\partial\Gamma g^{-1}+2\sigma^2\Gamma^2g^{-1} \\
&\nonumber\\
&= -\partial\mu - \sigma^2 g^{-1}\left(\partial\Gamma - 2\Gamma^2\right)\;.
\end{align}
That combination is a relief, but before I get to that let me combine all of the terms. 
%\\\\
\subsubsection{Normalization and Evolution}
Since $t_{N} = t_{N-1}+\Delta t$, I can give some attention to the left-hand side of the incremental evolution equation:
\begin{equation}
\psi(x_N,t_N) = \psi(x_N,t_{N-1}+\Delta t) = \psi(x_N,t_{N-1}) + \Delta t\,\frac{\partial}{\partial t_{N-1}}\psi(x_N,t_{N-1})\;.
\end{equation}
Demanding that the right-hand side match the left-hand side to zeroth order fixes the normalization factor:
\begin{align}
&C_{N,N-1}\,\sigma\sqrt{\Delta t} \sqrt{g(x_N)}\;\sqrt{\frac{2\pi}{g(x_N)}} = C_{N,N-1}\, \sigma\sqrt{2\pi\,\Delta t} \equiv 1 \nonumber\\
&\implies C_{N,N-1} = \frac{1}{\sigma \sqrt{2\pi\,\Delta t}}\;,
\end{align}
the same expression as before. It is important that the metric, which is now a function of the field, has dropped out. That verifies that the inclusion of $\sqrt{g(x(\cdot))}$ in the functional measure is essential. 
\\\\
Matching terms of $O(\Delta t)$ gives the evolution law:
\begin{equation}
\frac{\partial}{\partial t_{N-1}} \psi(x_N,t_{N-1}) = H\!\left(x_N,\frac{\partial}{\partial x_N}\right)\psi(x_N,t_{N-1})\;,
\end{equation}
where the Kolmogorov forward operator, or Hamiltonian, is
\begin{align}
H\!\left(x,\partial_x\right)\psi(x,t) &= \half\sigma^2 g^{-1}\partial_x^2\psi(x,t) -\left(\mu + 2\sigma^2 g^{-1}\Gamma\right)\partial_x\psi(x,t) \nonumber\\
&- \left[\partial_x\mu + \sigma^2 g^{-1}\left(\partial_x\Gamma - 2\Gamma^2\right)\right]\psi(x,t) \\
&\nonumber\\
&= \half\sigma^2 g^{-1}\partial_x^2\psi(x,t) - \partial_x\!\left[\mu\,\psi(x,t)\right] \nonumber\\
&-2\sigma^2g^{-1}\Gamma\,\partial_x\psi(x,t) - \sigma^2 g^{-1}\left(\partial_x\Gamma-2\Gamma^2\right)\psi(x,t)\;. \label{eq:almostKolmogorovForwardOperator}
\end{align}
The drift terms collected into the expected form, so that is encouraging. To make sense of the remaining terms, take some derivatives of the inverse metric:
\begin{equation}
\partial_x(g^{-1}) = -g^{-2}\partial_x g = -2g^{-1}\Gamma\;.
\end{equation}
\begin{align}
\partial_x^2(g^{-1}) &= -2\partial_x(g^{-1}\Gamma) = -2(-g^{-2}\partial_x g \Gamma + g^{-1}\partial_x\Gamma) \nonumber\\
&= -2(-2g^{-1}\Gamma^2 + g^{-1}\partial_x\Gamma) = -2g^{-1}(\partial_x\Gamma - 2\Gamma^2)\;.
\end{align}
Therefore:
\begin{align}
\partial_x^2(g^{-1}\psi) &= \partial_x(\partial_x g^{-1}\psi + g^{-1}\partial_x\psi) \\
&= \partial_x^2 g^{-1} + 2\partial_x g^{-1}\partial_x\psi + g^{-1}\partial_x^2\psi \\
&= -2g^{-1}(\partial_x\Gamma - 2\Gamma^2)\psi - 4g^{-1}\Gamma\partial_x\psi + g^{-1}\partial_x^2\psi\;.
\end{align}
Multiplying that by $\half\sigma^2$, I find precisely the combination in Eq.~(\ref{eq:almostKolmogorovForwardOperator}). Therefore:
\begin{equation}\label{eq:kolmogorovForwardOperatorForNLSM}
H(x,\partial_x)\,\psi(x,t) = \half \sigma^2\partial_x^{\,2}\!\left[g(x)^{-1}\psi(x,t)\right] - \partial_x\!\left[\mu(x)\,\psi(x,t)\right]\;.
\end{equation}
The path integral for the Ito-regularized nonlinear sigma model obeys the Kolmogorov forward equation. 
\subsection{Backward Equation}\label{sec:NLSMBackwardEquation}
The goal is to show that the function (cf. Eq.~(\ref{eq:NLSM_psi}))
\begin{equation}\label{eq:NLSM_phi}
\phi(x_0,t_0) = \int_{-\infty}^\infty\!\! dx_f\;U(x_f,t_f|x_0,t_0)\;\phi(x_f,t_f)
\end{equation}
satisfies the Kolmogorov backward equation (cf. Eq.~(\ref{eq:NLSM_psiEquation}))
\begin{equation}\label{eq:NLSM_phiEquation}
-\partial_t\phi(x,t) = \frac{\sigma^2}{2}g(x)^{-1}\partial_x^{\,2}\phi(x,t) + \mu(x)\,\partial_x\phi(x,t)\;.
\end{equation}
Again I will follow Feynman, except this time with respect to the backward variables:\footnote{The reader interested in optimization could equally well follow Bellman \cite{bellman-AdaptiveControlProcesses}.}
\begin{equation}\label{eq:incrementalCompositionForNLSM_backward}
\phi(x_0,t_0) = C_{1,0}\int_{-\infty}^{\infty}\!\! dx_1\;\sqrt{g(x_0)}\;e^{-S_{1,0}}\;\phi(x_1,t_1)\;,
\end{equation}
with the incremental action 
\begin{align}
S_{1,0} &= \Delta t\;\frac{1}{2\sigma^2}\;g(x_0)\left[\frac{x_1-x_0}{\Delta t} - \mu(x_0) \right]^2 \\
&= \frac{1}{2}\;g(x_0)\left[\frac{x_1-x_0}{\sigma\sqrt{\Delta t}} - \frac{\sqrt{\Delta t}}{\sigma}\;\mu(x_0)\right]^2\;.
\end{align}
For the umpteenth time, the fluctuations are $O(\sqrt{\Delta t})$.\footnote{That is because this entire section, being physically elementary if not mathematically so, is organized around free-field theory.} But this time, the point around which the expansion is taken is different, leading to the opposite sign in the expansion parameter (cf. Eq.~(\ref{eq:NLSM_fluctuation})):
\begin{equation}\label{eq:NLSM_fluctuation_backward}
\xi \equiv \frac{x_{1}-x_{0}}{\sigma\sqrt{\Delta t}} \implies x_{1} = x_0+\sigma\sqrt{\Delta t}\,\xi\;.
\end{equation}
This time, the functions $g$ and $\mu$ are evaluated at $x_0$, while the integration is over $x_1$. So in this case, the mathematics reduces to that of ordinary Brownian motion, in which all parameters are constant. The derivation is almost the same as that in Sec.~\ref{sec:backwardEvolutionForBrownianMotion}, except that this time I am including drift. 
\\\\
The incremental action is (let $g \equiv g(x_0)$ and $\mu \equiv \mu(x_0)$):
\begin{align}
S_{1,0} &= \half g\left(\xi - \s^{-1}\mu\sqrt{\Delta t}\right)^2 \\
&= \half g\left(\xi^2 - 2\sigma^{-1}\mu\sqrt{\Delta t}\,\xi + \sigma^{-2}\mu^2\Delta t\right)\;.
\end{align}
(Instead of writing down that second line, one could shift integration variables and shuffle the additional terms into $\phi(x_1,t_1)$.) Therefore, $e^{-S_{1,0}}$ is:
\begin{align}
e^{-S_{1,0}} &= e^{-\half g\xi^2}e^{+g\sigma^{-1}\mu\sqrt{\Delta t}\,\xi - \half g\sigma^{-2}\mu^2\Delta t} \\
&= e^{-\half g\xi^2}\left[1 + g\sigma^{-1}\mu\sqrt{\Delta t}\,\xi - \half g\sigma^{-2}\mu^2\Delta t +\half \left(g\sigma^{-1}\mu\sqrt{\Delta t}\,\xi\right)^2 + o(\Delta t)\right] \\
&= e^{-\half g\xi^2}\left[1 + g\sigma^{-1}\mu\sqrt{\Delta t}\,\xi + \half g\sigma^{-2}\mu^2\left(-1 + g\xi^2\right) \Delta t\right]\;.\label{eq:e^-S_10}
\end{align}
%\\\\
The function $\phi$ at the later time step is
\begin{align}
\phi(x_1,t_1) &= \phi(x_0 + \sigma\sqrt{\Delta t}\,\xi,\,t_1) \\
&= \phi(x_0,t_1) + \sigma\sqrt{\Delta t}\,\xi\frac{\partial}{\partial x_0}\phi(x_0,t_1) + \half\sigma^2\Delta t\,\xi^2\,\frac{\partial^2}{\partial x_0^2}\phi(x_0,t_1) + o(\Delta t)\;.\label{eq:phi(x1,t1)}
\end{align}
Multiplying Eqs.~(\ref{eq:e^-S_10}) and~(\ref{eq:phi(x1,t1)}) gives (let $\partial\phi \equiv \frac{\partial}{\partial x_0}\phi(x_0,t_1)$, etc.):
\begin{align}
e^{-S_{1,0}}\phi(x_1,t_1) &= e^{-\half g\xi^2}\left[1 + g\sigma^{-1}\mu\sqrt{\Delta t}\,\xi + \half g\sigma^{-2}\mu^2\left(-1 + g\xi^2\right) \Delta t\right] \times \nonumber\\
&\qquad\qquad\left[\phi + \sigma\sqrt{\Delta t}\,\xi\,\partial\phi + \half\sigma^2\Delta t\,\xi^2\,\partial^2\phi\right] \\
&\nonumber\\
&= e^{-\half g\xi^2}\left\{ \phi + \sqrt{\Delta t}\cdot(\text{linear in $\xi$}) \right. \nonumber\\
&\qquad\left. + \Delta t\left[\half\sigma^2\xi^2\,\partial^2\phi + \mu\,g\xi^2\,\partial\phi + \half g\sigma^{-2}\mu^2\left(-1+g\xi^2\right)\phi \right] \phantom{\sqrt{\Delta t}}\!\!\!\!\!\!\!\!\!\!\!\!\!\right\}\;.
\end{align}
Recalling the Gaussian integrals of Eq.~(\ref{eq:basicGaussianIntegral}) and~(\ref{eq:otherGaussianIntegrals}), I can evaluate the right-hand side of the incremental evolution equation in Eq.~(\ref{eq:incrementalCompositionForNLSM_backward}):
\begin{align}
\phi(x_0,t_0) &= C_{1,0}\sqrt{g}\,\s\sqrt{\Delta t}\sqrt{\frac{2\pi}{g}}\left\{\phi + \Delta t\left[\half\sigma^2\frac{1}{g}\partial^2\phi + \mu\partial\phi + \half g\sigma^{-2}\mu^2\left(-1+1\right)\phi \right]  \right\} \\
&= C_{1,0}\sqrt{2\pi\sigma^2\Delta t}\left\{ \phi(x_0,t_1) + \Delta t\left[\half\sigma^2g(x_0)^{-1}\frac{\partial^2}{\partial x_0^2}\phi(x_0,t_1) + \mu(x_0)\frac{\partial}{\partial x_0}\phi(x_0,t_1) \right] \right\}\;.
\end{align}
Now for the left-hand side:
\begin{equation}
\phi(x_0,t_0) = \phi(x_0,t_1-\Delta t) = \phi(x_0,t_1) - \Delta t\;\frac{\partial}{\partial t_1}\phi(x_0,t_1) + o(\Delta t)\;.
\end{equation}
Matching up the $O(\Delta t^0)$ terms again gives the familiar normalization factor, 
\begin{equation}
C_{1,0} = \frac{1}{\sqrt{2\pi\sigma^2\Delta t}}\;,
\end{equation}
while matching the $O(\Delta t)$ terms gives a Kolmogorov backward equation:
\begin{equation}
-\partial_t \phi(x,t) = K(x,\partial_x)\,\phi(x,t)\;,
\end{equation}
with 
\begin{equation}\label{eq:kolmogorovBackwardOperatorForNLSM}
K(x,\partial_x)\,\phi(x,t) = \half\sigma^2 g(x)^{-1}\partial_x^{\,2}\phi(x,t) + \mu(x)\,\partial_x\phi(x,t)\;.
\end{equation}
The operator $K$ in Eq.~(\ref{eq:kolmogorovBackwardOperatorForNLSM}) is indeed the adjoint of the operator $H$ in Eq.~(\ref{eq:kolmogorovForwardOperatorForNLSM}).
\subsection{Ito vs. Stratonovich Regularization}\label{sec:itoVsStratonovich}
As I have emphasized, the paths that enter the Brownian path integral are not differentiable. Because those paths are not differentiable, the choice of how to define each increment matters. 
\\\\
In the incremental update rule, or stochastic differential equation, written in the continuum notation
\begin{equation}\label{eq:continuumSDE}
dx(t) = B(x(t),t)\,\sqrt{dt}\,Z(t) + A(x(t),t)\,dt\;,
\end{equation}
there is a mathematical ambiguity of where precisely, in each interval $[t,t+dt]$, to evaluate the functions $B$ and $A$. As always in physical modeling, mathematical ambiguities signal not an absence of mathematical rigor, but an absence of physical rigor. One has to think more carefully about which physical details need to be expressed in the mathematics. 
\\\\
On a temporal lattice of the kind used in the previous calculations, the general choice of how to regularize Eq.~(\ref{eq:continuumSDE}) can be expressed as:
\begin{align}
&\Delta x(t_i) \equiv x(t_{i+1})-x(t_i) = B(x_{\alpha}(t_i,t_{i+1}),t_i)\,\sqrt{\Delta t}\,Z(t_i) + A(x_{\alpha}(t_i,t_{i+1})),t_i)\,\Delta t\;, \\
&x_\alpha(t_i,t_{i+1}) \equiv \alpha\, x(t_{i+1}) + (1-\alpha)\,x(t_i)\;,\;\;\alpha \in [0,1]\;.
\end{align}
The choice $\alpha = 0$ is called Ito regularization or forward regularization, and the choice $\alpha = \half$ is called Stratonovich regularization or midpoint regularization. The less common choice $\alpha = 1$ is called anti-Ito regularization, or backward regularization. 
\\\\
Which regularization should be chosen? The answer, again, is dictated by physical rigor, not mathematical rigor. 
%\\\\
\subsubsection{Quantum Mechanics}
In quantum mechanics, one fundamental physical principle is gauge invariance.\footnote{Unlike special relativity and unitarity, one could split hairs about whether invariance under local transformations is really a ``physical'' principle. I would say it is. For example, one could speak of electromagnetism without introducing a photon, but only at the cost of locality. So regardless of the mathematical fact that gauge invariance is a way to \textit{throw out} states instead of \textit{relate} states, physical principles are always involved. Classically, the primary physical content of general relativity is the equivalence principle, not general coordinate invariance \cite{zeeGR}. But the world is quantum mechanical, and one still has to make sure that energy cannot appear and disappear at will \cite{gravitationalAnomalies}.} For example, if you want to model electric charges, you had better not find that electric charge spontaneously leaks out of the universe.\footnote{Charged particles can come and go, but total charge has to be conserved. Global symmetries can be anomalous, but gauge symmetries cannot. This topic is way too intricate to get into here; see a discussion about anomalies in any textbook on quantum field theory.} The choice of regularization that preserves gauge invariance is Stratonovich regularization, so that is the regularization appropriate for that context.\footnote{See, for example, Ch.~5 of Schulman \cite{schulman}. In field theory, one imposes renormalization conditions to relate theory to experiment, so that the predictions of theory are insensitive to the choice of regularization. That involves various limiting procedures and is not pertinent to this paper. What is pertinent is that one could, in principle, adopt a regularization that violates gauge invariance, as long as one shows that gauge invariance is restored after the calculation is done. That approach is sometimes practical (e.g., adding a mass term for the photon), and other times impractical (e.g., adding a mass term for the gluons).}
\\\\
What the detailed calculations of the previous sections demonstrate is that, whatever the regularization, there are two instructions to follow. 
\begin{enumerate}
\item Establish the scale of fluctuations. Rescaling the increment by a factor of $\sqrt{\Delta t}$ so that the kinetic term becomes dimensionless expresses that fluctuations are to be measured relative to the noninteracting model of Brownian motion. That is the meaning of ``$(dx)^2 \sim dt$'' in the path integral.\footnote{That relation is violated for interacting fixed points \cite{fBm, JLProbabilityOld}.}
\item Once the scale of fluctuations has been established, drop all terms that go to zero with $\Delta t$ faster than $\Delta t$ as $\Delta t \to 0$. Keeping only terms of order $\Delta t$ is the fundamental rule of path-integral calculus \cite{mcLaughlinSchulman}. 
\end{enumerate}
Feynman's procedure to derive an evolution law for the distribution of values from the incremental evolution law for the distribution of paths will then establish the relation between the functions that appear in the action and the functions that appear in the Kolmogorov equation \cite{vanN, langouche}.
%\\\\
\subsubsection{Financial Mechanics}
\begin{quotation}
``As another instance of early questioning of the core model, a paper I refereed argued that Black-Scholes must be fundamentally flawed because a different valuation formula is derived from the replication argument if the R. L. Stratonovich (1968) stochastic calculus is used for modeling instead of the Ito calculus. My report showed that while the paper's mathematics was correct, its economics was not: A Stratonovich-type formulation of the underlying price process implies that traders have a partial knowledge about future asset prices that the nonanticipating character of the Ito process does not. The `paradox' is thus resolved because the assumed information sets are essentially different and hence, so should the pricing formulas.''---Merton \cite{mertonNobel}
\end{quotation}
In financial mechanics, the fundamental physical principle is that you act in response to a feed. One must use Ito regularization.\footnote{I disagree with Matacz \cite{matacz} that only the Stratonovich rule makes sense, and I support Dash's \cite{dash} usage of Ito regularization. Moreover, it would make no sense to choose one regularization for portfolio construction but a different regularization for fitting models to data.}
\\\\
Merton's phrasing of the logic that enters the option-pricing formula invokes the assumption that traders are not crooks and do not trade on insider information.\footnote{Try not to laugh.} This might be the correct reasoning for that context, but in the context I have in mind the reasoning is a bit different. It is important to nail down exactly where the possibility of insider trading may leak into the formalism.\footnote{I thank Seth Blumberg for a conversation about this.}
\\\\
What I have in mind is explained in a companion paper \cite{portfolioConstruction}, but let me give the following brief overview. A robot acts according to a model, a market feed, and a performance criterion. You input the model and the performance criterion, and the robot decides how to incrementally rebalance the portfolio as it receives the latest market update from the feed: New price, new position; new price, new position; and so on---the crucial point being that the decisions happen \textit{in that order}. The robot acts in response to a feed. 
\\\\
In that context, insider information has nothing to do with the regularization. If you have a better feed---maybe it updates every second instead of every 15 seconds, or has lower latency because you move the firm from California to New York---then you tell the robot to use that feed instead. Either way, the robot always acts in response to a feed. It does not make any sense whatsoever to regularize anywhere but the start of the interval. 
\\\\
Where does the possibility of insider information enter?
\\\\
To answer that question, let me reiterate my opening line: The fundamental assumption of theoretical quantitative finance is that percentage changes in asset prices behave as if they were stochastic variables. And let me repeat the corresponding footnote: The companion empirical assumption is that you can get by using only tick data. If those combined can be called the ``pure quant'' hypothesis, then the following can be called the pure quant oath:
\begin{quotation}
\textit{I hereby swear that all priors are constructed solely from symmetry principles or the data itself.}
\end{quotation}
Priors that are constructed by other means fall into two broad categories: Market research, which is above board, and insider trading, which is illegal. The pure quant oath precludes both. 
\\\\
In practice, it may be worth stipulating a handful of unobjectionable violations of the pure quant oath in an otherwise pure quant system. For example, suppose I want to trade MSFT and that its next ex-dividend date is soon. It is standard practice to include a jump contribution in the model to compensate, as in Eq.~(\ref{eq:distributionOfPathsForPureJumpProcess}).\footnote{See p. 591 of Dash \cite{dash}.}
\\\\
Say I include in the model of price an impulse term of size $-D$ at a time $t_D$. That is part of model-building, no problem. What would violate the pure quant oath would be to write down priors for $D$ and $t_D$ that are peaked around the amount of the dividend and the ex-dividend date. Insisting that the robot know nothing about that particular nugget of public information seems more obstinate than virtuous.
\newpage
\section{The Log-Field}\label{sec:logField}
Financial mechanics models the percentage change in price, whereas the market displays the price itself. Accordingly, one is forced to engage with the nonlinear change of variables
\begin{equation}\label{eq:logField}
x \equiv \ln X\;\iff\;\; X \equiv e^x\;.
\end{equation}
Nonlinear transformations in path-integral calculus are subtle, but I have already subjected you to most of the requisite machinery. 
\subsection{Correction of the Drift}
First I will recapitulate the standard treatment of this problem by transforming the characteristics. Consider the stochastic differential equation
\begin{equation}
dx(t) \equiv x(t+dt) - x(t) = \sqrt{dt}\,Z(t)\;,
\end{equation}
with $Z(t)$ drawn from $\rho(Z(t)) = \frac{1}{\sqrt{2\pi}}e^{-\half Z(t)^2}$. At each $t$, define $X(t) = e^{x(t)}$, then subtract that definition at successive times:
\begin{align}
dX(t) &\equiv X(t+dt)-X(t) = e^{x(t+dt)}-e^{x(t)} = e^{x(t)+dx(t)}-e^{x(t)} = e^{x(t)}\left(e^{dx(t)}-1\right) \\
&= e^{x(t)}\left[dx(t) + \half \left(dx(t)\right)^2 + \ldots\right]\;. \label{eq:expansion_in_dx(t)}
\end{align}
Inside expectations over $\{Z(t)\}_t$ taken with respect to $\prod_t \rho(Z(t))$, the operator $(dx(t))^2$ can be replaced by $dt$; and higher powers of $dx(t)$ can be dropped. To see that, consider, instead of Eq.~(\ref{eq:expansion_in_dx(t)}), the differential
\begin{equation}\label{eq:SDEwithf(Z)}
dX(t) = X(t)\left[\sqrt{dt}\,Z(t) + f(Z(t))\,dt\right]\;.
\end{equation}
The goal is to show that, because $f(Z(t))$ multiplies $dt$, no information would be lost inside path integrals by replacing $f(Z(t))$ with its expectation. 
\\\\
Constructing the path integral according to the procedure of Sec.~\ref{sec:actionFromSDE} leads to
\begin{equation}
e^{-S(X(\cdot);\,t_f|t_0)} = \prod_{i\,=\,0}^{N-1}\int_{-\infty}^\infty\!\! dZ_i\,\rho(Z_i)\,\delta\!\left(\Delta X_i - X_i\left[\sqrt{\Delta t}\,Z_i + f(Z_i)\,\Delta t\right]\right)\;.
\end{equation}
Each delta function could be rewritten in Fourier space:
\begin{equation}
\delta\!\left(\Delta X - X\left[\sqrt\Delta t\,Z + f(Z)\,\Delta t\right]\right) = \int_{-\infty}^\infty\frac{dk}{2\pi}\;e^{\,ik\left(\Delta X - X\left[\sqrt\Delta t\,Z + f(Z)\,\Delta t\right]\right)}\;.
\end{equation}
Bringing the integral over $Z$ inside the integral over $k$ and completing the square according to 
\begin{equation}
\half Z^2 + ik X\sqrt{\Delta t}\,Z = \half\left[\left(Z+ik X\sqrt{\Delta t}\right)^2 + k^2 X^2 \Delta t\right]
\end{equation}
gives
\begin{align}
\int_{-\infty}^{\infty}\!\! dZ\,\rho(Z)&\,\delta\!\left(\Delta X - X\left[\sqrt{\Delta t}\,Z + f(Z)\Delta t\right] \right) \nonumber \\
&= \int_{-\infty}^{\infty}\frac{dk}{2\pi}\;e^{\,ik\,\Delta X - \half k^2 X^2\Delta t}\int_{-\infty+ic}^{\infty+ic}\!\! dz\,\rho(z)\,e^{-ikX f(z-ikX\sqrt{\Delta t})\,\Delta t}\;. \label{eq:afterCompletingTheSquare}
\end{align}
Because $f$ multiplies $\Delta t$ in the exponent, I can write $f(z-ikX\sqrt{\Delta t})\,\Delta t = f(z)\,\Delta t + o(\Delta t)$. Assuming that no poles are crossed in the complex plane, I can drop the $+ic$, relabel $z$ as $Z$, and recognize the result as a series of expectations:
\begin{align}
\int_{-\infty+ic}^{\infty+ic}\!\! dz\,\rho(z)\,e^{-ikX f(z-ikX\sqrt{\Delta t}) \Delta t} &= \int_{-\infty}^\infty\!\! dZ\,\rho(Z)\,e^{-ikX f(Z)\Delta t \,+\, o(\Delta t)} \\
&= \int_{-\infty}^\infty\!\! dZ\,\rho(Z)\left[ 1 - ikX f(Z)\,\Delta t + o(\Delta t)\right] \\
&= 1 - ikX\,E_Z(f(Z))\,\Delta t + o(\Delta t) \\
&= e^{-ikX\,E_Z(f(Z))\,\Delta t \,+\, o(\Delta t)}\;.
\end{align}
Inserting that back into Eq.~(\ref{eq:afterCompletingTheSquare}) gives 
\begin{align}
\int_{-\infty}^{\infty}\!\! dZ\,\rho(Z)\,\delta\!\left(\Delta X - X\left[\sqrt{\Delta t}\,Z + f(Z)\Delta t\right] \right) &= \int_{-\infty}^{\infty}\frac{dk}{2\pi}\;e^{\,ik\left[\Delta X - X\,E_Z(f(Z))\Delta t\right] - \half k^2 X^2 \Delta t} \\
&= \frac{1}{\sqrt{2\pi \Delta t}}\,\frac{1}{X}\;e^{-\frac{1}{2\Delta t\,X^2}\left[\Delta X - X\,E_Z(f(Z))\,\Delta t\right]^2}\;.
\end{align}
The path integral would then be 
\begin{align}
e^{-S(X(\cdot);\,t_f|t_0)} &= \prod_{i\,=\,0}^{N-1} \frac{1}{\sqrt{2\pi \Delta t}}\, \frac{1}{X_i}\;e^{-\Delta t \frac{1}{2 X_i^2}\left[\frac{\Delta X_i}{\Delta t} - X_i\,E_Z(f(Z)) \right]^2} \\
&= \text{constant}\times \left(\prod_{t\,=\,t_0}^{t_f-0^+}\frac{1}{X(t)}\right)\;e^{-\int_{t_0}^{t_f}\! dt\,\frac{1}{2X(t)^2}\left[\dot X(t) - X(t)\, E_Z(f(Z))\right]^2}\;.
\end{align}
That is the result I would have obtained had I replaced $f(Z)$ by $E_Z(f(Z))$ in Eq.~(\ref{eq:SDEwithf(Z)}). 
\\\\
In the present case, $f(Z) = \half Z^2$, and $E_Z(f(Z)) = \half$, leading to the standard rule
\begin{equation}\label{eq:itoCorrection}
\frac{dX(t)}{X(t)} = dx(t) + \half dt\;.
\end{equation}
Drift-free Brownian motion in $x$ leads to geometric Brownian motion in $X$ with drift $\half$.\footnote{There is some ambiguity in the terminology, depending on whether you want to reserve the term ``drift'' for the right-hand of $dX(t)/X(t)$ or $dX(t)$. For the former, the drift would be the constant value $\half$, while for the latter, the drift would be the linear function $\half X(t)$. When $X(t)$ is interpreted as a price, a better term for the constant value $\half$ is ``appreciation rate.''} In stochastic mechanics, that phenomenon is called the Ito correction. Restoring the volatility parameter $\sigma = \sqrt{\hbar}$, one can recognize the correction as an $O(\hbar)$ effect.\footnote{I remain uncertain whether to call this an ``anomaly,'' in the precise sense of stochasticity breaking a symmetry of the deterministic model. Regardless of whether the correction should be thought of as purely an artifact of the regularization, the temporal lattice in financial mechanics is as real as a brokerage account.}
\subsubsection{Nonlinear Sigma Model}\label{sec:ItoCorrectionInPathIntegral}
The steps of the previous section lead to the distribution of paths
\begin{align}
P(X(\cdot)) &\equiv E_{Z(\cdot)}\left[dX(\cdot)-X(\cdot)\left(\half dt + \sqrt{dt}\,Z(\cdot)\right)\right] \\
%&= \int \left(\prod_{i\,=\,0}^{N-1}\frac{dZ_i}{2\pi}\,e^{-\half Z_i^2}\right)\prod_{i\,=\,0}^{N-1} \delta\!\left[\Delta X_i - X_i\left(\half \Delta t + \sqrt{\Delta t}\,Z_i \right)\right] \\
%&= \text{const}\times \prod_{i\,=\,0}^{N-1}\int_{-\infty}^{\infty}\!\! dZ_i\;e^{-\half Z_i^2}\;\delta\!\left[\left(\Delta X_i - \half X_i\Delta t \right) - X_i\sqrt{\Delta t}\,Z_i\right] \\
%&= \text{const}\times \prod_{i\,=\,0}^{N-1}\int_{-\infty}^{\infty}\!\! dZ_i\;e^{-\half Z_i^2}\; \frac{1}{X_i\sqrt{\Delta t}}\;\delta\!\left[\frac{\Delta t}{X_i\sqrt{\Delta t}}\left(\frac{\Delta X_i}{\Delta t} - \half X_i\right) - Z_i \right] \\
%&= \text{const}\times \prod_{i\,=\,0}^{N-1}\;\frac{1}{X_i}\;e^{-\half\left[\frac{\sqrt{\Delta t}}{X_i}\left(\frac{\Delta X_i}{\Delta t} - \half X_i \right) \right]^2} \\
&= \text{const}\times \left(\prod_{t}\frac{1}{X(t)}\right)\;e^{-\int\! dt\;\frac{1}{2X(t)^2}\left[\dot X(t) - \half X(t)\right]^2}\;,
\end{align}
a nonlinear sigma model with
\begin{equation}
g(X) = \frac{1}{X^2}\;,\;\;\mu(X) = \half X\;.
\end{equation}
\subsection{Derivation within the Path Integral}
Let me see if I can provide an alternative derivation of the Ito correction by studying the Brownian path integral,
\begin{equation}
\mathscr Z = \int \mathscr Dx(\cdot)\; P(x(\cdot))\;,\;\; P(x(\cdot)) = e^{-\int\! dt\,\half \dot x(t)^2}\;.
\end{equation}
\subsubsection{Evolution Equation}
I already showed in Sec.~\ref{sec:NLSMForwardEquation} that the path integral
\begin{equation}
\psi(x_f,t_f) = \int_{-\infty}^\infty \!\!dx_0\,\psi(x_0,t_0) \int_{x(t_0)\,=\,x_0}^{x(t_f)\,=\,x_f}\!\!\mathscr Dx(\cdot)\sqrt{g(x(\cdot))}\;e^{-\int_{t_0}^{t_f}\!dt\,\half g(x(t))\,\left[\dot x(t)-\mu(x(t))\right]^2}
\end{equation}
satisfies the Kolmogorov forward equation,
\begin{equation}
\partial_t \psi(x,t) = \half\,\partial_x^2\!\left[g(x)^{-1}\psi(x,t)\right]-\partial_x\!\left[\mu(x)\,\psi(x,t)\right]\;.
\end{equation}
That means that the path integral $\psi(x,t)$ with action
\begin{equation}
S = \int_{t_0}^{t_f}\!\! dt\;\half\,\dot x(t)^2\qquad(g(x) = 1\;,\;\;\mu(x) = 0)
\end{equation}
and measure\footnote{The $dx(t)$s appearing there are the differentials of ordinary calculus, not to be confused with the $dx(t) = x(t+dt)-x(t)$ appearing in the stochastic differential equation.}
\begin{equation}
\mathscr Dx(\cdot) = \!\!\!\!\prod_{t\,=\,t_0\,+\,dt}^{t_f\,-\,dt}\!\!\!dx(t)
\end{equation}
satisfies the evolution equation
\begin{equation}\label{eq:evolutionEquationInx}
\partial_t\psi(x,t) = \half\,\partial_x^2\psi(x,t)\;,
\end{equation}
while the path integral $\Psi(X,t)$ with action
\begin{equation}
S = \int_{t_0}^{t_f}\!\!dt\;\half \left(\frac{\dot X(t)}{X(t)}-\frac{1}{2}\right)^2\qquad(g(X) = \frac{1}{X^2}\;,\;\;\mu(X) = \half X)
\end{equation}
and measure 
\begin{equation}
\mathscr DX(\cdot) = \!\!\!\!\prod_{t\,=\,t_0\,+\,dt}^{t_f\,-\,dt}\!\!\!\frac{dX(t)}{X(t)}
\end{equation}
satisfies the evolution equation
\begin{equation}\label{eq:evolutionEquationInX}
\partial_t\Psi(X,t) = \half\,\partial_X^2\!\left[X^2\Psi(X,t)\right]-\half\,\partial_X\!\left[X\Psi(X,t)\right]\;.
\end{equation}
\subsubsection{Comparison of Evolution Equations}
Something might seem amiss. Consider the change of variables 
\begin{equation}
X = e^x \implies \partial_x X = X\;.
\end{equation}
The derivatives of a function $f(x)$ will satisfy, by the familiar chain rule of ordinary calculus:
\begin{align}
&\partial_x f = \partial_X f \partial_x X = X\partial_X f\;,\\
&\partial_x^2 f = \partial_X(X\partial_X f)\partial_x X = X^2\partial_X^2 f + X\partial_X f\;.
\end{align}
So the evolution equation in Eq.~(\ref{eq:evolutionEquationInx}) is
\begin{equation}\label{eq:evolutionEquationInxAfterChainRule}
\partial_t\psi(x,t) = \half X^2\partial_X^2 \psi(\ln X,t) + X\partial_X\psi(\ln X,t)\;.
\end{equation}
But expanding the partial derivatives on the right-hand side of the evolution equation in Eq.~(\ref{eq:evolutionEquationInX}) leads instead to
\begin{equation}\label{eq:evolutionEquationInXExpanded}
\partial_t\Psi(X,t) = \half\,X^2\partial_X^2\Psi(X,t) + \tfrac{3}{2}X\partial_X\Psi(X,t) + \half\Psi(X,t)\;.
\end{equation}
What gives?
\subsubsection{Integration Bounds}
One place to hunt for a problem might be in the bounds of integration. That is not where the problem will lie, but it is worth explaining why. 
\\\\
Recall that a key step in the derivation of the Kolmogorov forward equation in Sec.~\ref{sec:NLSMForwardEquation} was the explicit evaluation of Gaussian integrals. The integration bounds of those integrals were $-\infty$ to $\infty$, inherited from the range of allowed values of the field at each time, namely $x_i\in(-\infty,\infty)$ for each $i = 1,\ldots,N-1$.
\\\\
Each transformation $x_i = \ln(X_i)$ means that the range of allowed values of the transformed field at each time is $X_i \in [0,\infty)$, only a semi-infinite interval. Let me explore the consequence of that by examining the incremental evolution equation
\begin{equation}\label{eq:incrementalCompositionForLogField}
\Psi(X_N,t_N) = \frac{1}{\sqrt{2\pi\Delta t}}\int_0^\infty \frac{dX_{N-1}}{X_{N-1}}\;e^{-\frac{1}{2X_{N-1}^2}\,\frac{(X_N-X_{N-1})^2}{\Delta t}}\Psi(X_{N-1},t_{N-1})
\end{equation}
on its own terms. (For this argument it will be sufficient to consider the action without drift.)
\\\\
Let
\begin{equation}\label{eq:fluctuationOfLogField}
\xi \equiv \frac{X_N-X_{N-1}}{X_{N-1}\sqrt{\Delta t}} = \frac{1}{\sqrt{\Delta t}}\left(\frac{X_N}{X_{N-1}}-1\right)\;, 
\end{equation}
so that the kinetic term will once again take the $\Delta t$-independent form $\half \xi^2$. Inverting Eq.~(\ref{eq:fluctuationOfLogField}), making sure to keep all terms of the requisite order (remember that there is an overall $\frac{1}{\sqrt{\Delta t}}$ on the right-hand side of Eq.~(\ref{eq:incrementalCompositionForLogField})), and so on, will eventually lead to the pertinent Kolmogorov forward equation. 
\\\\
The important takeaway from Eq.~(\ref{eq:fluctuationOfLogField}) is the range of allowed values of the fluctuation:
\begin{equation}
\xi(X_{N-1} \to 0) \to \infty\;,\;\; \xi(X_{N-1} \to \infty) \to \frac{-1}{\sqrt{\Delta t}}\;.
\end{equation}
That second one is the key. All of these derivations hold only in the limiting process $\Delta t \to 0$ with $N \to \infty$ such that $N\Delta t$ is held fixed; in that limit, the lower bound of $\xi$ indeed goes to $-\infty$, and the derivation of Sec.~\ref{sec:NLSMForwardEquation} carries through. 
\subsubsection{Normalization}\label{sec:logFieldNormalization}
The resolution of the discrepancy between Eqs.~(\ref{eq:evolutionEquationInxAfterChainRule}) and~(\ref{eq:evolutionEquationInXExpanded}) centers on the observation that 
\begin{equation}
\Psi(X,t) \neq \psi(\ln X,t)\;.
\end{equation}
That is because the distributions in this paper have been normalized in a noncovariant fashion:
\begin{equation}\label{eq:noncovariantNormalization}
\int_{-\infty}^\infty\!\! dx\;\psi(x,t) = 1\;.
\end{equation}
Under the change of variables $x = \ln X$, the measure (and bounds) of integration change, while the integral itself does not. Therefore:
\begin{align}
&1 = \int_{-\infty}^\infty\!\! dx\;\psi(x,t) = \int_0^\infty\!\! \frac{dX}{X}\;\psi(\ln X,t) = \int_0^\infty\!\! dX\;\Psi(X,t) \\
&\implies \psi(\ln X,t) = X\Psi(X,t)\;.\label{eq:psiVsPsi}
\end{align}
But not even that fully resolves the discrepancy, because inserting that into both sides of Eq.~(\ref{eq:evolutionEquationInxAfterChainRule}), then dividing both sides by $X$, leads to
\begin{equation}\label{eq:oops}
\partial_t\Psi(X,t) = \half X^2\partial_X^2\Psi(X,t) + 2X\partial_X\Psi(X,t) + \Psi(X,t)\;.
\end{equation}
Oops. Path integrals are bunk, call the president of physics.
\subsubsection{Invariant Laplacians for Densities}\label{sec:logFieldLaplacian}
Just kidding. The remaining problem is that, because of Eq.~(\ref{eq:psiVsPsi}), there is an additional transformation from turning ``$D^2 = \partial_x^{\,2}$'' acting on $\psi(x,t)$ into the correct ``$D^2$'' acting on $\Psi(X,t)$. The missing piece will subtract $\half(X\partial_X + 1)\Psi(X,t)$ from the right-hand side of Eq.~(\ref{eq:oops}), leading correctly to Eq.~(\ref{eq:evolutionEquationInXExpanded}). 
\\\\
As I explain in App.~\ref{sec:laplacianForDensities}, the Kolmogorov forward equation takes the form
\begin{equation}
\partial_t\psi(x,t) = \half\partial_x^2\psi(x,t)
\end{equation}
in the one basis, and 
\begin{equation}
\partial_t\Psi(X,t) = \half D^2_X \Psi(X,t)\;,
\end{equation}
with 
\begin{equation}
D^2_X \Psi(X,t) = X^2\partial_X^2\Psi(X,t) + 3X\partial_X \Psi(X,t) + \Psi(X,t)
\end{equation}
in the other basis. Therefore, 
\begin{equation}
\partial_t\Psi(X,t) = \half\left(X^2\partial_X^2\Psi + 3X\partial_X + \Psi\right)\;,
\end{equation}
the result in Eq.~(\ref{eq:evolutionEquationInXExpanded}). 
\subsection{Remarks}
\begin{quotation}
``The path integral method is often considered as of heuristic value only, with the understanding that all results derived in this approach are to be checked by parallel calculations in the operator formalism. There is also a belief that path integrals can safely be used only through their correspondence with the diagram technique, so that manipulations with path integrals just represent manipulations with Feynman diagrams.\\
\indent We do not accept such limited definitions of the path integral method, especially in view of the fact that this method proves to be extremely useful in developing non-perturbative techniques.''---Gervais and Jevicki \cite{gervaisJevicki}
\end{quotation}
I set out in Sec.~\ref{sec:ItoCorrectionInPathIntegral} to derive the Ito correction purely within the path-integral formalism. What I ended up showing was not quite up to the standard of Gervais and Jevicki. 
\\\\
What I should have done is start from the regularized version of Eqs.~(\ref{eq:P=sqrtDetG_e^-S}) and~(\ref{eq:S=NLSM}), namely
\begin{equation}
P(x(\cdot)) = \prod_{i\,=\,0}^{N-1}\sqrt{g(x_i)}\;e^{-\Delta t\,\half g(x_i)\left[\frac{\Delta x_i}{\Delta t} -\mu(x_i)\right]^2}\;,
\end{equation}
and attempt to perform the point transformation 
\begin{equation}
x_i \to X_i = e^{\,x_i}
\end{equation}
without any recourse to the work I did in Sec.~\ref{sec:nonlinearSigmaModel}. Many of Gervais and Jevicki's expressions look similar to mine, and many of the steps in my Sec.~\ref{sec:nonlinearSigmaModel} would largely be repeated in such a derivation. I decided to just reuse my work from Sec.~\ref{sec:nonlinearSigmaModel} and make sure it is consistent with the normalization in Eq.~(\ref{eq:noncovariantNormalization}). 
\newpage
\section{The Model as a Gauge Theory}\label{sec:gaugeTheory}
Remember that preview of invariance principles from all the way back in Sec.~\ref{sec:poissonPathIntegral}? Now you are ready. 
\\\\
Everything heretofore discussed can be summarized by the action
\begin{equation}\label{eq:basicModel}
S(x(\cdot)) = \int\! dt\;\frac{1}{2c(t)}\left[Dx(t)\right]^2\;,\;\;Dx(t) = \dot x(t) + a(t)\,x(t)+b(t)\;.
\end{equation}
That form is guided by covariance and invariance under local dilations, translations, and temporal reparametrizations. I will explain each in turn. 
%
% \section{Local Transformations}
%
%
\subsection{Dilation}\label{sec:dilation}
Consider a local rescaling of the field, called a dilation:
\begin{equation}
x(t) \to x'(t) = \Lambda(t)\,x(t)\;.
\end{equation}
Under that transformation, the derivative of the field transforms as:
\begin{align}
\dot x(t) &\to \dot x'(t) = \frac{d}{dt}\left[\Lambda(t)\,x(t)\right] \\
&= \Lambda(t)\,\dot x(t) + \dot\Lambda(t)\,x(t)\;.
\end{align}
If $\dot\Lambda(t)$ were zero, then the derivative of the field would transform as the field itself. A standard procedure in physics is to introduce an additional degree of freedom, $a(t)$, to cancel off the term $\dot\Lambda(t)\,x(t)$. Let:
\begin{equation}
Dx(t) \equiv \dot x(t)+a(t)\,x(t)\;.
\end{equation}
Fix the transformation law of $a(t)$ so that $Dx(t)$ would transform as $\dot x(t)$ would if $\Lambda$ were constant:
\begin{align}
Dx(t) \to (Dx)'(t) &= \Lambda(t)\,\dot x(t) + \dot\Lambda(t)\, x(t) + a'(t)\,\Lambda(t)\,x(t) \\
&\equiv \Lambda(t)\,Dx(t) = \Lambda(t)\,\dot x(t) + \Lambda(t)\,a(t)\,x(t)
\end{align}
Solving for $a'(t)$ gives the transformation law
\begin{equation}
a(t) \to a'(t) = a(t) - \Lambda(t)^{-1}\dot \Lambda(t) = a(t) - \frac{d}{dt}\ln\left[\Lambda(t)\right]\;.
\end{equation}
The combination $Dx(t)$ is said to transform covariantly under local dilations; the additional degree of freedom, $a(t)$, is called the corresponding gauge field. The procedure of introducing a gauge field to promote a global transformation into a local one is called ``gauging the transformation.'' 
\\\\
The gauge field $a(t)$ could be taken to be a fixed background or given its own dynamics. I will demonstrate in Sec.~\ref{sec:gaugeTheoryApplied} that the mean-reversion speed of the Ornstein-Uhlenbeck model should be thought of as a constant background value for the field $a(t)$.
\\\\
In any case, given the covariant derivative $Dx(t)$, the combination that appears in the action (with $b = 0$) transforms as
\begin{equation}
[Dx(t)]^2 \to [(Dx)'(t)]^2 = \Lambda(t)^2\,[Dx(t)]^2\;.
\end{equation}
For the action to be invariant, the field $c(t)$ will have to transform with two powers of $\Lambda(t)$:
\begin{equation}\label{eq:c'}
c(t) \to c'(t) = \Lambda(t)^2\,c(t)\;.
\end{equation}
For a constant $c(t)$---i.e., for constant squared volatility---the action would not be invariant under local rescalings of the field. That makes sense, because the volatility sets the scale of fluctuations. Only when the scale of fluctuations itself compensates for a local rescaling of the field will the distribution of paths be invariant. 
\\\\
The question then arises as to whether the field $c(t)$ has its own dynamics. Models in which it does are called ``stochastic volatility models'' in the financial literature; a standard example is the Heston model (see Gatheral \cite{gatheral}).
\subsection{Translation}\label{sec:translation}
This time, set $a = 0$, and consider a local shifting of the field, called a translation:
\begin{equation}
x(t) \to x'(t) = x(t) + \xi(t)\;.
\end{equation}
Fix the transformation law of $b(t)$ so that $Dx(t)$ would transform as $\dot x(t)$ would if $\xi$ were constant:
\begin{equation}
b(t) \to b'(t) = b(t) - \dot \xi(t)\;.
\end{equation}
In that case, 
\begin{equation}
[Dx(t)]^2 \to [Dx(t)]^2\;,
\end{equation}
and the action in Eq.~(\ref{eq:basicModel}) is invariant. This is the mathematical construction needed to implement the artifice of moving planks for a configuration-space description of a Poisson-driven process (recall Sec.~\ref{sec:planksAsMath}). 
\subsection{Dilation and Translation}\label{sec:dilationAndTranslation}
To organize the mathematics of a simultaneous dilation and translation, it is convenient to introduce a $2\times 2$ matrix formalism. Introduce an auxiliary degree of freedom, $y(t)$,\footnote{This has no relation whatsoever to the fluctuation notation in Eq.~(\ref{eq:decompositionOfGeneralPath}).} and define
\begin{equation}
X(t) \equiv \left(\begin{matrix} x(t) \\ y(t) \end{matrix}\right)\;.
\end{equation}
Consider the local transformation
\begin{equation}
X(t) \to X'(t) = M(t)\,X(t)\;,\;\; M(t) = \left(\begin{matrix} \Lambda(t) & \xi(t) \\ 0 & 1  \end{matrix}\right)\;.
\end{equation}
The components of $X(t)$ transform as\footnote{Because the transformation law for $y(t)$ does not involve $x(t)$, the representation is reducible. But because the transformation law for $x(t)$ does involve $y(t)$, the representation does not split. This is a hallmark of semidirect-product groups (I thank Alexei Kitaev for explaining this to me in an unrelated context years ago). Moreover, there are two inequivalent ways to compose translations and dilations; the alternative to Eq.~(\ref{eq:combinedTranslationAndDilation}) is $x(t) \to x'(t) = \hat\Lambda(t)\left[x(t) + \hat\xi(t)\,y(t)\right]$. As emphasized by Jaynes \cite{jaynes} (his Sec.~12.4), the groups are different and describe different prior information in the context of inference. For my purpose here, I do not think it makes a difference which group I use, and it is unclear to me whether the existence of both invariances has additional consequences.}
\begin{align}
&x(t) \to x'(t) = \Lambda(t)\,x(t) + \xi(t)\,y(t)\;, \label{eq:combinedTranslationAndDilation}\\
&y(t) \to y'(t) = y(t)\;.
\end{align}
For $y(t) = 1$, one recovers the desired combination of dilation and translation:
\begin{equation}\label{eq:transformedField}
x(t) \to x'(t) = \Lambda(t)\,x(t) + \xi(t)\;.
\end{equation}
Define
\begin{equation}
DX(t) = \dot X(t) + A(t)\,X(t)\;,\;\;A(t) = \left(\begin{matrix} a(t) & b(t) \\ 0 & 0 \end{matrix}\right)\;.
\end{equation}
Demanding that $DX(t)$ transform as $\dot X(t)$ would if $M(t)$ were constant leads to the transformation law
\begin{equation}
A(t) \to A'(t) = M(t)\left(I\frac{d}{dt}+A(t)\right)M^{-1}(t)\;,
\end{equation}
where the inverse matrix is
\begin{equation}
M^{-1}(t) = \left(\begin{matrix} \Lambda(t)^{-1} & -\Lambda(t)^{-1}\xi(t) \\ 0 & 1\end{matrix}\right)\;.
\end{equation}
The transformation law of $a(t)$ does not depend on the translation:
\begin{equation}\label{eq:transformationLawForDilationGaugeField}
a(t) \to a'(t) = \Lambda(t)\left[\frac{d}{dt}+a(t)\right]\Lambda(t)^{-1} = a(t)-\frac{d}{dt}\ln\left[\Lambda(t)\right]
\end{equation}
But the transformation law of $b(t)$ does depend on the dilation:
\begin{align}
b(t) \to b'(t) &= \Lambda(t)\left\{ b(t) - \left[\frac{d}{dt}+a(t) \right]\left[\Lambda(t)^{-1}\xi(t)\right]\right\} \label{eq:transformationLawForTranslationGaugeField} \\
&= \Lambda(t)\,b(t) - \left\{\dot\xi(t) + \left[a(t)-\frac{d}{dt}\ln\left(\Lambda(t)\right)\right]\xi(t)\right\}\;.
\end{align}
The purpose of the $2\times 2$ matrix formalism is to organize that mouthful. Note that, in the absence of a local translation, the translational gauge field transforms in the way that $x(t)$ does:
\begin{equation}
b(t) \to b'(t) = \Lambda(t)\,b(t)\qquad(\text{for }\; \xi = 0)\;.
\end{equation}
The translational gauge field, in the gauge $\xi(t) = 0$, transforms as a frame field \cite{grignaniNardelli}.
\subsection{Temporal Reparametrization}\label{sec:temporalReparametrization}
Consider a reparametrization of time:
\begin{equation}\label{eq:temporalReparametrization}
t \equiv f(\tau) \implies dt = \partial_\tau f(\tau)\, d\tau\;.
\end{equation}
Let\footnote{One is not typically that pedantic, but I have already demonstrated in Sec.~\ref{sec:logField} what might happen if one forgets that invariance of a scalar field is not a foregone conclusion.}
\begin{equation}\label{eq:fieldInReparametrizedTime}
\widetilde x(\tau) \equiv x(t = f(\tau))\;.
\end{equation}
The derivative is
\begin{equation}
\dot x(t) \equiv \frac{dx(t)}{dt} = \frac{d\tau}{dt}\frac{d\widetilde x(\tau)}{d\tau} = \frac{1}{\partial_\tau f(\tau)}\partial_\tau\widetilde x(\tau)\;,
\end{equation}
which means that:
\begin{align}
Dx(t) &\equiv \dot x(t) + a(t)\,x(t)+b(t) \nonumber\\
&= \frac{1}{\partial_\tau f(\tau)}\left\{\partial_\tau \widetilde x(\tau) + \partial_\tau f(\tau)\left[a(t)\,\widetilde x(\tau) + b(t) \right]\right\}\;.
\end{align}
Evidently I should define
\begin{equation}\label{eq:gaugeFieldsInReparametrizedTime}
\widetilde a(\tau) \equiv \partial_\tau f(\tau)\,a(f(\tau))\;,\;\; \widetilde b(\tau) \equiv \partial_\tau f(\tau)\,b(f(\tau))\;,
\end{equation}
which could have been anticipated from the definition of the gauge fields as part of the covariant derivative. In terms of the new covariant derivative
\begin{equation}\label{eq:reparametrizedCovariantDerivative}
\widetilde{Dx}(\tau) \equiv \frac{d}{d\tau}\widetilde x(\tau) + \widetilde a(\tau)\,\widetilde x(\tau) + \widetilde b(\tau)\;,
\end{equation}
the original covariant derivative is
\begin{equation}\label{eq:relationBetweenCovariantDerivatives}
Dx(t) = \frac{1}{\partial_\tau f(\tau)}\,\widetilde{Dx}(\tau)\;.
\end{equation}
Therefore, 
\begin{equation}
dt\left[Dx(t)\right]^2 = \partial_\tau f(\tau)\;d\tau\left(\frac{1}{\partial_\tau f(\tau)}\right)^2\left[\widetilde{Dx}(\tau)\right]^2 = \frac{1}{\partial_\tau f(\tau)}\;d\tau\left[\widetilde{Dx}(\tau)\right]^2\;.
\end{equation}
For the action to be invariant under a general temporal reparametrization, the field $c(t)$ must be rescaled in the manner of $a(t)$ and $b(t)$:
\begin{equation}\label{eq:dynamicalScaleOfTime}
\widetilde c(\tau) \equiv \partial_\tau f(\tau)\,c(f(\tau))\;.
\end{equation}
In reparametrized time, the action is
\begin{equation}
S = \int\! d\tau\;\frac{1}{2\widetilde c(\tau)}\left[\widetilde{Dx}(\tau)\right]^2\;,
\end{equation}
with the covariant derivative in Eq.~(\ref{eq:reparametrizedCovariantDerivative}). 
\\\\
Consider the particular reparametrization for which 
\begin{equation}\label{eq:c=1}
\widetilde c(\tau) = 1\;.
\end{equation}
In that case, the action will reduce to
\begin{equation}
S = \int\! d\tau\;\half\left[\widetilde{Dx}(\tau)\right]^2\;,
\end{equation}
which describes a canonically normalized stochastic process that evolves under a modified clock. 
\\\\
This is gravitational time-dilation. It is known in the mathematical and financial literature as subordination \cite{mandelbrotTaylor}. The field $c(t)$ is therefore a dynamical scale of time, variously known as a gravitational field, dilaton, subordinator, local volatility, or trading time. 
\subsection{Further Remarks about Field-Dependent Volatility}\label{sec:nonlinearSigmaModel2}
Let me briefly revisit the nonlinear sigma model, generalized further with time-dependent parameters:
\begin{equation}\label{eq:NLSMwithTimeDependentParameters}
S(x(\cdot)) = \int\!dt\;\frac{1}{2}\,g(x(t),t)\left[\dot x(t) - \mu(x(t),t)\right]^2\;.
\end{equation}
Eq.~(\ref{eq:NLSMwithTimeDependentParameters}) will reduce to the model in Eq.~(\ref{eq:basicModel}) in the special case
\begin{equation}
g(x(t),t) = \frac{1}{c(t)}\;,\;\;\mu(x(t),t) = a(t)\, x(t) + b(t)\;.
\end{equation}
An alternative special case could be
\begin{equation}
g(x(t),t) = g(x(t))\;,\;\;\mu(x(t),t) = \mu(x(t))\;,
\end{equation}
which is what I discussed in Sec.~\ref{sec:nonlinearSigmaModel}.
\\\\
In principle, the first special case could be turned into the second special case by introducing independent dynamics for the functions $a(t)$, $b(t)$, and $c(t)$---those functions would then be promoted from deterministic background fields to independent stochastic degrees of freedom. Upon integrating them out, one would (again, in principle) obtain an effective action for $x(t)$ in the form of a nonlinear sigma model, plus corrections. 
\\\\
I will not discuss that here. I point it out because in the financial literature both types of models appear, and it took me some time to disentangle them. Let me provide an example, in the form of the multicomponent stochastic differential equation
\begin{equation}\label{eq:multicomponentBrownianMotion}
dx^\mu(t) = \sum_{a\,=\,1}^d \sigma_a^{\,\mu}(x(t))\,Z^a(t)\,\sqrt{dt}\;;\;\;\mu = 1,\ldots,d\;.
\end{equation}
Eq.~(\ref{eq:multicomponentBrownianMotion}) describes $d$-component, correlated Brownian motion. If the Gaussian variates have the unconditional two-point correlation, for each $t$,
\begin{equation}
E\left[Z^a(t)\,Z^b(t)\right] = \eta^{ab}\;,
\end{equation}
then the conditional expectation of the squared increment is
\begin{align}
g^{\mu\nu}(x) \equiv E\left[\left.\frac{dx^\mu(t)\,dx^\nu(t)}{dt}\,\right|\left\{x^\rho(t)\,=\,x^\rho\right\}_{\rho\,=\,1}^d\right] = \sum_{a,b\,=\,1}^d\sigma_a^{\,\mu}(x)\,\sigma_b^{\,\nu}(x)\,\eta^{ab}\;.
\end{align}
In differential geometry, that is the standard decomposition of a metric into a local frame. The action that follows from turning Eq.~(\ref{eq:multicomponentBrownianMotion}) into a path integral is
\begin{equation}
S(x(\cdot)) = \int\!dt\;\frac{1}{2}\!\sum_{\mu,\,\nu\,=\,1}^dg_{\mu\nu}(x(t))\,\dot x^\mu(t)\,\dot x^\nu(t)\;,
\end{equation}
where $g_{\mu\nu}$ is the matrix-inverse of $g^{\mu\nu}$. That means that the volatility matrix is the frame for the \textit{sigma-model} metric, not the gravitational field. The gravitational field is derived from $c(t)$, which is what can be interpreted as a dynamical scale of time. 
\newpage
\section{Gauge Theory Applied}\label{sec:gaugeTheoryApplied}
The takaway of Sec.~\ref{sec:gaugeTheory} is that the action 
\begin{equation}\label{eq:actionWeWant}
S = \int_{t_0}^{t_f}\!\! dt\;\frac{1}{2c(t)}\left[\frac{d}{dt}x(t)+a(t)\,x(t) + b(t)\right]^2
\end{equation}
can be transformed into the action
\begin{equation}\label{eq:actionWeKnow}
S = \int_{\tau_0}^{\tau_f}\!\! d\tau\;\frac{1}{2\widetilde c(\tau)}\left[\frac{d}{d\tau}\widetilde x(\tau) + \widetilde a(\tau)\,\widetilde x(\tau) + \widetilde b(\tau)\right]^2
\end{equation}
by a combination of local dilation, local translation, and temporal reparametrization. The purpose of doing so would be to transform a conditional probability I want into a conditional probability I know, then transform back. In view of that intended application, I have restored the bounds of integration in Eqs.~(\ref{eq:actionWeWant}) and~(\ref{eq:actionWeKnow}).
\subsection{Pure Gauge}
The simplest action is that of Brownian motion. So let me explore the consequence of attempting to transform the general action in Eq.~(\ref{eq:actionWeWant}) into the simplest possible form of Eq.~(\ref{eq:actionWeKnow}):
\begin{equation}
S = \int_{\tau_0}^{\tau_f}\!\! d\tau\;\frac{1}{2}\left[\frac{d}{d\tau}\widetilde x(\tau)\right]^2\;,\;\;\text{i.e.,}\;\;\;\; \widetilde a(\tau) = 0\;,\;\;\widetilde b(\tau) = 0\;,\;\;\widetilde c(\tau) = 1\;.
\end{equation}
I will refer to gauge fields that are gauge transformations of zero as ``pure gauge.''\footnote{In field theory, the term ``pure gauge'' describes gauge fields whose field strengths are zero. In $0+1$ dimensions there is no field strength.} The transformation law for the dilation gauge field is independent of the transformation law for the translation gauge field, so I will start with Eq.~(\ref{eq:transformationLawForDilationGaugeField}), with $a'(t) = 0$:
\begin{equation}\label{eq:wilsonLine}
a(t) - \frac{d}{dt}\ln[\Lambda(t)] = 0 \implies \Lambda(t) = \Lambda(t_0)\,e^{\int_{t_0}^t\! dt'a(t')}\;.
\end{equation}
The exponentiated integral of a gauge field is called a Wilson line.\footnote{I do not know the precise reference in which Wilson, or anyone else, first defined this concept. For a brief overview, see Ch.~82 of Srednicki \cite{srednicki}. There is a long history of trying to formulate nonabelian gauge theory in a gauge-invariant way as a dynamical theory of Wilson loops, but I know very little about it.} It describes the procedure of covariant parallel transport from one instant to another instant. If the integral $\int_{t_0}^t\!dt'\, a(t')$ can be evaluated, then the transformation function $\Lambda(t)$ will be known. 
\\\\
One down, two to go. Under the transformation in Eq.~(\ref{eq:wilsonLine}), the transformation for the translational gauge field simplifies. For $b'(t) = 0$ and $a(t) = (d/dt)\ln[\Lambda(t)]$, Eq.~(\ref{eq:transformationLawForTranslationGaugeField}) becomes:
\begin{equation}\label{eq:wilsonLineForTranslations}
\Lambda(t)\, b(t) - \frac{d}{dt}\xi(t) = 0 \implies \xi(t) = \xi(t_0) + \int_{t_0}^t \! dt'\,\Lambda(t')\,b(t')\;.
\end{equation}
I will call that the residual Wilson line for translations. If $\Lambda(t)$ is known, and if the integral $\int_{t_0}^t\! dt'\,\Lambda(t')\,b(t')$ can be evaluated, then the transformation function $\xi(t)$ will be known. 
\\\\
Given the Wilson line, the transformed dynamical scale of time is known from Eq.~(\ref{eq:c'}), repeated below for convenience:
\begin{equation}\label{eq:c'again}
c'(t) = \Lambda(t)^2c(t)\;.
\end{equation}
The idea is then to reparametrize time and consult Eqs.~(\ref{eq:dynamicalScaleOfTime}) and~(\ref{eq:c=1}), except with $c(t)$ replaced by $c'(t)$:
\begin{equation}\label{eq:solveThisForProperTime}
\widetilde c(\tau) = \partial_\tau f(\tau)\,c'(f(\tau)) \equiv 1\;.
\end{equation}
Since temporal reparametrizations just rescale the gauge fields (recall Eq.~(\ref{eq:gaugeFieldsInReparametrizedTime})), $a'$ and $b'$ will remain zero. So all that remains is to solve Eq.~(\ref{eq:solveThisForProperTime}) for the reparametrization function $f(\tau)$. The success of the method can be judged on the extent to which that is possible. 
%If that integral can be done, then the proper time $\tau$ will be known as a function of coordinate time; if the result can be inverted, then the reparametrization function $f(\tau)$ will be known. **the example below just showed me that i treated $c$ and $\widetilde c$ backwards: the gauge transformation for constant $c$ will INDUCE a nonconstant $\widetilde c$, which then has to be compensated by a temporal reparametrization. fix it.**
%
\subsection{Mean Reversion}\label{sec:meanReversion}
A special case in which everything can be calculated is where all of the external fields are constant:
\begin{equation}
a(t) = \theta\;,\;\;b(t) = -\mu\;,\;\; c(t) = \sigma^2\;.
\end{equation}
The corresponding action
\begin{equation}\label{eq:OUaction}
S = \int_{t_0}^{t_f}\!\!dt\;\frac{1}{2\sigma^2}\left[\dot x(t) + \theta x(t) -\mu\right]^2
\end{equation}
describes the Ornstein-Uhlenbeck model of mean reversion, with mean-reversion speed $\theta$, drift $\mu$, volatility $\sigma$, and asymptotic mean zero. Alternatively, as is more standard, you can think of Eq.~(\ref{eq:OUaction}) as describing a process with drift zero and asymptotic mean $m = \mu/\theta$. 
\\\\
Given everything in Sec.~\ref{sec:gaugeTheory}, I interpret this model as Brownian motion in the presence of constant background gauge fields for dilations and translations. 
\\\\
With the initial condition $\Lambda(t_0) = 1$, the Wilson line of Eq.~(\ref{eq:wilsonLine}) for a constant gauge field $a(t) = \theta$ becomes
\begin{equation}
\Lambda(t) = e^{\,\theta\cdot (t-t_0)}\;.
\end{equation}
With the initial condition $\xi(t_0) = 0$, the residual translational Wilson line of Eq.~(\ref{eq:wilsonLineForTranslations}) for $b(t) = -\mu$ becomes
\begin{equation}
\xi(t) = -\mu\int_{t_0}^t\! dt'\,e^{\,\theta\cdot(t'-t_0)} = \frac{\mu}{\theta}\left(1-e^{\,\theta\cdot(t-t_0)}\right)\;.
\end{equation}
Inserting the Wilson line into Eq.~(\ref{eq:c'again}) gives 
\begin{equation}
c'(t) = e^{\,2\theta\cdot(t-t_0)}\sigma^2\;.
\end{equation}
Eq.~(\ref{eq:solveThisForProperTime}) is then
\begin{equation}
1 = \partial_\tau f(\tau) c'(f(\tau)) = \frac{dt}{d\tau}c'(t) \implies d\tau = dt\,c'(t)\;,
\end{equation}
which can be integrated:
\begin{equation}\label{eq:properTimeForOU}
\tau-\tau_0 = \int_{t_0}^{t}\!dt'\,c'(t') = \sigma^2\int_{t_0}^{t}\!dt'\,e^{\,2\theta\cdot(t'-t_0)} = \frac{\sigma^2}{2\theta}\left(e^{\,2\theta\cdot(t-t_0)}-1\right)\;.
\end{equation}
%\\\\
Through a sequence of transformations, the action of Eq.~(\ref{eq:OUaction}) has been successfully brought into the form
\begin{equation}
S = \int_{\tau_0}^{\tau_f}\!\!d\tau\;\frac{1}{2}\left[\frac{d}{d\tau}\widetilde x(\tau)\right]^2\;,
\end{equation}
with all transformation functions known. 
\subsubsection{Conditional Probability in Transformed Variables}\label{sec:conditionalProbabilityInTransformedVariables}
The conditional probability in the transformed variables is old trusty:
\begin{equation}\label{eq:conditionalProbabilityInTransformedVariables}
\widetilde P(\widetilde x_f,\tau_f|\widetilde x_0,\tau_0) = \frac{1}{\sqrt{2\pi(\tau_f-\tau_0)}}\;e^{-\frac{1}{2(\tau_f-\tau_0)}(\widetilde x_f-\widetilde x_0)^2}\;.
\end{equation}
The values of the field at the endpoints can be calculated from Eqs.~(\ref{eq:transformedField}) and~(\ref{eq:fieldInReparametrizedTime}):
\begin{equation}
\widetilde x_i = \Lambda(t_i)\,x_i + \xi(t_i)\;,\;\;i \in \{0,f\}\;.
\end{equation}
The initial conditions $\Lambda(t_0) = 1$ and $\xi(t_0) = 0$ were chosen so that $\widetilde x_0 = x_0$. That leaves the terminal value:
\begin{align}
\widetilde x_f &= \Lambda(t_f)\,x_f+\xi(t_f) \label{eq:terminalValueGeneralForm}\\
&= e^{\,\theta\cdot(t_f-t_0)}x_f+\frac{\mu}{\theta}\left(1-e^{\,\theta\cdot(t_f-t_0)}\right)\\
&= e^{\,\theta\cdot(t_f-t_0)}\left[x_f-\frac{\mu}{\theta}\left(1-e^{-\theta\cdot(t_f-t_0)}\right)\right]\;.\label{eq:terminalValueSpecificForm}
\end{align}
Therefore:\footnote{The parentheses around $x_f-\frac{\mu}{\theta}$ are just to guide the eye.}
\begin{align}
\widetilde x_f - \widetilde x_0 &= e^{\,\theta\cdot(t_f-t_0)}\left[x_f-x_0\,e^{-\theta\cdot(t_f-t_0)}-\frac{\mu}{\theta}\left(1-e^{-\theta\cdot(t_f-t_0)}\right)\right]\;, \\
&= e^{\,\theta\cdot(t_f-t_0)}\left[\left(x_f-\frac{\mu}{\theta}\right)-\left(x_0-\frac{\mu}{\theta}\right)e^{-\theta\cdot(t_f-t_0)}\right]\;,\label{eq:totalDisplacement}
\end{align}
and:
\begin{align}
\frac{(\widetilde x_f-\widetilde x_0)^2}{\tau_f-\tau_0} &= \frac{1}{\frac{\sigma^2}{2\theta}\left(1-e^{-2\theta\cdot(t_f-t_0)}\right)}\left[\left(x_f-\frac{\mu}{\theta}\right)-\left(x_0-\frac{\mu}{\theta}\right)e^{-\theta\cdot(t_f-t_0)}\right]^2\;.\label{eq:exponentOfTransformedConditionalProbability}
\end{align}
A brief remark is in order in case you want to compare Eq.~(\ref{eq:exponentOfTransformedConditionalProbability}) to other references, for example Wolfram's documentation on the Ornstein-Uhlenbeck process \cite{wolframOU}.
\\\\
The symbol ``$\mu$'' has at least three distinct and incompatible standard usages in statistics: The mean of a Gaussian distribution, the drift of a Brownian motion, and the asymptotic mean of an Ornstein-Uhlenbeck process. It is in the second sense that the symbol ``$\mu$'' is used in Eq.~(\ref{eq:OUaction}), while it is in the third sense that the symbol ``$\mu$'' is used in the Wolfram documentation. My $\theta$ is the same as theirs. Finally, note that I work with the boundary conditions $x_f,t_f$ and $x_0,t_0$, whereas they work with $x,t$ and $x_0,0$. 
\\\\
When I first stared at Wolfram's assertion that ``OrnsteinUhlenbeckProcess[$\mu,\sigma,\theta,x_0$] value at time $t$ follows NormalDistribution[$\mu+(x_0-\mu)exp(-\theta t),\sqrt{\sigma^2(1-exp(-2\theta t))/(2\theta)}$],'' I found it difficult to understand where those combinations of parameters came from. My derivation demonstrates that they come from gauge invariance. A derivation that does not rely on gauge invariance is in App.~\ref{sec:generatingFunctionForOU}.
\subsubsection{Conditional Probability in Original Variables}
Finally, there is the question of normalization. The conditional probability in Eq.~(\ref{eq:conditionalProbabilityInTransformedVariables}) is normalized in the sense that
\begin{equation}
\int_{-\infty}^{\infty}\!\! d\widetilde x_f\;\widetilde P(\widetilde x_f,\tau_f|\widetilde x_0,\tau_0) = 1\;.
\end{equation}
Meanwhile, the conditional probability in the original variables should also be normalized:
\begin{equation}
\int_{-\infty}^{\infty}\!\! dx_f\; P(x_f,t_f|x_0,t_0) = 1\;.
\end{equation}
From Eqs.~(\ref{eq:terminalValueGeneralForm}) and~(\ref{eq:terminalValueSpecificForm}), I know that there is a Jacobian for the terminal value:
\begin{equation}\label{eq:jacobianForTheTerminalValue}
d\widetilde x_f = \Lambda(t_f)\,dx_f = e^{\,\theta\cdot(t_f-t_0)}\,dx_f\;.
\end{equation}
So it must be the case that, once again, the conditional probability transforms as a density in the forward variable:
\begin{equation}
P(x_f,t_f|x_0,t_0) = e^{-\theta\cdot(t_f-t_0)}\,\widetilde P(\widetilde x_f,\tau_f|\widetilde x_0,\tau_0)\;.
\end{equation}
Using the explicit solution for the proper time from Eq.~(\ref{eq:properTimeForOU}), I find the conditional probability
\begin{equation}
P(x_f,t_f|x_0,t_0) = \frac{1}{\sqrt{2\pi \frac{\sigma^2}{2\theta}\left( 1-e^{-2\theta\cdot(t_f-t_0)}\right)}}\;e^{-K}\;,
\end{equation}
with $K$ the right-hand side of Eq.~(\ref{eq:exponentOfTransformedConditionalProbability}):
\begin{equation}
K = \frac{1}{\frac{\sigma^2}{2\theta}\left(1-e^{-2\theta\cdot(t_f-t_0)}\right)}\left[\left(x_f-\frac{\mu}{\theta}\right)-\left(x_0-\frac{\mu}{\theta}\right)e^{-\theta\cdot(t_f-t_0)}\right]^2\;.
\end{equation}
Why did the Jacobian in Eq.~(\ref{eq:jacobianForTheTerminalValue}) need to be fixed by hand? I showed that the action is gauge invariant, but in doing so I did not include the transformation of the path-integral measure. The transformations of this section did not depend on the field, so the Jacobian could always be fixed after the fact by normalizing the path integral. That is the procedure I adopted here. 
\subsection{Jumps}\label{sec:meanReversionWithJumps}
Adding jumps to the mean-reverting process would amount to replacing the constant $b(t)$ with a sequence of impulses,
\begin{equation}
b(t) = \sum_{\alpha}\gamma_\alpha\,\delta(t-t_\alpha^*)\;,
\end{equation}
then averaging over the jump heights $\gamma_\alpha$ and jump times $t_\alpha^*$. The new local translation would be
\begin{equation}
\xi(t) = \sum_\alpha \gamma_\alpha\, e^{\,\theta\cdot(t_\alpha^*-t_0)}\,\Theta(t-t_\alpha^*)\;.
\end{equation}
So this case would just amount to Brownian motion on a constant background dilation field as well as a background of translational defects, as I emphasized in earlier sections. 
\newpage
\section{Discussion}\label{sec:discussion}
The basic model of stochastic mechanics is diffusion with drift and jumps. I have provided an introductory but detailed take on the path-integral description of that model.\footnote{Given that Wiener's paper was written over 100 years ago, I am hardly the first to do this and will not be the last. I offer a choice remark by Schulman \cite{schulman} (p.~39): ``Methods for handling the quadratic Lagrangian are legion and have been well developed since the earliest work on path integrals. Oddly enough papers on the subject continue to appear and may give some historian of science material for a case history on the nondiffusion of knowledge.'' I, for one, knew none of the material I developed in Sec.~\ref{sec:poissonPathIntegral} before trying to understand the pertinent chapter of Gatheral \cite{gatheral}.} I will discuss where to go from here, especially in the context of financial mechanics. 
\subsection{Phenomenology}\label{sec:phenomenology}
A cursory look at the price history of a typical collection of stocks will demonstrate that, as a starting point, diffusion with drift and jumps is pretty good. (Remember that it is the percentage change in price, not the absolute change in price, that diffuses.) The extent to which that is all there is to it constitutes the gladiator games between hedge funds.
\\\\
I will offer some thoughts about more realistic phenomenology, taking various empirical assertions about financial time series at face value. 
\subsubsection{Interactions}\label{sec:interactions}
As should be familiar by now, the conditional probability for Brownian motion with constant volatility $\sigma$ and constant drift $\mu$ is
\begin{equation}\label{eq:conditionalProbabilityForBrownianMotionWithConstantVolatilityAndDrift}
U_B(x_f,t_f|x_0,t_0) = \int_{x(t_0)\,=\,x_0}^{x(t_f)\,=\,x_f} \!\!\!\!\mathscr Dx(\cdot)\;e^{-\int_{t_0}^{t_f}\! dt\;\frac{1}{2\sigma^2}\left[\dot x(t)-\mu\right]^2}\;.
\end{equation}
One way to generalize that model would be to add a potential, $V(x)$. A quadratic polynomial would keep the action quadratic, and hence more or less equivalent to the Ornstein-Uhlenbeck model.\footnote{I have a few bones to pick with the common take that the Lagrangian $\la = \half\left(\dot x + \theta x\right)^2$ is ``like a harmonic oscillator'' because it equals $\half\dot x^2 + \half\theta^2x^2$ up to a boundary term. First, the important object in finance is the conditional probability, which is the path integral with specified boundary conditions---the boundary terms matter. Second, the concept of stochastic fluctuations around deterministic exponential decay is too different from quantum fluctuations around determinsitic oscillations to be lumped into the same basket of ideas. Third, as I explained in the main text, the mean-reversion speed is a constant background profile for the gauge field of dilations, not a mass term.} To add bona fide interactions, one has to go beyond quadratic order. 
\\\\
Perturbation theory along those lines is the bread and butter of field theory. I will explain how to set up the problem and leave the details for another day. 
\\\\
Consider a quartic potential,
\begin{equation}
V(x) = \tfrac{1}{4!}\lambda\,x^4\;,
\end{equation}
with $\lambda$ a new constant. Instead of Eq.~(\ref{eq:conditionalProbabilityForBrownianMotionWithConstantVolatilityAndDrift}), the conditional probability will be:
\begin{equation}\label{eq:quarticInteraction}
U(x_f,t_f|x_0,t_0) = \int_{x(t_0)\,=\,x_0}^{x(t_f)\,=\,x_f} \!\!\!\!\mathscr Dx(\cdot)\;e^{-\int_{t_0}^{t_f}\! dt\;\left\{\frac{1}{2\sigma^2}\left[\dot x(t)-\mu\right]^2 + \frac{1}{4!}\lambda\,x(t)^4\right\}}\;.
\end{equation}
To proceed perturbatively in $\lambda$, one reads Eq.~(\ref{eq:quarticInteraction}) as the biconditional expectation of $e^{-\int_{t_0}^{t_f}\!dt\;\frac{1}{4!}\lambda\,x(t)^4}$ with respect to the Brownian action. Expanding in $\lambda$ therefore gives a sequence of autocorrelation functions of Brownian motion:
\begin{align}\label{eq:perturbationTheoryForQuarticInteraction}
&U(x_f,t_f|x_0,t_0) = \nonumber\\
&\quad\sum_{n\,=\,0}^{\infty}\frac{1}{n!}\left(\frac{\lambda}{4!}\right)^n\!\!\int_{t_0}^{t_f}\!\! dt_1\ldots \int_{t_0}^{t_f}\!\! dt_n\int_{x(t_0)\,=\,x_0}^{x(t_f)\,=\,x_f} \!\!\!\!\mathscr Dx(\cdot)\;e^{-\int_{t_0}^{t_f}\! dt\;\frac{1}{2\sigma^2}\left[\dot x(t)-\mu\right]^2}x(t_1)^4\ldots x(t_n)^4\;.
\end{align}
One way to compute those expectations is to introduce a correlation-generating function.  Generalize Eq.~(\ref{eq:conditionalProbabilityForBrownianMotionWithConstantVolatilityAndDrift}) by introducing a linear coupling to a local source:
\begin{equation}\label{eq:generatingFunctionForBrownianMotion}
U_B(J(\cdot);x_f,t_f|x_0,t_0) = \int_{x(t_0)\,=\,x_0}^{x(t_f)\,=\,x_f} \!\!\!\!\mathscr Dx(\cdot)\;e^{-\int_{t_0}^{t_f}\! dt\;\left\{\frac{1}{2\sigma^2}\left[\dot x(t)-\mu\right]^2 - J(t)\,x(t)\right\}}\;.
\end{equation}
Each functional derivative with respect to $J(t)$ will bring down a power of $x(t)$ inside the path integral, enabling construction of the series in Eq.~(\ref{eq:perturbationTheoryForQuarticInteraction}). And since the action in Eq.~(\ref{eq:generatingFunctionForBrownianMotion}) remains quadratic, one can compute $U_B(J(\cdot);x_f,t_f|x_0,t_0)$ in terms of the Green function for Brownian motion. See App.~\ref{sec:generatingFunctionForOU} for a derivation of the correlation-generating function for the Ornstein-Uhlenbeck model.
%\\\\
\subsubsection{Effective Field Theory}\label{sec:effectiveFieldTheory}
\begin{quote}
``The renormalization group is not a descriptive theory of nature but a general method for constructing theories.''---Wilson \cite{wilsonManyScales}
\end{quote}
By adding local interactions of the type in Sec.~\ref{sec:interactions}, one can systematically explore the corner of model space perturbatively related to Brownian motion. This probably amounts to a more systematic formulation of GARCH models (see Mandelbrot et al. \cite{MMAR1} for an overview). I do not have a sense of which types of potentials, if any, might lead to profitable arbitrage opportunities, and I think it is worth finding out. 
\\\\
Wilson \cite{wilsonHamiltonians} argued that one must generically add all possible terms allowed by symmetry, organized according to the scale of fluctuations around a hypothesized fixed point. The reason to add all terms is that any term not forbidden by symmetry will be generated at some order in perturbation theory.\footnote{That is the mathematics behind `t Hooft's naturalness principle. See Seiberg \cite{seibergNonnaturalness} for an explanation and a counterexample.} For the same reason, parameters that were already included will receive corrections.
\\\\
The purpose of theory is twofold: To find new models, and to improve existing models. The principles of Wilson should help with both. 
\subsubsection{Dependent Increments and Fractional Brownian Motion}\label{sec:dependentIncrements}
\begin{quote}
``It is known that economic time series `typically' exhibit cycles of all orders of magnitude, the slowest cycles having periods of duration comparable to the total sample size.''---Mandelbrot and van Ness \cite{fBm}
\end{quote}
The hallmark of the models considered in the main text is that the increments are independent: The change in price today gives no information about the change in price tomorrow. Mandelbrot proposed a radically different view: The change in price tomorrow is influenced by all changes in price since the dawn of time.\footnote{The hypothesis is of course only asserted within an intermediate regime of validity \cite{mandelbrot-FractalGeometryOfNature}.} 
\\\\
The proposed model is fractional Brownian motion \cite{fBm}. The lattice regularization of fractional Brownian motion is known in the financial literature as a type of ARFIMA model \cite{fractionalCalculus}. To the best of my understanding, fractional Brownian motion is the general parametrization of an interacting Gaussian fixed point that is invariant under continuous global dilations and translations, and has continuous sample paths. 
\\\\
I have not yet completely understood the path-integral description of fractional Brownian motion. A good starting point is Sebastian \cite{fBm_sebastian}, and an important integral relation can be found in Pipiras and Taqqu \cite{pipirasTaqqu}.\footnote{Also worth noting, in the same volume, is Taqqu's discussion of renormalization \cite{taqquRG}.} Lim has written or co-written many papers on this topic that look useful (e.g., \cite{fBm_lim}), but I have not worked through them. What I have understood is how to use gauge invariance to augment the model by mean-reversion and jumps. 
\\\\
The model of Mandelbrot and van Ness is\footnote{True to my physicist roots, I have placed the system into a temporal box. To obtain the form written down by Mandelbrot and van Ness, choose a reference time, $t_1 \in (t_0,t)$, consider the difference $B_H(t) - B_H(t_1)$, and rearrange some terms under the integral sign. It is important to understand that, in my notation, $t_0$ is a regularization of the dawn of time, not an initial time in the sense of linear response. To use the terminology of Mandelbrot and van Ness, I am regularizing the Weyl fractional integral, not resorting to the Holmgren-Riemann-Liouville integral.}
\begin{equation}\label{eq:fBm}
B_H(t) = B_H(t_0) + \int_{t_0}^{t}\! dt'\;\frac{(t-t')^{H-\half}}{\Gamma(H+\half)}\;\dot B(t')\;,
\end{equation}
where $\dot B(t) = Z(t)/\sqrt{dt}$ with $Z(t)$ drawn from a Gaussian distribution, $H \in (0,1)$, and $t_0 \to -\infty$. Inversion by means of a trial solution gives, for the antipersistent case,
\begin{equation}\label{eq:antipersistent}
\dot B(t) = \int_{t_0}^{t}\! dt'\;\frac{(t-t')^{-\half-H}}{\Gamma(\half-H)}\;\dot B_H(t')\;,\;\; H \in (0,\half)\;,
\end{equation}
and, for the persistent case, 
\begin{equation}\label{eq:persistent}
\dot B(t) = \frac{(t-t_0)^{-(H-\half)}}{\Gamma(\tfrac{3}{2}-H)}\,\dot B_H(t_0) + \int_{t_0}^{t}\! dt'\;\frac{(t-t')^{-(H-\half)}}{\Gamma(\tfrac{3}{2}-H)}\,\ddot B_H(t')\;,\;\; H \in (\half,1)\;.
\end{equation}
The action for fractional Brownian motion would then by given by\footnote{This looks at least superficially consistent with the results of Meerson et al. \cite{fBm_meerson}} inserting either Eq.~(\ref{eq:antipersistent}) or Eq.~(\ref{eq:persistent}) into 
\begin{equation}
S = \int_{t_0}^{t_f}\!\!dt\;\half\;\dot B(t)^2\;.
\end{equation}
To augment the model by mean-reversion and jumps, I insist on a description that is covariant under local dilations and translations. Following Secs.~\ref{sec:gaugeTheory} and~\ref{sec:gaugeTheoryApplied}, I introduce gauge fields, promote the ordinary time derivative to a covariant derivative, and insert Wilson lines. 
\\\\
Eq.~(\ref{eq:antipersistent}), for example, would become\footnote{I am upgrading the right-hand side only; the left-hand side is just notation for a Gaussian variate.}
\begin{equation}
\dot B(t) = \int_{t_0}^{t}\! dt'\left[W(t)^{-1}\,\frac{(t-t')^{-\half-H}}{\Gamma(\half-H)}\,W(t')\right] DB_H(t')\;,
\end{equation}
with $DB_H(t)  = \dot B_H(t) + a(t)\,B_H(t) + b(t)$, as in Eq.~(\ref{eq:basicModel}), and $W(t) = e^{\,\int_{t_0}^{t}dt'\,a(t')}$, as in Eq.~(\ref{eq:wilsonLine}). In Eq.~(\ref{eq:persistent}), the second derivative would have to be promoted to $DDB_H(t)$, with the composition of derivatives understood covariantly. 
\\\\
Specializing to a constant $a(t)$ and a sequence of impulses for $b(t)$ would then incorporate mean-reversion and jumps. Further details will have to wait. 
\subsubsection{Critical Exponents and the Renormalization Group}
\begin{quote}
``Reliance upon a single time scale leads to inefficiency, or worse, forecasts that vary with the time-scale of the chosen data.''---Mandelbrot, Fisher, and Calvet \cite{MMAR1}
\end{quote}
Guided by Secs.~\ref{sec:interactions} and~\ref{sec:effectiveFieldTheory}, one is led to search for a dynamical explanation of an interacting fixed point of the kind described in Sec.~\ref{sec:dependentIncrements}. See Cassandro and Jona-Lasinio for examples in probability theory \cite{JLProbabilityOld}.
\\\\
Dash has proposed to model Hurst behavior using Reggeon field theory \cite{dash}. I too hoped to reproduce Hurst behavior from the renormalization of a local interacting model, but I have not yet understood how to do so. I am not satisfied with Dash's analysis.\footnote{With respect to Dash as a pioneer in the field, and from one former Caltech physicist to another, I must push back on some of his assertions. I agree that finance is phenomenology. But it is wrong to say that simply copying results from physics to finance is the way physics in finance has worked for over 100 years: Mandelbrot predated Kadanoff and Wilson \cite{mandelbrotRG}. Derman, another former QCD guy, reminisced about learning stochastic calculus from scratch on the New York subway \cite{dermanMemoir}. And Merton adopted the mathematics of dynamic programming to work out myriad new results in portfolio construction \cite{merton}, results that are nowhere to be found in a physics education. More importantly, the core mission is risk management: How am I supposed to assess the reliability of a model if I throw around uncontrolled approximations and hope for the best?} I suspect that Dyson's hierarchical model \cite{dyson, JLProbabilityOld, JLProbabilityNew} and a few chapters of Zinn-Justin's book \cite{zinnJustin} would help, but I have not cracked the code. 
\\\\
It is clear that Mandelbrot understood the concepts behind the renormalization group and developed those concepts independently \cite{mandelbrot-FractalGeometryOfNature, mandelbrotRG} from Wilson \cite{wilsonManyScales, wilsonNobel}.\footnote{I find the following throwaway line from Wilson's Nobel lecture \cite{wilsonNobel} fascinating: ``There is a murky connection between scaling ideas in critical phenomena and Mandelbrot's (1982) `fractals' theory---a description of scaling of irregular geometrical structures (such as coastlines).'' The connection is not murky: Mandelbrot's theory of fractals is the geometry of the renormalization group.}
\subsubsection{Multifractal Model of Asset Returns}\label{sec:MMAR}
\begin{quote}
``The salient feature of trading time is explicit modelling of the relationship between unobserved natural time-scale of the returns process, and clock time, which is what we observe.''---Mandelbrot, Fisher, and Calvet \cite{MMAR1}
\end{quote}
Mandelbrot distilled his career experience into a model of fractional Brownian motion in multifractal time \cite{MMAR1, MMAR2, MMAR3}. (See also the followup book by Calvet and Fisher \cite{calvetFisher}.) I have already mentioned fractional Brownian motion; in this section I want to discuss the trading time. 
\\\\
The general procedure of separating an overall model of increments into a primitive model evolving under a fixed clock, then stretching and squeezing those increments by modifying that clock, is the same as the mathematics of coupling quantum fields to gravity.\footnote{The gravitational field provides ``an arena for the other fields to play in.''---Zee \cite{zeeGR}, p.~286.} (I mentioned that back in Sec.~\ref{sec:temporalReparametrization}.)
\\\\
Multifractality can be defined in a ``microscopic'' way, as a property of a probability distribution, or in a ``macroscopic'' way, as a scaling hypothesis for moments. The former implies the latter under a saddle-point approximation, in the way that the thermodynamic relation between free energy, internal energy, and entropy can be derived from a saddle-point approximation of the partition function for the microcanonical ensemble.\footnote{I thank Igor Aleiner for teaching this to the class back in 2007.} See Mandelbrot \cite{mandelbrot-geophysics} for an explicit demonstration using the binomial cascade.
\\\\
What I want to clarify is what Mandelbrot et al.~\textit{did} in their model: They proposed a hypothetical configuration for the trading time, without making any attempt to derive that configuration as the extremal solution of a dynamical theory. That is in precise analogy with the history of gravity: Newton's ``$GM/r$'' was proposed far earlier than Einstein's equivalence principle and the corresponding action. 
\\\\
Understanding that Newton's theory of gravity is the weak-field limit of Einstein's theory, and that Einstein's theory is the starting point for a Wilson expansion, allows physicists to systematically improve Newton's theory and explore beyond Einstein's theory. 
\\\\
If the multifractality of trading time is responsible for volatility, and if the primary obstacle to risk management is the difficulty of forecasting volatility \cite{mandelbrot-misbehavior}, then it is of paramount importance to find a dynamical motivation for Mandelbrot's hypothesized saddle point; or to propose a different one. In doing so, it would also become clear how to systematically compute corrections to that saddle point, and to compute the effects of trading time on any price process that couples to it. 
\\\\
A promising place to start is Schramm-Loewner theory \cite{pommerenke, schramm}.\footnote{I thank Alexei Kitaev for that recommendation.} Mandelbrot's parameter of cascade depth might then be replaced by the central charge of a conformal field theory \cite{duplantier}. Going down that rabbit hole seems to lead to string theory.\footnote{Since I have undertaken only a cursory investigation in this direction, I will not bother citing references. Instead, I will quote Polyakov \cite{fineStructureOfStrings}: ``We can say that, in some sense, strings lead not only to unification of interactions but to unification of ideas.''} 
\subsubsection{Correlations between Assets}\label{sec:correlations}
Everything in this paper, except for a brief interlude in Sec.~\ref{sec:nonlinearSigmaModel2}, was about the dynamical structure of a single asset. It is straightforward, in principle, to add more assets and repeat the analysis. 
\\\\
But the more assets considered, the more terms there will be in the action, and the harder keeping track of all of those parameters will be. I suspect that a fruitful approach would be to impose symmetries with the zeal of a theorist, and then check whether the model is satisfactory; as opposed to starting with the largest collection of arbitrary parameters and hoping to wrangle them. 
\\\\
For instance, there are times when an asset class, say commodities, generally behaves in a certain way, but then one such asset, say corn, behaves in the opposite way. Traders refer to that kind of phenomenology as ``divergence''---to me, it sounds like pions.\footnote{See Sec.~VI.4 of Zee \cite{zeeQFT} and Ch.~83 of Srednicki \cite{srednicki}.} Imposing flavor symmetry among a bunch of assets and then classifying market regimes into broken and unbroken phases with respect to that symmetry might be useful.\footnote{With that in mind, revisit Sec.~\ref{sec:MMAR}. Unlike in gravity, there may be no reason to expect universal coupling to trading time. Indeed, if trading time has its physical origin in the volume of trades \cite{mandelbrotTaylor, clark}, then different types of assets might evolve under different clocks; and assets that trade together may still couple with different ``$G_{\text{Newton}}$''s. Once again, I expect that a reasonable place to start would be to assume universality, at least per asset class, until proven otherwise.}
\subsection{Inference and Optimization}
Finally, I want to emphasize that modeling is only one third of the theoretical foundation of quantitative risk management: Models must be supplemented, on the one hand, by fitting those models to data, and on the other hand, by formulating strategies for how to trade those models. The former is the provenance of probability theory \cite{jeffreys, jaynes, bretthorst}, and the latter of decision theory \cite{bellman-AdaptiveControlProcesses, bellman-AppliedDynamicProgramming, merton}. For further discussion of those topics, see my companion paper \cite{portfolioConstruction}.
\newpage
\begin{center}\textit{Acknowledgments}\end{center}
I thank Alex Rasmussen, Joe Swearngin, Justin Wilson, Seth Blumberg, Alexei Kitaev, and A. Zee for helpful discussions. I thank Joe Swearngin in particular for a discussion about Sec.~\ref{sec:poissonPathIntegral}. I thank the IQIM at Caltech for hosting me during the summer of 2024. I thank Keith McCullough of Hedgeye Risk Management for emphasizing the importance of Mandelbrot, and for private conversations. 
\newpage
\appendix
\section{Legendre Transformation}\label{sec:legendre}
To someone familiar with the relation between the Lagrangian and Hamiltonian in classical mechanics, some of the signs in stochastic mechanics may look a bit wonky. In this appendix I will work out the consequences of the conventions of this paper for the corresponding relation in stochastic mechanics. 
\subsection{Kolmogorov Equation}
The starting point will be a Kolmogorov forward equation of the form 
\begin{equation}\label{eq:kolmogorovEquationForLegendreTransformation}
\partial_t \psi(x,t) = H(x,\partial_x,t)\,\psi(x,t)\;,
\end{equation}
with the particular example of interest having a right-hand side
\begin{equation}\label{eq:specialCaseForLegendreTransformation}
H(x,\partial_x,t)\,\psi(x,t) = \frac{\s^2}{2}\partial_x^{\,2}\psi(x,t) - \mu\,\partial_x\psi(x,t) - V(x)\,\psi(x,t)\;.
\end{equation}
Cf. Eq.~(\ref{eq:kolmogorovForwardOperatorForNLSM}) with $g(x) = 1$ and $\mu(x) = $ constant. I have included a potential, in anticipation of work along the lines of Sec.~\ref{sec:interactions}. 
\\\\
Rewriting Eq.~(\ref{eq:kolmogorovEquationForLegendreTransformation}) in integral form by following the procedure of Sec.~\ref{sec:phaseSpaceFromKolmogorov} leads to the path-integral expression 
\begin{equation}\label{eq:integralRewritingOfKolmogorovEquationForLegendreTransform}
\psi(x_f,t_f) = \int_{-\infty}^{\infty}\!\! dx_0\int_{x(t_0)\,=\,x_0}^{x(t_f)\,=\,x_f}\!\!\!\!\!\!\!\!\mathscr Dx(\cdot)\int\mathscr Dk(\cdot)\;e^{\,\int_{t_0}^{t_f}\!dt\,\left[\,ik(t)\,\dot x(t) + H(x(t),\,ik(t),\,t)\, \right]}\;\psi(x_0,t_0)\;.
\end{equation}
The configuration-space action, $S$, is defined from Eq.~(\ref{eq:integralRewritingOfKolmogorovEquationForLegendreTransform}) by formally integrating over the Fourier variable, leaving behind a functional of the path $x(t)$:
\begin{equation}\label{eq:definitionOfActionForLegendreTransform}
e^{-S(x(\cdot);\,t_f|t_0)} \equiv \int\mathscr Dk(\cdot)\;e^{\,\int_{t_0}^{t_f}\!dt\,\left[\,ik(t)\,\dot x(t) + H(x(t),\,ik(t),\,t)\, \right]}\;.
\end{equation}
\subsection{Saddle-Point Approximation}
The saddle point, or most-probable configuration, of the \textit{Fourier variable}, for any value of the path, is defined by 
\begin{equation}\label{eq:mostProbablek}
i \dot x(t) + \left.\frac{\delta}{\delta k(t)}\int_{t_0}^{t_f}\!\! dt'\; H(x(t'),\,ik(t'),\,t')\;\right|_{k(\cdot)\,=\,k_*(\cdot)} \equiv 0\;.
\end{equation}
If the action for the path can be expressed as the integral of a local Lagrangian, 
\begin{equation}
S(x(\cdot);\,t_f|t_0) \equiv \int_{t_0}^{t_f}\!\! dt\;\mathscr L(x(t))\;,
\end{equation}
Then an expansion of Eq.~(\ref{eq:definitionOfActionForLegendreTransform}) around the solution of Eq.~(\ref{eq:mostProbablek}) will determine the Lagrangian in terms of the Hamiltonian:
\begin{equation}\label{eq:legendre}
\mathscr L(x(t)) \approx -\left[ik_*(t)\,\dot x(t) + H(x(t),\,ik_*(t),\,t)\right]\;.
\end{equation}
That is the Legendre transformation in stochastic mechanics.\footnote{It is often important to treat the Lagrangian as a function of $x$ and $\dot x$, such as in dynamic programming \cite{bellman-DynamicProgramming}. Here I am only interested in the property that $\mathscr L$ is a function of the single field $x(t)$ and not the two independent fields $x(t)$ and $k(t)$.}
\subsection{Self-Interacting Diffusion with Drift}
Now consider the special case of Eq.~(\ref{eq:specialCaseForLegendreTransformation}). The Hamiltonian is 
\begin{equation}
H(x(t),\,ik(t),\,t) = -\,\frac{\s^2}{2}\,k(t)^2 -i\mu\, k(t) - V(x(t))\;.
\end{equation}
Eq.~(\ref{eq:mostProbablek}) becomes 
\begin{equation}
i\dot x(t) - \s^2\,k_*(t) - i\mu = 0 \implies k_*(t) = -\frac{i}{\s^2}\left(\dot x(t)-\mu\right)\;.
\end{equation}
The saddle-point solution is imaginary. That is the standard situation in statistical mechanics, where one derives the thermodynamic relation $F = E-TS$ from the partition function of the microcanonical ensemble.
\\\\
Inserting that into Eq.~(\ref{eq:legendre}) gives
\begin{equation}
\mathscr L(x(t)) = \frac{1}{2\sigma^2}\left[\dot x(t)-\mu\right]^2 + V(x(t))\;,
\end{equation}
consistent with Eqs.~(\ref{eq:brownianPathIntegralWithDrift}) and~(\ref{eq:quarticInteraction}).
\section{Coordinate-Invariant Laplacian for Densities}\label{sec:laplacianForDensities}
In this appendix I will show how to generalize the coordinate-invariant Laplacian to scalar fields that transform as densities. I will presume familiarity with differential geometry \cite{nakahara, zeeGR}. 
\subsection{Coordinate Transformations}
The reasoning is more organized in higher dimensions, so consider a $d$-dimensional space with coordinates $\{x^\mu\}_{\mu\,=\,1}^d$. The situation of interest in the main text is $d = 1$. 
\\\\
A general coordinate transformation is a linear, space-dependent transformation of the basis forms:
\begin{equation}\label{eq:generalCoordinateTransformation}
dx^\mu \to dx'^\mu = \Lambda^\mu_{\;\;\nu}(x)\,dx^\nu\;.
\end{equation}
The space is presumed to have a metric tensor $g_{\mu\nu}(x)$, which transforms under Eq.~(\ref{eq:generalCoordinateTransformation}) as
\begin{equation}\label{eq:transformationOfMetric}
g_{\mu\nu}(x) \to g'_{\mu\nu}(x') = g_{\rho\sigma}(x)\,(\Lambda^{-1})^\rho_{\;\;\mu}(x)\,(\Lambda^{-1})^\sigma_{\;\;\nu}(x)\;.
\end{equation}
Of particular importance in this section is the consequence of Eq.~(\ref{eq:transformationOfMetric}) for the determinant, or its square root, 
\begin{equation}\label{eq:sqrtDetg}
s(x) \equiv \sqrt{\det[g(x)]}\;.
\end{equation}
Let
\begin{equation}
\lambda(x) \equiv \det[\Lambda(x)]\;.
\end{equation}
The transformation law 
\begin{equation}\label{eq:transformationOfSqrtDetg}
s(x) \to s'(x') = \lambda(x)^{-1}s(x)
\end{equation}
implies that the combination 
\begin{equation}
d^dx\,s(x)
\end{equation}
is invariant under Eq.~(\ref{eq:generalCoordinateTransformation}). Consequently, any field that transforms as Eq.~(\ref{eq:transformationOfSqrtDetg}) is said to transform ``as a density.''
\subsection{Scalar Fields}
Let $\phi_p(x)$ be a scalar field---the term ``scalar'' means, in this context, a single number at each point $x$, but it does not necessarily mean ``invariant.'' The field $\phi_p(x)$ will be defined to transform under Eq.~(\ref{eq:generalCoordinateTransformation}) in the same way as $s(x)^p$:
\begin{equation}\label{eq:p-density}
\phi_p(x) \to \phi'_p(x') = \lambda(x)^{-p}\,\phi_p(x)\;.
\end{equation}
It is occasionally helpful to express that transformation law by defining 
\begin{equation}
\phi_p(x) \equiv s(x)\,\widetilde\phi_p(x)\;,
\end{equation}
with $\widetilde\phi_p(x)$ defined to be invariant. I will refer to a scalar field that transforms as in Eq.~(\ref{eq:p-density}) as a ``$p$-density.'' An ordinary density would have $p = 1$; an invariant scalar would have $p = 0$.
\subsection{Covariant Derivative}
By now I am hard-wired to carry out the procedure of Sec.~\ref{sec:gaugeTheory}: When I see a local transformation, I introduce a new degree of freedom to define a modified derivative that transforms covariantly. 
\\\\
Under Eq.~(\ref{eq:generalCoordinateTransformation}), the dual basis transforms as 
\begin{equation}\label{eq:dualBasis}
\partial_\mu \to \partial\,'_\mu = (\Lambda^{-1}(x))^\nu_{\;\;\mu}(x)\,\partial_\nu\;.
\end{equation}
For the special case in which $\Lambda^\mu_{\;\;\nu}(x)$ is constant, the partial derivatives of $\phi_p(x)$ would transform covariantly:
\begin{equation}
\partial_\mu\phi_p(x) \to \partial'_\mu \phi'_p(x) = (\Lambda^{-1})^\nu_{\;\;\mu}\,\lambda^{-p}\,\partial_\nu\phi_p(x)\qquad (\text{for}\;\;\partial_\rho\, \Lambda^\mu_{\;\;\nu}(x) = 0)\;.
\end{equation}
Let
\begin{equation}\label{eq:covariantDerivativeForp-density}
D_\mu\phi_p(x) \equiv \partial_\mu\phi_p(x) - p\,\Gamma_\mu(x)\,\phi_p(x)\;,
\end{equation}
with $\{\Gamma_\mu(x)\}_{\mu\,=\,1}^d$ a new collection of fields. Demanding that the derivative in Eq.~(\ref{eq:covariantDerivativeForp-density}) transform as 
\begin{equation}
D_\mu\phi_p(x) \to D'_\mu\phi'_p(x') = (\Lambda^{-1}(x))^\nu_{\;\;\mu}\,\lambda(x)^{-p}\,D_\nu\phi_p(x)
\end{equation}
under general, nonconstant transformations fixes the transformation law for $\Gamma_\mu(x)$:
\begin{equation}
\Gamma_\mu(x) \to \Gamma'_\mu(x') = (\Lambda^{-1}(x))^\nu_{\;\;\mu}\left[\Gamma_\nu(x) - \lambda(x)^{-1}\partial_\nu\lambda(x)\right]\;.
\end{equation}
\subsection{Terminology and Shorthand}
I can summarize the scalar transformation laws above using the group-theoretic shorthand of particle physics:
\begin{align}
&d^dx \sim 1\;,\\
&s \sim -1\;,\\
&\phi_p \sim -p\;.\label{eq:chargep}
\end{align}
You can consider Eq.~(\ref{eq:covariantDerivativeForp-density}) to be the covariant derivative for an $\mathbb R$ gauge theory (the same algebra as $U(1)$, but globally not compact), and in that sense describe the shorthand in Eq.~(\ref{eq:chargep}) by saying that ``$\phi_p$ has charge $p$.'' (Or $-p$, depending on your linguistic convention.) 
\subsection{Metric Compatibility}
So far, the gauge field $\Gamma_\mu(x)$ has no relation to the metric, $g_{\mu\nu}(x)$. You could fix a relation between them by demanding that the covariant derivative of the square-root of the determinant of the metric is zero. Because $s(x)$ has charge $p = 1$, its covariant derivative is:
\begin{equation}\label{eq:metricCompatibility}
D_\mu s(x) = \partial_\mu s(x) - \Gamma_\mu(x)\,s(x) \equiv 0 \implies \Gamma_\mu(x) = \frac{1}{s(x)}\partial_\mu s(x)\;.
\end{equation}
This is \textit{partial} or \textit{incomplete} metric compatibility. Full metric compatibility would imply the usual relation for the full Christoffel symbol, which I have no need to introduce here.\footnote{The reason I did not introduce full metric compatibility is that I could get away with it. The relation between my $\Gamma_\mu$ and the usual $\Gamma^\mu_{\nu\rho}$ is the trace: $\Gamma_\mu = \Gamma^\nu_{\mu\nu}$. I am not sure whether there would be any interesting geometrical or field-theoretic consequence of imposing only the dilation part of the equation of motion for the spin-connection in the Cartan formalism. Note that this is different from a partial fixing of the gauge: It is a partial imposition of one of the equations of motion, in any gauge.}
\subsection{Invariant Actions and Adjoint Operators}
The adjoint of a differential operator is defined by integration by parts, dropping boundary terms. It is standard to combine that definition with an invariant action to derive the coordinate-invariant generalization of the Laplacian acting on an invariant scalar field. In this section I will set up the analogous procedure for a $p$-density. 
\\\\
Let $\phi_p(x)$ be a $p$-density and $\chi_q(x)$ be a $q$-density. The integral 
\begin{equation}\label{eq:invariantCrossTerm}
S \equiv \int\! d^dx\, s(x)^{1-p-q}\,g^{\mu\nu}(x)\,D_\mu\phi_p(x)\,D_\nu\chi_q(x)
\end{equation}
is invariant under Eq.~(\ref{eq:generalCoordinateTransformation}). Integration by parts so as to move all derivatives off of $\chi_q(x)$ will lead to a covariant notion of ``$D^2$'' that, by definition, generalizes the Laplacian to act on $p$-densities for $p \neq 0$:
\begin{equation}\label{eq:definitionOfGeneralizedLaplacian}
S \equiv - \int\! d^dx\,s(x)^{1-p-q}\,\left[ D^2\phi_p(x)\right] \chi_q(x) + \text{boundary terms}\;.
\end{equation}
The definition symbol in Eq.~(\ref{eq:invariantCrossTerm}) should be read as defining $S$, while the definition symbol in Eq.~(\ref{eq:definitionOfGeneralizedLaplacian}) should be read as defining $D^2$ from Eq.~(\ref{eq:invariantCrossTerm}). The overall sign matches the standard convention in physics, so that $D^2$ will reduce to the usual $\nabla^2$ when acting on a scalar field with $p = 0$. 
\subsection{Dependence on $q$}
One check on the subsequent work is that $D^2\phi_p(x)$ must be independent of $q$. In Eq.~(\ref{eq:invariantCrossTerm}), there are two sources of dependence on $q$---the explicit linear dependence from $D_\nu\chi_q$ must cancel the factors of $q$ that will be brought down by moving the $\partial_\nu$ onto $s^{1-p-q}g^{\mu\nu}D_\mu\phi_p$. For temporary convenience, let me give a name to the explicit dependence:
\begin{equation}
Q \equiv s^{1-p-q} g^{\mu\nu} D_\mu\phi_p \,\Gamma_\nu\;,
\end{equation}
in terms of which the action reads 
\begin{align}
S &= \int\! d^dx \left[ s^{1-p-q}g^{\mu\nu}D_\mu\phi_p\partial_\nu\chi_q - q Q \chi_q\right] \\
&= \int \! d^dx\left[-\partial_\nu\left(s^{1-p-q}g^{\mu\nu}D_\mu\phi_p\right) - q Q\right]\chi_q\; + \text{boundary terms}\;. \label{eq:qTerms}
\end{align}
Because of the partial metric compatibility imposed in Eq.~(\ref{eq:metricCompatibility}), I know that 
\begin{equation}
\partial_\nu s^{-q} = -q s^{-q-1}\partial_\nu s = -q s^{-q}\Gamma_\nu\;.
\end{equation}
The overall additional minus sign from integration by parts means that, as required, there is a $+qQ$ to cancel off the $-qQ$ in Eq.~(\ref{eq:qTerms}). What remains is
\begin{align}
S &= -\int\! d^dx\,s^{-q}\, \partial_\nu\!\left(s^{1-p} g^{\mu\nu} D_\mu\phi_p\right)\chi_q \\
&= -\int\! d^dx\,s^{1-p-q}\,\frac{1}{s^{1-p}}\partial_\nu\!\left(s^{1-p} g^{\mu\nu} D_\mu\phi_p\right)\chi_q\;,
\end{align}
and therefore
\begin{equation}\label{eq:generalizedLaplacian}
D^2\phi_p = \frac{1}{s^{1-p}}\partial_\nu\!\left(s^{1-p} g^{\mu\nu} D_\mu\phi_p\right)\;.
\end{equation}
For an invariant scalar ($p = 0$), Eq.~(\ref{eq:generalizedLaplacian}) reduces to the familiar expression 
\begin{equation}
D^2\phi_0 = \frac{1}{s}\partial_\nu\!\left(s\,g^{\mu\nu}\partial_\mu\phi_0\right)\;.
\end{equation}
\subsection{Log-Field}
Eq.~(\ref{eq:generalizedLaplacian}) could be analyzed in a number of ways; all I want to do is verify the analysis of the log-field in Sec.~\ref{sec:logField}. So first I will specialize to $d = 1$, then I will specialize to $p = 1$, and then I will specialize to the particular coordinate transformation in Eq.~(\ref{eq:logField}). 
\subsubsection{$d = 1$}
In one dimension, the metric is the same as its determinant, and the matrix inverse is the multiplicative inverse. For example, $g^{\mu\nu}$ becomes $g^{-1} = s^{-2}$. Eq.~(\ref{eq:generalizedLaplacian}) simplifies to:
\begin{equation}\label{eq:generalizedLaplacianOneDimension}
D^2\phi_p = \frac{1}{s^{1-p}}\frac{d}{dx}\left[s^{-1-p}\left(\frac{d}{dx} - p\,\Gamma \right)\phi_p \right]\;,
\end{equation}
with 
\begin{equation}
\Gamma = s^{-1}\frac{d}{dx}s\;.
\end{equation}
\subsubsection{$p = 1$}
For an ordinary density ($p = 1$), Eq.~(\ref{eq:generalizedLaplacianOneDimension}) reduces to 
\begin{equation}\label{eq:generalizedLaplacianOneDimensionForDensity}
D^2\phi_{-1} = \frac{d}{dx}\left[s^{-2}\left(\frac{d}{dx} - \Gamma\right)\phi_{-1}\right]\;.
\end{equation}
\subsubsection{$X = e^x$}
The symbol $x$ has, up to this point, denoted an arbitrary coordinate system. But now let me specialize to the case of Sec.~\ref{sec:logField}: The symbol $x$ will denote the coordinate in which the metric is constant (and set to 1), and the symbol $X$ will denote the coordinate in which the metric is 
\begin{equation}
g(X) = \frac{1}{X^2}\;.
\end{equation}
The square root of the determinant is 
\begin{equation}
s(X) = \frac{1}{X}\;,
\end{equation}
and the gauge field is 
\begin{equation}
\Gamma(X) = X \frac{d}{dX}\left(\frac{1}{X}\right) = -\frac{1}{X}\;.
\end{equation}
If $\Psi(X)$ denotes the $p = 1$ density in the $X$-system, then Eq.~(\ref{eq:generalizedLaplacianOneDimensionForDensity}) will read:
\begin{align}
D^2\Psi(X) &= \frac{d}{dX}\left[ X^2\left( \frac{d}{dX} + X^{-1}\right)\Psi(X)\right] \\
&= X^2\left[ \frac{d^2}{dX^2}\Psi(X) + \frac{d}{dX}\left(X^{-1}\Psi(X)\right)\right] + 2X\left( \frac{d}{dX} + X^{-1}\right)\Psi(X) \\
&= X^2\frac{d^2}{dX^2}\Psi(X) + X^2\left[ X^{-1}\frac{d}{dX}\Psi(X) - X^{-2}\Psi(X)\right] + 2X \frac{d}{dX}\Psi(X) + 2\Psi(X) \\
&= X^2\frac{d^2}{dX^2}\Psi(X) + 3X\frac{d}{dX}\Psi(X) + \Psi(X)\;.
\end{align}
That is twice the right-hand side of Eq.~(\ref{eq:evolutionEquationInXExpanded}), as required. Game, set, match. 
\section{Path Integral with Sources for the Ornstein-Uhlenbeck Model}\label{sec:generatingFunctionForOU}
Given an action $S(x(\cdot);\,t_f|t_0)$ and an external local source $J(t)$, define the conditional probability 
\begin{equation}\label{eq:generatingFunction}
U(J(\cdot);\,x_f,t_f|x_0,t_0) \equiv \int_{x(t_0)\,=\,x_0}^{x(t_f)\,=\,x_f}\!\!\!\!\mathscr Dx(\cdot)\;e^{-\left[S(x(\cdot);\,t_f|t_0) - \int_{t_0}^{t_f}\! dt\;J(t)\,x(t)\right]}\;.
\end{equation}
The source term $\int\!dt\, J(t)\,x(t)$ can be viewed in one of two ways. 
\\\\
First, the source $J(t)$ could be an additional physical or economic degree of freedom. For example, if $x$ describes log-AAPL, $J$ could describe log-MSFT. The additional term in the action would then describe a linear interaction between the two degrees of freedom, and performing the path integral over log-AAPL would produce additional terms in the effective action for log-MSFT. 
\\\\
Alternatively, $J$ could be viewed as an artifice for defining a generating function. Each derivative with respect to $J(t)$ will bring down a factor of $x(t)$; after taking as many derivatives as I want and then taking $J(t) \to 0$, I will be left with an autocorrelation function weighted by the action $S(x(\cdot);\,t_f|t_0)$. 
\\\\
Even if $J(t)$ were physical, I could still take derivatives with respect to it to generate correlation functions---just as I could calculate the integral $\int_{-\infty}^{\infty}\! dx\;x^2\;e^{-\half a x^2}$ either by adding a $+bx$ term to the argument of the exponential, taking two derivatives with respect to $b$, then setting $b$ to zero; or by taking one derivative with respect to $a$.\footnote{Jaynes \cite{jaynes} has an amusing anecdote along similar lines about differentiating with respect to $\pi$ (his footnote 17 on p. 194).}
\\\\
In the usual treatment of probability theory, one works with a distribution of \textit{values} and generates \textit{moments}. In the theory of stochastic processes, one works with a distribution of \textit{paths} and generates \textit{correlations}. Just as the log of the statistical generating function generates \textit{cumulants}, the log of Eq.~(\ref{eq:generatingFunction}) will generate \textit{connected correlations}.\footnote{These are also called ``Ursell functions'' or ``truncated correlation functions'' \cite{JLProbabilityOld}.}
\subsection{Action}
Now I will specialize to the action in Eq.~(\ref{eq:OUaction}):
\begin{equation}
S(x(\cdot);\,t_f|t_0) = \int_{t_0}^{t_f}\!\! dt\;\frac{1}{2\sigma^2}\left\{D[x(t)-m]\right\}^2\;,\;\; D = \frac{d}{dt} + \theta\;.
\end{equation}
The total action is 
\begin{equation}
S(x(\cdot),J(\cdot);\,t_f|t_0) = S(x(\cdot);\,t_f|t_0) - \int_{t_0}^{t_f}\!\! dt\;J(t)\,x(t)\;.
\end{equation}
I will proceed along the lines of Sec.~\ref{sec:mostProbablePath}. The concepts are identical, but the algebra becomes substantially more complicated by the source term.
\subsection{Factorization}
Factorize the integral over paths using the decomposition 
\begin{equation}\label{eq:decompositionOfPaths}
x(t) = x_*(t) + y(t)\;,\;\; y(t_0) = y(t_f) = 0\;.
\end{equation}
\begin{align}
\{D[x(t)-m]\}^2 &= \{D[x_*(t)-m] + Dy(t)\}^2 \nonumber \\
&= \{D[x_*(t)-m]\}^2 + 2D[x_*(t)-m]Dy(t) + [Dy(t)]^2\;.
\end{align}
The middle term contains a total derivative:\footnote{I never know how many steps to display for a routine calculation like this, but each step introduces a potential point of failure. I would rather err on the side of showing too much work instead of too little.\label{ft:reproducibility}}
\begin{align}
&D[x_*(t)-m]Dy(t) = D[x_*(t)-m]\dot y(t) + D[x_*(t)-m]\theta y(t) \\
&= \frac{d}{dt}\left\{D[x_*(t)-m]y(t)\right\} - \left\{\frac{d}{dt}D[x_*(t)-m]\right\}y(t) + D[x_*(t)-m]\theta y(t) \\
&= \frac{d}{dt}\left\{D[x_*(t)-m]y(t)\right\} - \left\{ \frac{d}{dt}\left[\dot x_*(t) + \theta[x_*(t)-m]\right]\right\}y(t) + D[x_*(t)-m]\theta y(t) \\
&= \frac{d}{dt}\left\{D[x_*(t)-m]y(t)\right\} - \left[ \ddot x_*(t) + \theta \dot x_*(t)\right]y(t) + D[x_*(t)-m]\theta y(t) \\
&= \frac{d}{dt}\left\{D[x_*(t)-m]y(t)\right\} - \left[ \ddot x_*(t) + \theta \dot x_*(t)\right]y(t) + \left[\dot x_*(t) + \theta(x_*(t)-m)\right] \theta y(t) \\
&= \frac{d}{dt}\left\{D[x_*(t)-m]\,y(t)\right\} + \left\{-\ddot x_*(t) + \theta^2\,[x_*(t)-m]\right\} y(t)\;.\label{eq:Dx^2}
\end{align}
Because of the boundary conditions in Eq.~(\ref{eq:decompositionOfPaths}), the total derivative in Eq.~(\ref{eq:Dx^2}) drops out of the action, leaving behind:
\begin{align}
S(x(\cdot),J(\cdot);\,t_f|t_0) &= \int_{t_0}^{t_f}\!\! dt\left[\;\frac{1}{2\sigma^2}\left(\phantom{\frac{}{}}\left\{D[x_*(t)-m]\right\}^2  + [Dy(t)]^2 \right.\right. \nonumber\\
&\qquad\qquad\qquad \left.+ 2\left\{-\ddot x_*(t) + \theta^2\,[x_*(t)-m]\right\} y(t) \phantom{\frac{}{}}\right)\nonumber\\
&\qquad\qquad \left.- J(t)\,[x_*(t) + y(t)]\phantom{\frac{1}{2\sigma^2}}\!\!\!\!\!\!\!\!\right]\;.
\end{align}
If the path $x_*(t)$ is defined to be a solution of 
\begin{equation}\label{eq:equationOfMotion}
\frac{1}{\sigma^2}\left\{-\ddot x_*(t) + \theta^2\left[x_*(t) - m\right]\right\} - J(t) = 0\;,
\end{equation}
then the terms linear in $y(t)$ will drop out of the action, enabling a complete factorization:
\begin{equation}
S(x(\cdot),J(\cdot);\,t_f|t_0) = S(x_*(\cdot),J(\cdot);\,t_f|t_0) + \int_{t_0}^{t_f}\!\! dt\;\frac{1}{2\sigma^2}\left[Dy(t)\right]^2\;.
\end{equation}
The generating function defined in Eq.~(\ref{eq:generatingFunction}) takes the form 
\begin{equation}
U(J(\cdot);\,x_f,t_f|x_0,t_0) = C(t_f|t_0)\;e^{-S(x_*(\cdot),J(\cdot);\,t_f|t_0)}\;,
\end{equation}
where the overall constant is 
\begin{equation}
C(t_f|t_0) = \int_{y(t_0)\,=\,0}^{y(t_f)\,=\,0}\!\!\!\!\mathscr Dy(\cdot)\;e^{-\int_{t_0}^{t_f}\! dt\;\frac{1}{2\sigma^2}\left[Dy(t)\right]^2}\;.
\end{equation}
The next steps are:
\begin{enumerate}
\item Solve for the most-probable path, $x_*(t)$;
\item Insert $x_*(t)$ into the action and collect terms in powers of $J$; 
\item Fix $C(t_f|t_0)$ by imposing the composition law.
\end{enumerate}
\subsection{Most-Probable Path}
The goal is to solve Eq.~(\ref{eq:equationOfMotion}) subject to the boundary conditions 
\begin{equation}\label{eq:endpointConditions}
x_*(t_0) = x_0\;,\;\; x_*(t_f) = x_f\;.
\end{equation}
I will invoke linearity, causality, and stationarity to start with the trial solution
\begin{equation}\label{eq:trialSolution}
x_*(t) = \alpha\!\int_{t_0}^{t}\!\! dt'\;G(t-t')\,J(t') + x_h(t)\;,
\end{equation}
where $x_h(t)$ solves the homogeneous or source-free equation, 
\begin{equation}
-\ddot x_h(t) + \theta^2\left[x_h(t) - m\right] = 0\;,
\end{equation}
$G(t)$ is, at this stage, an arbitrary function of a single variable, and $\alpha$ is a constant to be fixed such that $G(t)$ will end up taking a standardized form. (Causality was invoked by cutting off the integral at $t$.)
\subsubsection{General Solution}\label{sec:generalSolution}
One derivative of Eq.~(\ref{eq:trialSolution}) gives:
\begin{equation}
\dot x_*(t) = \alpha\,G(0)\,J(t) + \alpha\!\int_{t_0}^{t}\!\! dt'\;\dot G(t-t')\,J(t') + \dot x_h(t)\;.
\end{equation}
To avoid inducing a $\dot J(t)$ term upon taking an additional derivative, I will have to impose the first condition on $G(t)$:
\begin{equation}\label{eq:G(0)}
G(0) \equiv 0\;.
\end{equation}
A second derivative gives 
\begin{equation}
\ddot x_*(t) = \alpha\,\dot G(0)\,J(t) + \alpha\!\int_{t_0}^{t}\!\! dt'\;\ddot G(t-t')\,J(t') + \ddot x_h(t)\;.
\end{equation}
Comparison with Eq.~(\ref{eq:equationOfMotion}) shows that $\alpha\,\dot G(0) = -\sigma^2$. To obtain the standardized form I am used to, I want to take 
\begin{equation}\label{eq:Gdot(0)}
\dot G(0) \equiv 1\;,
\end{equation}
which is why I introduced the constant $\alpha = -\sigma^2$. With the $J(t)$-term matched and $x_h(t)$ defined to satisfy the homogeneous equation, what remains is to make the coefficient of $J(t')$ inside the integral drop out. That fixes the defining equation for $G(t)$:
\begin{equation}\label{eq:greenFunctionEquation}
-\ddot G(t) + \theta^2\,G(t) = 0\;.
\end{equation}
The solution of Eq.~(\ref{eq:greenFunctionEquation}) subject to Eqs.~(\ref{eq:G(0)}) and~(\ref{eq:Gdot(0)}) is 
\begin{equation}
G(t) = \frac{1}{\theta}\sinh(\theta t)\;.
\end{equation}
The homogeneous solution, meanwhile, has the general form 
\begin{equation}
x_h(t) = m + c_1\,\cosh(\theta t) + c_2\,\sinh(\theta t)\;,
\end{equation}
with undetermined constant coefficients $c_1$ and $c_2$. It is convenient to shift the argument of the homogeneous solution with respect to $t_0$ and redefine the coefficients such that the general solution of Eq.~(\ref{eq:equationOfMotion}) takes the form 
\begin{equation}\label{eq:generalSolution}
x_*(t) = -\sigma^2\!\int_{t_0}^{t}\!\! dt'\; G(t-t')\,J(t') + m + A\,\dot G(t-t_0) + B\,G(t-t_0)\;.
\end{equation}
\subsubsection{Particular Solution}\label{sec:particularSolution}
The general solution in Eq.~(\ref{eq:generalSolution}) is to be specialized to a particular solution by imposing the boundary conditions in Eq.~(\ref{eq:endpointConditions}). 
\\\\
The first condition is, by construction, simple:
\begin{equation}
x_*(t_0) = 0 + m + A \equiv x_0 \implies A = x_0 - m\;.
\end{equation}
For convenience, let $T \equiv t_f-t_0$. The second condition is:
\begin{align}
&x_*(t_f) = -\sigma^2\int_{t_0}^{t_f}\!\! dt'\;G(t_f-t')\,J(t') + m + (x_0-m)\,\dot G(T) + B\,G(T) \equiv x_f \\
&\implies B = \frac{1}{G(T)}\left[(x_f-m)-(x_0-m)\,\dot G(T) + \sigma^2\int_{t_0}^{t_f}\!\! dt'\;G(t_f-t')\,J(t')\right]\;.
\end{align}
Inserting that into Eq.~(\ref{eq:generalSolution}) and using the hyperbolic-trigonometric identity 
\begin{equation}
G(T)\,\dot G(t-t_0) - \dot G(T)\, G(t-t_0) = G(t_f-t)
\end{equation}
gives the particular solution:
\begin{align}
x_*(t) &= m + \frac{1}{G(T)}\left[(x_f-m)\,G(t-t_0) + (x_0-m)\,G(t_f-t)\right] \nonumber\\
&+ \frac{\sigma^2}{G(T)}\int_{t_0}^{t_f}\!\! dt'\left[ G(t_f-t')\,G(t-t_0) - \Theta(t-t')\,G(t-t')\,G(T)\right]J(t')\;,  \label{eq:particularSolution}
\end{align}
where $\Theta(t) \equiv 1$ for $t > 0$ and $0$ for $t \leq 0$, a shorthand notation for capturing the interior region of each integral.\footnote{I trust you to parse the standard inconsistency in parenthetical notation: $\Theta(t-t')$ denotes $\Theta$ as a function of $t-t'$, whereas $\theta(t-t')$ denotes $\theta$ times the quantity $t-t'$.}
\subsection{Extremal Action}
Inserting Eq.~(\ref{eq:particularSolution}) into the action and collecting terms in powers of $J$ is by far the most irritating aspect of this calculation. I have not found a way to streamline it, so buckle up and enjoy the ride. 
\subsubsection{$D(x_*-m)$}
The quantity $D(x_*-m) = \dot x_* + \theta (x_*-m)$ will simplify somewhat because 
\begin{equation}\label{eq:exp}
\dot G(t) \pm \theta\,G(t) = e^{\pm\theta t}\;.
\end{equation}
Note that $\frac{d}{dt}G(t_f-t) = -\dot G(t_f-t)$, where the notation on the right-hand side is understood to be the function $\dot G(s) = \cosh(\theta s)$ evaluated at $s = t_f-t$. (If you want to verify some of this work in Wolfram, I suggest that you define separate functions \verb|Gdot[t_] := Cosh[t]| and \verb|G[t_] := Sinh[t]| instead of trying to compute $\frac{d}{dt}G(t)$ from $G(t)$.) With that and Eq.~(\ref{eq:exp}), I find:
\begin{align}
&D[x_*(t)-m] = \frac{1}{G(T)}\left\{(x_f\!-\!m)[\dot G(t\!-\!t_0) + \theta\,G(t\!-\!t_0)] - (x_0\!-\!m)[\dot G(t_f\!-\!t) - \theta\,G(t_f\!-\!t)]\right\} \nonumber\\
&\;\;+ \frac{\sigma^2}{G(T)} \int_{t_0}^{t_f}\!\! dt'\left\{ G(t_f-t')\left[ \dot G(t-t_0) + \theta\,G(t-t_0)\right] \right. \nonumber\\
&\qquad\qquad\qquad\qquad \left.-\, \Theta(t-t')\left[\dot G(t-t') + \theta\, G(t-t')\right] G(T)\right\}J(t') \\
&= \frac{1}{G(T)}\left[(x_f-m)\,e^{\,\theta(t-t_0)} - (x_0-m)\,e^{-\theta(t_f-t)}\right] \nonumber\\
&\;\;+\frac{\sigma^2}{G(T)} \int_{t_0}^{t_f}\!\! dt'\left[G(t_f-t')\,e^{\,\theta(t-t_0)} - \Theta(t-t')\;e^{\,\theta(t-t')}\;G(T)\right]J(t') \\
&= \frac{1}{G(T)}\,e^{\,\theta t}\left\{\phantom{\int}\!\!\!\!\! (x_f-m)\,e^{-\theta t_0} - (x_0-m)\,e^{-\theta t_f} \right. \nonumber\\
&\;\;\;\;\left. +\;\sigma^2 \int_{t_0}^{t_f}\!\! dt'\left[ G(t_f-t')\,e^{-\theta t_0} - \Theta(t-t')\,e^{-\theta t'}\,G(T)\right] J(t')\;\right\}\;. \label{eq:D(x-m)}
\end{align}
\subsubsection{$[D(x_*-m)]^2$}
Note that, because of the step-function term in Eq.~(\ref{eq:D(x-m)}), it is not quite the case that all dependence on $t$ has been factored out. In any case, the next step is to square it and collect terms in powers of $J$:
\begin{align}
&[D(x_*(t)-m)]^2 = \frac{e^{\,2\theta t}}{G(T)^2}\left\{ \left[(x_f-m)\,e^{-\theta t_0} - (x_0-m)\,e^{-\theta t_f}\right]^2 \right. \nonumber \\
&\;\; + 2\sigma^2 \left[(x_f-m)\,e^{-\theta t_0} - (x_0-m)\,e^{-\theta t_f}\right]\int_{t_0}^{t_f}\!\! dt'\left[ G(t_f-t')\,e^{-\theta t_0} - \Theta(t-t')\,e^{-\theta t'}\,G(T)\right] J(t') \nonumber\\
&\;\; + \sigma^4 \int_{t_0}^{t_f}\!\! dt'\left[ G(t_f-t')\,e^{-\theta t_0} - \Theta(t-t')\,e^{-\theta t'}\,G(T)\right] J(t')\;\times \nonumber\\
&\qquad\qquad\int_{t_0}^{t_f}\!\! dt''\left[ G(t_f-t'')\,e^{-\theta t_0} - \Theta(t-t'')\,e^{-\theta t''}\,G(T)\right] J(t'')\;. \label{eq:[D(x-m)]^2}
\end{align}
Before expanding the quadratic part, let me turn to an important property of the linear part.
\subsubsection{$\int\!dt\,[D(x_*-m)]^2$}\label{sec:int(Dx)^2}
Upon integration with respect to $t$, the linear part of Eq.~(\ref{eq:[D(x-m)]^2}) will go to zero. One integral is
\begin{equation}
\int_{t_0}^{t_f}\!\! dt\;e^{\,2\theta t} = \frac{1}{2\theta}\left(e^{\,2\theta t_f} - e^{\,2\theta t_0}\right)\;,
\end{equation}
and the other integral is 
\begin{equation}
\int_{t_0}^{t_f}\!\! dt\;e^{\,2\theta t}\;\Theta(t-t') = \int_{t'}^{t_f}\!\! dt\;e^{\,2\theta t} = \frac{1}{2\theta}\left(e^{\,2\theta t_f} - e^{\,2\theta t'}\right)\;.
\end{equation}
The first integral is multiplied by $G(t_f-t')\,e^{-\theta t_0}$, resulting in a term:
\begin{align}
\frac{1}{2\theta}\left(e^{\,2\theta t_f-\theta t_0} - e^{\,\theta t_0}\right) G(t_f-t') &= \frac{1}{2\theta}\,e^{\,\theta t_f}\left(e^{\,\theta(t_f-t_0)} - e^{-\theta(t_f-t_0)}\right)G(t_f-t') \nonumber \\
&= e^{\,\theta t_f} G(T)\,G(t_f-t')\;.
\end{align}
The second integral is multiplied by $e^{-\theta t'}G(T)$, resulting in an identical term:
\begin{align}
\frac{1}{2\theta}\left(e^{\,2\theta t_f-t'}-e^{\,\theta t'}\right) G(T) &= \frac{1}{2\theta}\,e^{\,\theta t_f}\left(e^{\,\theta(t_f-t')} - e^{-\theta(t_f-t')}\right) G(T) \nonumber \\
&= e^{\,\theta t_f} G(t_f-t')\,G(T)\;.
\end{align}
Those terms appear with a relative minus sign in the part of Eq.~(\ref{eq:[D(x-m)]^2}) linear in $J$, and therefore drop out. This is a consequence of the general relation between the average path and the most-probable path, which I will discuss in Sec.~\ref{sec:averageVsSaddle}. 
\\\\
What remains upon integrating Eq.~(\ref{eq:[D(x-m)]^2}) with respect to $t$ is
\begin{align}
\int_{t_0}^{t_f}\!\! dt\;[D(x_*(t)-m)]^2 &= \frac{e^{\,\theta T}}{G(T)}\left[\,(x_f\!-\!m) - (x_0\!-\!m)\,e^{-\theta T}\,\right]^2 \nonumber\\
&\qquad + \frac{\sigma^4}{G(T)^2}\int_{t_0}^{t_f}\!\!\! dt'\!\int_{t_0}^{t_f}\!\!\! dt''\, M(t',t'')\,J(t')\,J(t'')\;, \label{eq:int[D(x-m)]^2}
\end{align}
where
\begin{align}
M(t',t'') &= \int_{t_0}^{t_f}\!\! dt\;e^{\,2\theta t} \left[ G(t_f-t')\,e^{-\theta t_0} - \Theta(t-t')\,e^{-\theta t'}\,G(T)\right]\;\times \nonumber\\
&\qquad\qquad\qquad\qquad\qquad \left[ G(t_f-t'')\,e^{-\theta t_0} - \Theta(t-t'')\,e^{-\theta t''}\,G(T)\right]\;. \label{eq:M}
\end{align}
\subsubsection{$\int\!dt\;\{\frac{1}{2\sigma^2}[D(x_*-m)]^2 - Jx_*\}$}
Since there is no linear term in $\int_{t_0}^{t_f} dt [D(x_*(t)-m)]^2$, and since $\int_{t_0}^{t_f}\!dt\,J(t)\,x_*(t)$ has an explicit overall factor of $J$, the only work remaining is to simplify all of the quadratic terms. Moreover, the function $M(t',t'')$ in Eq.~(\ref{eq:M}) is manifestly symmetric in $t'$ and $t''$, so before simplifying all of the quadratic terms, I should write the quadratic portion of $\int_{t_0}^{t_f}\!dt\,J(t)\,x_*(t)$ in a manifestly symmetric form, after relabeling $t$ as $t''$ to match Eq.~(\ref{eq:M}). 
\\\\
Returning to the particular solution in Eq.~(\ref{eq:particularSolution}), I can see that the $JJ$-contribution from $\int_{t_0}^{t_f}\! dt\,J(t)\,x_*(t)$ is
\begin{align}
N(t',t'') &\equiv \left[G(t_f-t')G(t''-t_0) - \Theta(t''-t') G(t''-t') G(T)\right]\; +\; (t'\leftrightarrow t'')\;. \label{eq:N}%\\
%&= \Theta(t''-t')\left[G(t_f-t')G(t''-t_0) + G(t_f-t'')G(t'-t_0) - G(t''-t')G(T)\right] \nonumber\\
%&+\Theta(t'-t'')\left[G(t_f-t'')G(t'-t_0) + G(t_f-t')G(t''-t_0) - G(t'-t'')G(T)\right]\;. 
\end{align}
Including all overall factors, what I need to simplify is the combination:\footnote{The overall $\half$ in the coefficient of $N$ came from the symmetrization in $t'$ and $t''$.}
\begin{equation}\label{eq:O}
O(t',t'') \equiv \frac{\sigma^2}{2G(T)^2}\,M(t',t'') - \frac{\sigma^2}{2G(T)}\,N(t',t'')\;.
\end{equation}
I have put off simplifying Eq.~(\ref{eq:M}) long enough:
\begin{align}
M(t',t'') &= G(t_f-t')\,G(t_f-t'')\,e^{-2\theta t_0}\!\int_{t_0}^{t_f}\!\! dt\,e^{\,2\theta t} \label{eq:M1}\\
&-G(T)\,G(t_f-t'')\,e^{-\theta(t'+t_0)}\!\int_{t'}^{t_f}\!\! dt\;e^{\,2\theta t} \label{eq:M2}\\
&-G(T)\,G(t_f-t')\,e^{-\theta(t''+t_0)}\!\int_{t''}^{t_f}\!\! dt\;e^{\,2\theta t} \label{eq:M3}\\
&+ G(T)^2\,e^{-\theta(t'+t'')}\!\int_{t_0}^{t_f}\!\! dt\;e^{\,2\theta t}\,\Theta(t-t')\,\Theta(t-t'')\;. \label{eq:M4}
\end{align}
The integral in the last line consists of two terms, one for $t' > t''$ and the other for $t'' > t'$:
\begin{equation}\label{eq:twoConditions}
\int_{t_0}^{t_f}\!\! dt\;e^{\,2\theta t}\,\Theta(t-t')\,\Theta(t-t'') = \Theta(t'-t'') \int_{t'}^{t_f}\!\! dt\;e^{\,2\theta t} + \Theta(t''-t')\int_{t''}^{t_f}\!\! dt\;e^{\,2\theta t}\;.
\end{equation}
The integral in Eq.~(\ref{eq:M1}) is:
\begin{align}
e^{-2\theta t_0}\int_{t_0}^{t_f}\!\! dt\;e^{\,2\theta t} &= e^{-2\theta t_0} \frac{1}{2\theta}\left(e^{\,2\theta t_f} - e^{\,2\theta t_0}\right) \\
&= e^{\,\theta (t_f-t_0)}\frac{1}{2\theta}\left(e^{\,\theta(t_f-t_0)} - e^{-\theta(t_f-t_0)}\right) \\
&= e^{\,\theta T} G(T)\;.
\end{align}
This dance of moving one power of $e^{\,\ldots}$ inside the parentheses and factoring one power out, then collecting terms into the form of $G$, is how to simplify Eqs.~(\ref{eq:M2}--\ref{eq:M4}).\footnote{One of those little tricks I remember from all the way back in the first physics class I ever took in college.} 
\\\\
The integral in Eq.~(\ref{eq:M2}) is:
\begin{align}
e^{-\theta(t'+t_0)} \int_{t'}^{t_f}\!\! dt\;e^{\,2\theta t} &= e^{-\theta(t'+t_0)}\frac{1}{2\theta}\left(e^{\,2\theta t_f} - e^{\,2\theta t'}\right) \\
&= e^{\,\theta(t_f-t_0)}\frac{1}{2\theta}\left(e^{\,\theta(t_f-t')} - e^{-\theta(t_f-t')}\right) \\
&= e^{\,\theta T} G(t_f-t')\;.
\end{align}
Eq.~(\ref{eq:M3}) is the same as Eq.~(\ref{eq:M2}) with $t'$ and $t''$ swapped. With Eq.~(\ref{eq:twoConditions}), the integral in Eq.~(\ref{eq:M4}) is:
\begin{align}
&e^{-\theta(t'+t'')} \int_{t_0}^{t_f}\!\! dt\;e^{\,2\theta t}\,\Theta(t-t')\,\Theta(t-t'') = \Theta(t'-t'')\,e^{-\theta(t'+t'')}\int_{t'}^{t_f}\!\! dt\;e^{\,2\theta t} \;+\; (t'\leftrightarrow t'') \\
&\;\;= \Theta(t'-t'')\,e^{-\theta(t'+t'')}\frac{1}{2\theta}\left(e^{\,2\theta t_f} - e^{\,2\theta t'}\right)  \;+\; (t'\leftrightarrow t'') \\
&\;\;= \Theta(t'-t'')\,e^{\,\theta(t_f-t'')}\frac{1}{2\theta}\left(e^{\,\theta(t_f-t')} - e^{-\theta(t_f-t')}\right) \;+\; (t'\leftrightarrow t'') \\
&\;\;= \Theta(t'-t'')\,e^{\,\theta(t_f-t'')}\,G(t_f-t')  \;+\; (t'\leftrightarrow t'')\;.
\end{align}
Now that all of those integrals have been evaluated and simplified, I can write $M$ as:
\begin{align}
M(t',t'') &= G(t_f-t')\,G(t_f-t'')\,e^{\,\theta T}\,G(T) \\
          &\phantom{=}\;- G(T)\,G(t_f-t'')\,e^{\,\theta T}\,G(t_f-t') \\
          &\phantom{=}\;- G(T)\,G(t_f-t')\,e^{\,\theta T}\,G(t_f-t'') \\
          &\phantom{=}\;+ G(T)^2\left[\Theta(t'-t'')\,e^{\,\theta(t_f-t'')}G(t_f-t') + \Theta(t''-t')\,e^{\,\theta(t_f-t')}G(t_f-t'')\right] \\
&\nonumber\\
&= - G(T)\,e^{\,\theta T}\,G(t_f-t')\,G(t_f-t'') \\
&\phantom{=}\;+ G(T)^2\left[\Theta(t'-t'')\,e^{\,\theta(t_f-t'')}G(t_f-t') + \Theta(t''-t')\,e^{\,\theta(t_f-t')}G(t_f-t'')\right]\;.
\end{align}
Now I can simplify Eq.~(\ref{eq:O}):
\begin{align}
&O(t',t'') = \frac{\sigma^2}{2G(T)^2} M(t',t'') - \frac{\sigma^2}{2G(T)}N(t',t'') \\
&= -\frac{\sigma^2 e^{\,\theta T}}{2G(T)}\,G(t_f-t')\,G(t_f-t'') \\
&\phantom{=}\;\; + \frac{\sigma^2}{2}\left[\Theta(t'-t'')\,e^{\,\theta(t_f-t'')}G(t_f-t') + \Theta(t''-t')\,e^{\,\theta(t_f-t')}G(t_f-t'')\right] \\
&\phantom{=}\;\; -\frac{\sigma^2}{2G(T)}\,\Theta(t'-t'')\left[G(t_f-t'')\,G(t'-t_0) + G(t_f-t')\,G(t''-t_0) - G(t'-t'')\,G(T)\right] \\
&\phantom{=}\;\; -\frac{\sigma^2}{2G(T)}\,\Theta(t''-t')\left[G(t_f-t')\,G(t''-t_0) + G(t_f-t'')\,G(t'-t_0) - G(t''-t')\,G(T)\right] \\
&\nonumber\\
&= -\frac{\sigma^2 e^{\,\theta T}}{2G(T)}\,G(t_f-t')\,G(t_f-t'') \label{eq:E1}\\
&\phantom{=}\;\; +\left\{\;\frac{\sigma^2}{2G(T)}\,\Theta(t'\!-\!t'')\left[ G(T)\,e^{\,\theta(t_f-t'')} G(t_f\!-\!t') - G(t_f\!-\!t'')G(t'\!-\!t_0) - G(t_f\!-\!t')G(t''\!-\!t_0) \right] \right. \label{eq:E2}\\
&\qquad\qquad\qquad \left. +\; (t'\leftrightarrow t'') \phantom{\frac{\sigma^2}{2G(T)}} \!\!\!\!\!\!\!\!\!\!\!\!\!\!\!\!\right\} \\
&\phantom{=}\;\; +\frac{\sigma^2}{2}\left[\Theta(t'-t'')\,G(t'-t'') + \Theta(t''-t')\,G(t''-t')\right]\;.
\end{align}
The way to make sense of that mess is to take one case at a time (i.e., first consider $t' > t''$, then write down $t'' > t'$ by symmetry), and focus on simplifying the combination of Eqs.~(\ref{eq:E1}) and~(\ref{eq:E2}). Three terms have a factor of $G(t_f-t')$, so I will focus on the combination:
\begin{align}
&P(t',t'') \equiv -e^{\,\theta T} G(t_f-t'') + G(T)\,e^{\,\theta(t_f-t'')} - G(t''-t_0) \\
&= -[\dot G(T) + \theta G(T)]\, G(t_f-t'') + G(T)\, [\dot G(t_f-t'') + \theta G(t_f-t'')] - G(t''-t_0) \\
&= [-\dot G(T)\,G(t_f\!-\!t'') + G(T)\,\dot G(t_f\!-\!t'')] + \theta[-G(T)\,G(t_f\!-\!t'') + G(T)\,G(t_f\!-\!t'')] - G(t''\!-\!t_0) \\
&= G(T-(t_f-t'')) + \theta\times 0 - G(t''-t_0) \\
&= G(t''-t_0) - G(t''-t_0) = 0\;.
\end{align}
Chef's kiss. All that is left from the $t' > t''$ case of Eqs.~(\ref{eq:E1}) and~(\ref{eq:E2}) is the term without $G(t_f-t')$, leading to the result:
\begin{align}
O(t',t'') &= \frac{\sigma^2}{2G(T)}\,\Theta(t'-t'')\,[-G(t_f-t'')G(t'-t_0)] + \frac{\sigma^2}{2}\Theta(t'-t'')\,G(t'-t'') \;+\; (t'\leftrightarrow t'') \\
&= \frac{\sigma^2}{2G(T)}\Theta(t'-t'')\left[G(T)G(t'-t'') - G(t_f-t'')G(t'-t_0)\right] \;+\; (t'\leftrightarrow t'') \\
&= \frac{\sigma^2}{2G(T)}\Theta(t'-t'')\,[-G(t_f-t')G(t''-t_0)] \;+\; (t'\leftrightarrow t'')\;.
\end{align}
Putting it all together, I find the extremal action 
\begin{align}
&S(x_*(\cdot),J(\cdot);\,t_f|t_0) = \int_{t_0}^{t_f}\!\! dt\left(\frac{1}{2\sigma^2}\left\{D[x_*(t)-m]\right\}^2 - J(t)\,x_*(t)\right) \\
&= \frac{e^{\,\theta T}}{2\sigma^2 G(T)}\left[\,(x_f\!-\!m) - (x_0\!-\!m)\,e^{-\theta T}\,\right]^2 \label{eq:extremalAction0}\\
&\phantom{=}\;\;-\int_{t_0}^{t_f}\!\!dt\left\{m + \frac{1}{G(T)}\left[(x_f\!-\!m)\,G(t\!-\!t_0) + (x_0\!-\!m)\,G(t_f\!-\!t)\right] \right\} J(t) \label{eq:extremalAction1}\\
&\phantom{=}\;\;-\frac{\sigma^2}{2G(T)}\int_{t_0}^{t_f}\!\! dt'\int_{t_0}^{t_f}\!\! dt''\left[\Theta(t'\!-\!t'')\,G(t_f\!-\!t')\,G(t''\!-\!t_0) \right. \nonumber\\
&\qquad\qquad\qquad\qquad\qquad\qquad \left. +\;\Theta(t''\!-\!t')\,G(t_f\!-\!t'')\,G(t'\!-\!t_0)\right]\,J(t')\,J(t'')\;. \label{eq:extremalAction2}
\end{align}
\subsection{Source-Free Conditional Probability}\label{sec:OUsourceFree}
From the work so far, I have the source-free conditional probability (i.e., the generating function in the limit $J \to 0$) in the form
\begin{equation}
U(x_f,t_f|x_0,t_0) = C(t_f|t_0)\,e^{-S_*(x_f,t_f|x_0,t_0)}\;,
\end{equation}
with $S_*(x_f,t_f|x_0,t_0) = S(x_*(\cdot),0;\,t_f|t_0)$ given by Eq.~(\ref{eq:extremalAction0}). Note that since 
\begin{equation}
G(T)\,e^{-\theta T} = \frac{1}{2\theta}\left(1-e^{-2\theta T}\right)\;,
\end{equation}
the result agrees with Eq.~(\ref{eq:exponentOfTransformedConditionalProbability}). I will fix the overall factor by imposing the composition law of Eq.~(\ref{eq:chapmanKolmogorov}), namely
\begin{equation}\label{eq:imposeCompositionLaw}
\int_{-\infty}^\infty\!\! dx_1\;U(x_2,t_2|x_1,t_1)\,U(x_1,t_1|x_0,t_0) = U(x_2,t_2|x_0,t_0)\;.
\end{equation}
Because $U(x,t|x_0,t_0)$ is a conditional probability, the quantity 
\begin{equation}
p(x,t) \equiv \int_{-\infty}^{\infty} \!dx_0\,U(x,t|x_0,t_0)\,f(x_0)
\end{equation}
is a probability distribution with initial condition $p(x,t_0) = f(x)$. So an alternative way to fix the overall factor would be to normalize the distribution. For the special case $f(x) = \delta(x-c)$ for some constant $c$, that would imply\footnote{This is where the difference between stochastic mechanics and quantum mechanics asserts itself: If $U(x,t|x_0,t_0)$ were a quantum-mechanical transition amplitude instead of a conditional probability, it would still satisfy the composition law, but $p(x,t)$ would be a wavefunction instead of a probability distribution, and would be normalized according to $\int_{-\infty}^\infty\!dx\,|p(x,t)|^2 \equiv 1$.}
\begin{equation}\label{eq:imposeNormalization}
\int_{-\infty}^{\infty}\!\! dx\;p(x,t) = \int_{-\infty}^{\infty}\!\! dx\;U(x,t|c,t_0) \equiv 1\;,
\end{equation}
independent of $c$. I will proceed with Eq.~(\ref{eq:imposeCompositionLaw}) and then verify Eq.~(\ref{eq:imposeNormalization}). 
\\\\
Let 
\begin{equation}\label{eq:convenience}
\s(t) \equiv \s\sqrt{G(t)\,e^{-\theta t}}\;,\;\; t_{ij} \equiv t_i - t_j\;,\;\; \sigma_{ij} \equiv \sigma(t_{ij})
\end{equation}
for notational convenience. Without the normalization factor, the left-hand side of Eq.~(\ref{eq:imposeCompositionLaw}) is 
\begin{equation}\label{eq:leftHandSideOfComposition}
\int_{-\infty}^{\infty}\!\! dx_1\;e^{-\frac{1}{2\sigma_{21}^2}[(x_2-m)-(x_1-m)\,e^{-\theta t_{21}}]^2}e^{-\frac{1}{2\sigma_{10}^2}[(x_1-m)-(x_0-m)\,e^{-\theta t_{10}}]^2} = \int_{-\infty}^{\infty}\!\! dx_1\;e^{-\frac{1}{2\s_{21}^2 \s_{10}^2} K}\;,
\end{equation}
where 
\begin{align}
K &= \s_{10}^2\left[(x_2-m)-(x_1-m)\,e^{-\theta t_{21}}\right]^2 + \s_{21}^2\left[(x_1-m)-(x_0-m)\,e^{-\theta t_{10}}\right]^2 \\
&\nonumber\\
&= \s_{10}^2\left[(x_2-m)^2 - 2(x_2-m)(x_1-m)\,e^{-\theta t_{21}} + (x_1-m)^2\,e^{-2\theta t_{21}}\right] \nonumber\\
&+\s_{21}^2\left[(x_1-m)^2-2(x_1-m)(x_0-m)\,e^{-\theta t_{10}} + (x_0-m)^2\,e^{-2\theta t_{10}}\right] \\
&\nonumber\\
&= \s_{10}^2(x_2-m)^2 + \s_{21}^2(x_0-m)^2\,e^{-2\theta t_{10}} \nonumber\\
&+(\sigma_{10}^2 e^{-2\theta t_{21}} + \s_{21}^2)(x_1-m)^2 - 2[(x_2-m)\s_{10}^2 e^{-\theta t_{21}} + (x_0-m)\s_{21}^2\,e^{-\theta t_{10}}](x_1-m) \\
&\nonumber\\
&= \s_{10}^2(x_2-m)^2 + \s_{21}^2(x_0-m)^2\,e^{-2\theta t_{10}} \nonumber\\
&+(\sigma_{10}^2 e^{-2\theta t_{21}} + \s_{21}^2)\left[(x_1-m)^2-2\left(\frac{(x_2-m)\s_{10}^2 e^{-\theta t_{21}} + (x_0-m)\s_{21}^2\,e^{-\theta t_{10}}}{\sigma_{10}^2 e^{-2\theta t_{21}} + \s_{21}^2}\right) (x_1-m)\right] \\
&\nonumber\\
&= \s_{10}^2(x_2-m)^2 + \s_{21}^2(x_0-m)^2\,e^{-2\theta t_{10}} \nonumber\\
&+(\sigma_{10}^2 e^{-2\theta t_{21}} + \s_{21}^2)\left[(x_1-m) - \left(\frac{(x_2-m)\s_{10}^2 e^{-\theta t_{21}} + (x_0-m)\s_{21}^2\,e^{-\theta t_{10}}}{\sigma_{10}^2 e^{-2\theta t_{21}} + \s_{21}^2}\right)\right]^2 \nonumber\\
&-\frac{\left[(x_2-m)\s_{10}^2 e^{-\theta t_{21}} + (x_0-m)\s_{21}^2\,e^{-\theta t_{10}}\right]^2}{\sigma_{10}^2 e^{-2\theta t_{21}} + \s_{21}^2}\;. \label{eq:simplifiedExponent}
\end{align}
The first and third lines of Eq.~(\ref{eq:simplifiedExponent}) do not depend on $x_1$, and the middle line can be shifted to obtain the usual Gaussian integrand. Eq.~(\ref{eq:leftHandSideOfComposition}) becomes:
\begin{equation}\label{eq:leftHandSideOfCompositionWithoutFactor}
\int_{-\infty}^{\infty}\!\! dx_1\;e^{-\frac{1}{2\s_{21}^2 \s_{10}^2} K} = 
e^{-\frac{1}{2\s_{21}^2\s_{10}^2(\s_{10}^2 e^{-2\theta t_{21}} + \s_{21}^2)}L}\sqrt{\frac{2\pi \s_{21}^2\s_{10}^2}{\s_{10}^2 e^{-2\theta t_{21}}+\s_{21}^2}}\;,
\end{equation}
where 
\begin{align}
L &= (\s_{10}^2 e^{-2\theta t_{21}} + \s_{21}^2)[\s_{10}^2(x_2-m)^2+\s_{21}^2(x_0-m)^2e^{-2\theta t_{10}}] \nonumber\\
&\;\;- \left[(x_2-m)\s_{10}^2 e^{-\theta t_{21}} + (x_0-m)\s_{21}^2\,e^{-\theta t_{10}}\right]^2 \\
&\nonumber\\
&= (\s_{10}^2 e^{-2\theta t_{21}} + \s_{21}^2)[\s_{10}^2(x_2-m)^2+\s_{21}^2(x_0-m)^2e^{-2\theta t_{10}}] \nonumber\\
&\;\;-\left[(x_2-m)^2\s_{10}^4e^{-2\theta t_{21}} + 2(x_2-m)(x_0-m)\s_{10}^2\s_{21}^2 e^{-\theta t_{20}} + (x_0-m)^2\s_{21}^4 e^{-2\theta t_{10}}\right] \\
&\nonumber\\
&= \s_{21}^2\s_{10}^2(x_2-m)^2 + \s_{10}^2\s_{21}^2 (x_0-m)^2e^{-2\theta_{21}}e^{-2\theta t_{10}} - 2(x_2-m)(x_0-m)\s_{10}^2\s_{21}^2 e^{-\theta t_{20}} \\
&\nonumber\\
&= \s_{21}^2\s_{10}^2\left[(x_2-m)^2 + (x_0-m)^2\,e^{-2\theta t_{20}} - 2(x_2-m)(x_0-m) e^{-\theta t_{20}}\right] \\
&\nonumber\\
&= \s_{21}^2\s_{10}^2\left[(x_2-m) - (x_0-m)\,e^{-\theta t_{20}}\right]^2\;.
\end{align}
That collected itself into the expected form. Meanwhile, recall the definitions in Eq.~(\ref{eq:convenience}) and observe that\footnote{Notwithstanding footnote~\ref{ft:reproducibility}, there are some steps I see no need to show. You can also verify this using Wolfram.}
\begin{align}
\s_{10}^2 e^{-\theta t_{21}} + \s_{21}^2 e^{+\theta t_{21}} %&= \s^2\left[ G(t_{10})e^{-\theta t_{10}}e^{-\theta t_{21}} + G(t_{21})e^{-\theta t_{21}} e^{\,\theta t_{21}}\right] \\
%&= \s^2\left[G(t_{10})e^{-\theta t_{20}} + G(t_{21})\right] \\
%&= \frac{\s^2}{2\theta}\left[\left(e^{\,\theta t_{10}} - e^{-\theta t_{10}}\right) e^{-\theta t_{20}} + e^{\,\theta t_{21}} - e^{-\theta t_{21}}\right] \\
%&= \frac{\s^2}{2\theta}\left( e^{-\theta t_{21}} - e^{-\theta t_{10}} e^{-\theta t_{20}} + e^{\,\theta t_{21}} - e^{-\theta t_{21}}\right) \\
%&= \frac{\s^2}{2\theta}\left(- e^{-\theta t_{10}} e^{-\theta t_{20}} + e^{\,\theta t_{21}}\right) \\
%&= \frac{\s^2}{2\theta}\,e^{-\theta t_{10}}\left(-e^{-\theta t_{20}} + e^{\,\theta t_{21}}e^{+\theta t_{10}}\right) \\
%&= \frac{\s^2}{2\theta}\,e^{-\theta t_{10}}\left(-e^{-\theta t_{20}} + e^{\,\theta t_{20}}\right) \\
&= \s^2 e^{-\theta t_{10}} G(t_{20})\;,
\end{align}
in which case 
\begin{equation}
\s_{10}^2 e^{-2\theta t_{21}} + \s_{21}^2 = e^{-\theta t_{21}}\s^2 e^{-\theta t_{10}} G(t_{20}) = \s^2\,e^{-\theta t_{20}} G(t_{20}) = \s_{20}^2\;,
\end{equation}
which is also of the expected form. So Eq.~(\ref{eq:leftHandSideOfCompositionWithoutFactor}) becomes
\begin{equation}
\int_{-\infty}^{\infty}\!\! dx_1\;e^{-\frac{1}{2\s_{21}^2\s_{10}^2}K} = e^{-\frac{1}{2\s_{20}^2}[(x_2-m)^2+(x_0-m)^2 e^{-2\theta t_{20}}]}\sqrt{\frac{2\pi \s_{21}^2\s_{10}^2}{\s_{20}^2}}\;,
\end{equation}
which has the form required by composition. Eq.~(\ref{eq:imposeCompositionLaw}) will be satisfied if the normalization factor obeys the equation 
\begin{equation}
C(t_2|t_1)\, C(t_1|t_0)\, \sqrt{\frac{2\pi \s_{21}^2\s_{10}^2}{\s_{20}^2}} = C(t_2|t_0)\;,
\end{equation}
whose solution is 
\begin{equation}
C(t_i|t_j) = \frac{1}{\sqrt{2\pi \s_{ij}^2}}\;.
\end{equation}
The conditional probability for the source-free mean-reverting model is therefore 
\begin{equation}\label{eq:sourceFreeConditionalProbabilityForMeanReversion}
U(x,t|x_0,t_0) = \frac{1}{\sqrt{2\pi \s^2 G(t-t_0)\, e^{-\theta(t-t_0)}}}\;e^{-\frac{1}{2\s^2 G(t-t_0)\,e^{-\theta(t-t_0)}} \left[(x-m) - (x_0-m)\,e^{-\theta(t-t_0)} \right]^2}\;.
\end{equation}
I remarked in Sec.~\ref{sec:conditionalProbabilityInTransformedVariables} that, because the mean-reverting model can be transformed into Brownian motion by a sequence of local transformations, the conditional probability will end up being a normal distribution, in the sense described by the Wolfram documentation. The calculation in this appendix leading up to Eq.~(\ref{eq:sourceFreeConditionalProbabilityForMeanReversion}) is an alternative derivation  that does not rely on covariance arguments. 
\\\\
The result is clearly normalized in the sense of Eq.~(\ref{eq:imposeNormalization}): $\int_{-\infty}^{\infty}\! dx\; U(x,t|x_0,t_0) = 1$, regardless of $x_0$.
\subsubsection{Phenomenology}\label{sec:OUphenomenology}
For the most part, the purpose of this paper is to convince myself that I understand the mathematics of stochastic processes. But it is important to occasionally stop and smell the science: What does the mean-reverting model, ahem, \textit{mean} anyway?
\\\\
The phenomenology of an individual trajectory defined by 
\begin{equation}
dx(t) = \s \sqrt{dt}\,Z(t) -\theta(x(t)-m)\,dt\;,
\end{equation}
with each $Z(t)$ an independent Gaussian variate with mean 0 and variance 1, is straightforward enough, in the sense that it is clear conceptually what each update will bring.
\\\\
Eq.~(\ref{eq:sourceFreeConditionalProbabilityForMeanReversion}) completes the story by providing the distribution of values at a fixed time. Whatever the initial condition is, as $t-t_0 \to \infty$, the conditional probability becomes 
\begin{equation}\label{eq:asymptoticConditionalProbability}
U(x,t|x_0,t_0) \to \frac{1}{\sqrt{2\pi\!\left(\frac{\s^2}{2\theta}\right) }}\;
e^{-\,\frac{1}{2\left(\frac{\s^2}{2\theta}\right)}(x-m)^2}\;,
\end{equation}
a Gaussian with standard deviation $\frac{\s}{\sqrt{2\theta}}$ around the asymptotic mean.\footnote{Note that, in the standard units of finance, $x$ and $m$ are dimensionless, $\sigma$ has dimensions of $1/\sqrt{t}$, and $\theta$ has dimensions of $1/t$, in which case $\frac{\s}{\sqrt{2\theta}}$ is correctly dimensionless.}
\\\\
In Sec.~\ref{sec:conditionalProbabilityInTransformedVariables} I noted that there are two ways to view the action for the Ornstein-Uhlenbeck model: Reversion to a mean of $m$, or reversion to a mean of zero under a drift $\mu$, with the relation between the two parameters being $\mu = \theta m$. You might quibble that Eq.~(\ref{eq:asymptoticConditionalProbability}) demonstrates that the second interpretation is favored, but I would counter that it is only an asymptotic result, and we are never in the asymptotic limit. 
\\\\
Moreover, it is useful to understand how to recover Brownian motion from mean-reversion: The appropriate limit is $\theta \to 0$ and $m \to \infty$, with $\mu = \theta m$ held fixed. That may sound obvious, but it confused me under the standard interpretation. From the point of view of normalizing the initial value of $x$ to zero, the mean-reverting model could equally well be thought of as describing the tendency to revert to the value at which you started observing the process, corrected by drift. 
\subsection{Correlation Functions}
The conditional probability with a source term has the form 
\begin{equation}
U(J(\cdot);\,x_f,t_f|x_0,t_0) = U(x_f,t_f|x_0,t_0)\;e^{\,\int_{t_0}^{t_f}\!\! dt\;x_*(t)\, J(t) + \half \int_{t_0}^{t_f}\!\! dt'\int_{t_0}^{t_f}\!\! dt''\;\Delta (t',t'')\;J(t')\,J(t'')}\;,
\end{equation}
with $\Delta(t',t'')$ defined by Eq.~(\ref{eq:extremalAction2}). As I mentioned earlier, regardless of whether I view $J$ as a physical source or an artifice, I may take derivatives with respect to it to generate correlation functions.
\subsubsection{Average Path vs. Most-Probable Path}\label{sec:averageVsSaddle}
I noted in Sec.~\ref{sec:int(Dx)^2} that the contributions linear in $J$ in $S(x_*)$ drop out, leaving behind only those from $-\int\!dt\, Jx_*$. Now I will briefly elaborate on the general relation between the average path and the most-probable path.
The average path, or the one-point function, is\footnote{Probability theory has a standard notation for conditional expectations. For example, $E[x(t)|x(t_0) = x_0]$ denotes the average path for $t > t_0$ conditioned on the value $x(t_0) = x_0$. But I am not aware of a standard notation for the two-point boundary-value problem considered here, with one condition imposed in the past and one condition imposed in the future. Quantum mechanics, however, does have a standard notation for transition amplitudes with fixed endpoints: $\bra x_f,t_f|\,\hat x(t)\,|x_0,t_0\ket$, with the hat over the $x(t)$ denoting the operator that corresponds to the field. I split the difference.}
\begin{align}
\overline x(t) \equiv E[x_f,t_f|x(t)|x_0,t_0] &= \frac{1}{U(J(\cdot);\,x_f,t_f|x_0,t_0)}\frac{\del}{\del J(t)}U(J(\cdot);\,x_f,t_f|x_0,t_0) \\
&= x_*(t) + \int_{t_0}^{t_f}\!\!dt'\,\Delta (t,t')\,J(t')\;.
\end{align}
The contribution from $J$ has the standard form of linear response theory, and when $J \to 0$, I find that $\overline x(t) = x_*(t)$. That equality is a consequence of the fact that I started with a quadratic action; if the model had interactions, there would be corrections of order $\sigma^2$. See Ch.~IV.3 of Zee \cite{zeeQFT}.
\subsubsection{Two-Point Function}\label{sec:twoPointFunction}
The correlation function, or two-point function, is\footnote{As I mentioned in footnote~\ref{ft:propagator}, this is also often called the propagator, but should not be confused with the evolution operator. In probability theory, there is a distinction between a correlation function and a covariance function, distinguished by the choice of normalization. I am used to field-theoretic terminology, where one just speaks of correlation functions.}
\begin{align}
&E[x_f,t_f|\,x(t_1)\,x(t_2)\,|x_0,t_0] = \frac{1}{U(J(\cdot);\,x_f,t_f|x_0,t_0)}\frac{\del^2}{\del J(t_1)\del J(t_2)}U(J(\cdot);\,x_f,t_f|x_0,t_0) \\
&\qquad = \Delta(t_1,t_2) + \left[ x_*(t_1) + \int_{t_0}^{t_f}\!\! dt\,\Delta (t_1,t)\,J(t)\right]\left[ x_*(t_2) + \int_{t_0}^{t_f}\!\! dt'\,\Delta (t_2,t')\,J(t')\right]\;.
\end{align}
This is the souped up version of the statistician's second moment. Subtracting off the contributions from the first moment gives the generalization of the second cumulant, or variance:
\begin{equation}
E[x_f,t_f|\left(x(t_1)-\overline x(t_1)\right)\left(x(t_2)-\overline x(t_2)\right)|x_0,t_0] = \Delta (t_1,t_2)\;.
\end{equation}
To conclude this discussion of the mean-reverting model, I will verify that the asymptotic limit of the function $\Delta(t_1,t_2)$ reproduces the standard expression of exponential decay with decay rate $\theta$. Eq.~(\ref{eq:extremalAction2}) gives 
\begin{equation}
\Delta(t',t'') = \frac{\sigma^2}{G(T)}\; \Theta(t'-t'')\,G(t_f-t')\,G(t''-t_0)\;+\;(t'\leftrightarrow t'')\;.
\end{equation}
I want to take the limit $t_0 \to -\infty$ and $t_f \to +\infty$ with fixed and finite $t'$ and $t''$, so as to retain a finite correlation function deep in the region of intermediate times, where the boundary conditions should not matter.\footnote{Note that this is a different limit from that in Sec.~\ref{sec:OUphenomenology}.} Since 
\begin{equation}
G(t) = \frac{\sinh(\theta t)}{\theta} = \frac{1}{2\theta}\left(e^{\,\theta t} - e^{-\theta t}\right)\;,
\end{equation}
one term will grow and one term will decay. The appropriate limit is
\begin{equation}
\frac{G(t_f-t')\,G(t''-t_0)}{G(t_f-t_0)} \approx \frac{1}{2\theta}\,\frac{e^{\,\theta(t_f-t')}\, e^{\,\theta(t''-t_0)}}{e^{\,\theta(t_f-t_0)}} = \frac{1}{2\theta}\,e^{-\theta(t'-t'')}\;.
\end{equation}
Therefore:
\begin{equation}
\lim_{\substack{t_0\, \to\, -\infty \\ \,t_f\, \to\, +\infty}}\Delta(t',t'') = \sigma^2\left[\Theta(t'-t'')\,\frac{1}{2\theta}\,e^{-\theta(t'-t'')}\;+\;(t'\leftrightarrow t'')\right] = \frac{\sigma^2}{2\theta}\,e^{-\theta|t'-t''|}\;.
\end{equation}
That is the correct result. 
%
% References
\bibliography{references}{}
\bibliographystyle{unsrt}  % ''unsrt'' instead of ''plain'' fixed references being out of order

\end{document}